\documentclass[reprint,amsmath,amssymb,aps,prb,
superscriptaddress,nobibnotes,nofootinbib]{revtex4-2}
\usepackage{bm}
\usepackage{braket}
\usepackage{comment}
\usepackage{graphicx}
\usepackage{afterpage}
\usepackage{hyperref}
\usepackage{xcolor}
\hypersetup{colorlinks,citecolor={blue},linkcolor={red},urlcolor={cyan}}

\begin{document}

\title{Quantum simulation of field-tunable spin spectroscopy of the quantum magnet Cs$_2$CoCl$_4$ on a trapped-ion quantum computer}

\author{Elias~Kokkas}
\thanks{These authors contributed equally to this work.}
\author{Nora~Bauer}
\thanks{These authors contributed equally to this work.}
\author{Justin~Provazza}
\author{Mark~E.~Nowakowski}
\affiliation{IonQ Inc, College Park, MD, 20740, USA}
\author{Kathleen~Hamilton}
\author{Gilles~Buchs}
\affiliation{Oak Ridge National Laboratory, Oak Ridge, TN, USA}
\author{Andrew~Sornborger}
\affiliation{Information Sciences, Los Alamos National Laboratory, Los Alamos, NM 87545}
\author{Travis~S.~Humble}
\affiliation{Oak Ridge National Laboratory, Oak Ridge, TN, USA}
\author{Ananth~Kaushik}
\author{Martin~Roetteler} 
\affiliation{IonQ Inc, College Park, MD, 20740, USA}
\date{\today}

\begin{abstract}

The collective excitation spectra of quantum magnets provide a direct test of quantum computers as tools for studying strongly correlated matter. Hardware noise limits the circuit depths available for these calculations, leaving the question of how much magnetic information can survive circuit compression. Using IonQ’s 36 qubit Forte Enterprise processor, we simulate the magnetic excitation spectra and spin dynamics of a 36 site Heisenberg XXZ chain describing the quantum magnet Cs$_2$CoCl$_4$ in a transverse field using quantum circuits containing just 420 two qubit gates. The workflow combines an ancilla free local kick protocol with classical tensor network compression, encoding both ground state preparation and time evolution. The reconstructed dynamical structure factors capture the evolution from a broad two spinon continuum toward a dispersive branch with magnon character near the expected critical field. We observe an interesting field driven sign reversal of the nearest neighbor spin correlations which shows a complementary signature of this magnetic evolution. Our error mitigation methods bring the reconstructed total spectra into close agreement with density matrix renormalization group calculations, with Pearson correlation and structural similarity above 0.9 and normalized mean squared error below 0.15. We also perform comparisons with experimental inelastic neutron scattering measurements to connect the recovered excitations to the material’s magnetic response. Our results demonstrate that circuit compression and error mitigation allow noisy quantum processors to recover both collective excitations and local magnetic correlations providing a quantitative benchmark for quantum spectroscopy and a basis for extending these methods to more complex magnetic systems.

\end{abstract}

\maketitle

\section{Introduction}

Predicting how microscopic interactions produce collective quantum behavior is a central challenge in condensed matter physics and a foundational motivation for quantum computing~\cite{feynman1982simulating}. For quantum materials, excitation spectra provide a direct test through observables accessible to both numerical calculations and experiment. Classical methods such as Density matrix renormalization group (DMRG) methods describe many one-dimensional systems accurately~\cite{white1992dmrg,white1993dmrg}, although entanglement growth limits their reach at longer evolution times and in higher dimensions~\cite{schollwock2011dmrg}. Quantum Monte Carlo methods face additional restrictions from the sign problem, particularly in frustrated systems~\cite{troyer2005computational}. Therefore, for current quantum processors, the practical question is how much reliable spectral information can be recovered within the circuit depths permitted by hardware noise while maintaining accuracy compared to classical benchmark calculations and experimental measurements.

Quantum magnets offer experimentally studied realizations of interacting spin models~\cite{vasiliev2018milestones}, with spin $1/2$ degrees of freedom that map directly onto qubits. Their dynamical spin structure factor (DSF), $S(q,\omega)$, describes correlations in momentum and frequency space and determines the magnetic response measured by inelastic neutron scattering (INS)~\cite{VanHove1954TimeDependentCB}. Dispersion relations and the distribution of spectral weight reveal how excitations propagate and how their character changes with the magnetic state~\cite{caux2005computation,lake2005longitudinal}. Quantum algorithms are well suited to access these spectra through measurements of spin correlations~\cite{francis2020quantum,sun2021quantum}. Lee \emph{et al.}~\cite{lee2026benchmarking} demonstrated quantitative comparisons with neutron measurements of KCuF$_3$ on superconducting processors, combining approximate quantum compilation with additional Trotter steps at longer evolution times.

Here, we demonstrate the effectiveness of quantum hardware as we study how much of the field evolution of a magnetic spectrum can be recovered while keeping the quantum circuit budget fixed. We consider Cs$_2$CoCl$_4$, whose dominant interactions are described by an anisotropic spin $1/2$ XXZ chain in a transverse field. The model supports a gapless Luttinger liquid at zero field, a gapped N\'eel ordered regime at intermediate fields, and a polarized phase above the critical field. Increasing the field redistributes spectral weight from a broad spinon continuum toward a dispersive branch with magnon character. To follow this evolution, we study a 36 site chain at four transverse fields on IonQ's 36 qubit Forte Enterprise processor (Appendix~\ref{sec:ion-traps-qpus}). Tensor network calculations and variational circuit compression~\cite{rudolph2024decomposition} encode the complete preparation of each time-evolved target state in a separate circuit. Using an ancilla free local kick protocol to compute the retarded spin response~\cite{baez2020dynamical}, we execute circuits on quantum hardware containing only 420 two qubit gates, including at the longest evolution of 20 Trotter steps, while maintaining a compression fidelity averaged over the spin channels above $0.6$. The resulting spectra test whether useful magnetic information survives appreciable errors in the many-body state.

Establishing that reliability requires checking both the recovered response and the analysis used to obtain it. We benchmark our quantum hardware-derived spectra (after error mitigation)  against correction-vector DMRG (DMRG-CV), which calculates the response directly in the frequency domain and reproduces the principal features of the neutron spectra~\cite{laurell2021quantifying}. A second classical benchmark uses a DMRG ground state followed by second-order Trotter time evolution (DMRG+Trotter), matching the chain length, time discretization, and observation window targeted by the hardware circuits. These comparisons assess the accuracy of the quantum solver, while comparisons with INS assess how well the model describes the material. Detailed mitigation comparisons examine how corrections to spectral shape and intensity affect agreement with these references. Our validation framework uses complementary spectral and physical metrics~\cite{buchs2026validation} and reports propagated measurement uncertainty separately from sensitivity to processing choices. At the two fields with a complete total response, error mitigation yields Pearson correlations and structural similarity above $0.9$ against DMRG-CV, with normalized mean squared errors below $0.15$ while Pearson correlations with INS exceed $0.85$ (Table~\ref{tab:metrics}). Agreement across these measures shows that the recovered excitation structure survives both circuit compression and hardware noise.

The field sweep also connects the changing excitation spectrum to local magnetic behavior. We also perform direct measurements of equal time correlations which reveal a sign reversal of the connected nearest neighbor $zz$ correlator, providing a test that does not require spectral reconstruction. Complementary analyses of local spin dynamics, static magnetic response, and entanglement examine what further physical information can be recovered from the measurements. Finite observation times nevertheless limit access to low-frequency structure and closely spaced excitations. We therefore investigate finite-memory extension (FME) of the measured records and compare its predictions with longer DMRG+Trotter calculations, identifying both improvements in spectral reconstruction and their dependence on field and observable. This tests a route to longer-time information without additional quantum circuits, while establishing the validation needed before extending compressed circuit spectroscopy to more complex spin systems.

\section{Model Hamiltonian}

The low-energy magnetic behavior of Cs$_2$CoCl$_4$ is described by an effective spin $1/2$ XXZ Heisenberg chain. Within the isolated chain approximation, we consider $L$ sites with open boundaries and label the sites by $j=0,\ldots,L-1$. The Hamiltonian is
\begin{equation}\label{eq:xxzhamiltonian}
\small{H = J\sum_{j=0}^{L-2} \left({S^x_jS^x_{j+1}+S^y_jS^y_{j+1}+\Delta S^z_jS^z_{j+1}}\right)+  h_x\sum_{j=0}^{L-1} S^x_j } \ ,
\end{equation}
Here $S^\alpha_j=\sigma^\alpha_j/2$ is the spin operator at site $j$, with $\sigma^\alpha_j$ the corresponding Pauli operator and $\alpha\in\{x,y,z\}$. The parameters $J>0$, $\Delta$, and $h_x$ specify the antiferromagnetic exchange, exchange anisotropy, and transverse field in energy units, respectively. With the positive field term in Eq.~\eqref{eq:xxzhamiltonian}, positive $h_x$ favors negative magnetization along $x$. We use $\hbar=1$ and set the lattice spacing to unity.

We take $J=0.23\,\mathrm{meV}$ and $\Delta=0.25$, following the model used to interpret the neutron scattering measurements of Cs$_2$CoCl$_4$~\cite{laurell2021quantifying}. For comparison with the experimental field $B$, we adopt the rescaling $h_x/J \simeq 0.764\,B/\mathrm{T}$, which aligns the model critical field $h_c\simeq1.604J$ with the experimental value $B_c\simeq2.1\,\mathrm{T}$~\cite{laurell2021quantifying}. Hardware measurements use $L=36$, with  classical calculations at $L=100$.

The phase diagram of the isolated chain provides the reference for interpreting the finite systems studied here. In the thermodynamic limit, the model at $h_x=0$ is a gapless Luttinger liquid with a continuum of spinon excitations. For $0<h_x<h_c$, it supports a gapped phase with N\'eel order along $y$. The excitation gap closes at $h_x=h_c$, beyond which the system enters a gapped phase polarized along the $x$ axis, with a spectrum dominated by a dispersing magnon branch~\cite{laurell2021quantifying}. The hardware field sweep extends from zero field to the vicinity of this expected critical point.

We characterize the magnetic excitations through the DSF. For the diagonal channels studied here, its inelastic component is related to the retarded Green's function (RGF) in thermal equilibrium by
\begin{equation}
\label{eq:dsf}
S^{\alpha\alpha}(q,\omega)
=
-\frac{1}{2\pi}
\left[1+n_B(\omega)\right]
\operatorname{Im}
G^{\alpha\alpha}_{\mathrm{ret}}(q,\omega),
\end{equation}
where $\omega>0$, $n_B(\omega)=[\exp(\omega/k_B T)-1]^{-1}$ is the Bose--Einstein distribution. Our calculations target zero temperature, for which $n_B(\omega)=0$ at positive frequencies. The normalization in Eq.~\eqref{eq:dsf} corresponds to the physical spin operators $S^\alpha_j=\sigma^\alpha_j/2$ and the Pauli operator convention for the RGF defined below.

The momentum and frequency response is estimated from the real space response to a perturbation at a central site $j_c$:
\begin{equation}
\label{eq:rgf}
\small{G^{\alpha\beta}_{\mathrm{ret}}(q,\omega)
\overset{\footnotemark}{=}
\Delta t
\sum_{j=0}^{L-1}
\sum_{k=0}^{N_{\mathrm{s}}-1}
e^{i\omega t_k}
e^{-iq(j-j_c)} 
\times
G^{\alpha\beta}_{\mathrm{ret}}(j,j_c,t_k)} \ ,
\end{equation}
\footnotetext[\value{footnote}]{Equation~\eqref{eq:rgf} gives the unwindowed finite record transform; the temporal window used in spectral reconstruction is described in the error mitigation Appendix~\ref{sec:appendix-a}.}

Here $N_{\mathrm{s}}$ is the number of sampled times $t_k$, with uniform spacing $\Delta t$, and $j_c=17$ for the $L=36$ chain. A source near the centre of the chain approximates the translationally invariant bulk response and reduces the sensitivity to the boundaries. An even-length open chain has no single central site, so we apply the kick on one of the two central sites, $j_c=17$, and compute the RGF from it. The DSF is then symmetrized using the reflection symmetry $j\rightarrow L-1-j$ of the Hamiltonian,
$S(q,\omega)\rightarrow \frac{1}{2}\left[S(q,\omega)+S(-q,\omega)\right]$
which is equivalent to averaging over the two central source sites, $17$ and $18$, at no additional circuit cost. 

At zero temperature, the real space RGF is
\begin{equation}
\label{eq:rgf-com}
G^{\alpha\beta}_{\mathrm{ret}}(j,j_c,t)
=
-\frac{i}{2}\Theta(t)
\braket{
\psi_{\mathrm{gs}}
|
[\sigma^\alpha_j(t),\sigma^\beta_{j_c}(0)]
|
\psi_{\mathrm{gs}}
},
\end{equation}
where $\ket{\psi_{\mathrm{gs}}}$ is the ground state of Eq.~\eqref{eq:xxzhamiltonian} and $\Theta(t)$ enforces causality. Operators evolve in the Heisenberg picture according to
$\sigma^\alpha_j(t)=e^{iHt}\sigma^\alpha_j e^{-iHt}$.
This retarded response is the quantity accessed by the local kick protocol described in the following section.

\section{Quantum computation of the spin response}
\label{sec: qsim}

We reconstruct the retarded spin response in
Eq.~\eqref{eq:rgf-com} using a local kick protocol combined
with classical circuit compression. The underlying
protocol prepares an approximation to the ground state,
applies a local spin rotation at the central site, and
evolves the resulting state before measuring the spin
response throughout the chain. Classical tensor network
calculations construct the target states at each sampled
time, and variational optimization maps each target to
a circuit with a fixed gate budget. The quantum processor
executes these compressed state preparation circuits and
provides the measurement samples used to reconstruct the
response.

\begin{figure*}[!htbp]
    \centering
    \includegraphics[width=\linewidth]{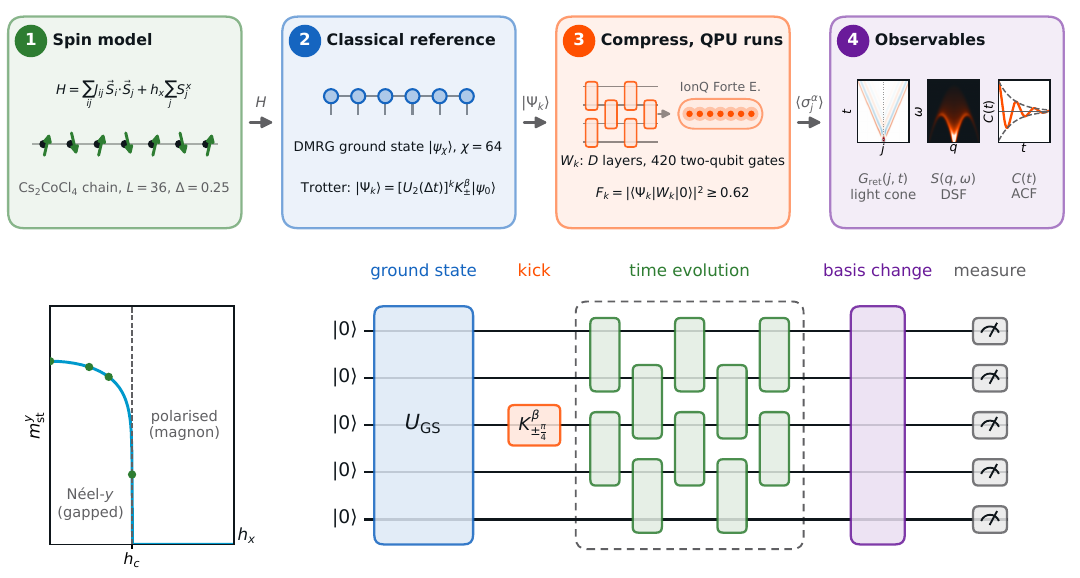}
    \caption{
    Workflow for calculating the magnetic response.
    Top: the model of Cs$_2$CoCl$_4$ is a spin $1/2$
    XXZ chain with $J=0.23\,\mathrm{meV}$,
    $\Delta=0.25$, and $L=36$ sites.
    Classical DMRG and second order Trotter evolution
    supply reference states, which are approximated
    by circuits composed of $D$ staircase layers of
    nearest neighbor two qubit blocks.
    The compressed circuits are executed on
    IonQ Forte Enterprise, and the measurements are
    used to reconstruct the retarded Green's function,
    DSF, and local
    autocorrelations.
    Bottom left: reference phase diagram of the model.
    The staggered order parameter $m^y_{\mathrm{st}}$
    vanishes at the critical field separating the
    N\'eel ordered regime from the phase polarized
    along $x$.
    Bottom right: the underlying circuit before
    compression, comprising ground state preparation,
    a local kick
    $K^\beta_\pm=\exp(\mp i\pi\sigma^\beta_{j_c}/4)$,
    and Trotter evolution.
    Measurements of the $x$, $y$, and $z$ components
    use basis rotations $H$, $HS^\dagger$, and
    $\mathbb{I}$, respectively, followed by measurement
    in the computational basis.
    }
    \label{fig:qa-pipeline}
\end{figure*}

Figure~\ref{fig:qa-pipeline} summarizes this workflow.
We consider $L=36$ sites and evolution times
$t_k=k\Delta t$, with $\Delta t=0.6/J$ and up to
$N_t=20$ steps, reaching $Jt=12$. A final basis rotation
allows all qubits to be measured in the selected spin
component. Measurements of the $zz$ response are available at all four transverse fields and the availability of the other channels is specified in Appendix~\ref{sec:appendix-b}.

\subsection{Ground state preparation}

We first obtain a classical ground state approximation
using DMRG~\cite{white1992dmrg,white1993dmrg,schollwock2011dmrg}. The optimization minimizes the energy over matrix
product states (MPS) $\ket{\psi_\chi}$ of bond dimension
$\chi$, with convergence monitored through the energy
and discarded weight during successive sweeps.
We use $\chi=64$ for the $L=36$ chain, and a more detailed analysis can be found in Figure \ref{fig:dmrg_con} in Appendix \ref{sec:appendix-a}.

An exact encoding of this MPS would require a circuit
whose sequential blocks act on up to
$\lceil\log_2\chi\rceil+1$
qubits~\cite{schon2005sequential,ran2020encoding}.
Decomposing these blocks into elementary gates imposes
a substantial cost. We therefore approximate the
reference state using a circuit of $D$ staircase layers
of nearest neighbor two qubit blocks,
\begin{equation}
    \ket{\psi_0}=W\ket{0}^{\otimes L},
    \qquad
    W=\prod_{d=1}^{D}\prod_j V^{(d)}_{j,j+1},
    \label{eq:prep-circuit}
\end{equation}
and maximize its fidelity with the DMRG reference,
\begin{equation}
    F_0=
    \left|
    \bra{\psi_\chi}W\ket{0}^{\otimes L}
    \right|^2.
    \label{eq:prep-fidelity_0}
\end{equation}
Here $V^{(d)}_{j,j+1}\in\mathrm{U}(4)$ acts on the
neighboring sites $(j,j+1)$, with the products ordered
according to the staircase circuit.

The circuit is optimized classically by updating each
block from its tensor network environment through a
polar decomposition. Layers are added sequentially,
with all existing layers optimized again after each
addition~\cite{evenbly2009algorithms,rudolph2024decomposition}.
Each block can be decomposed into at most three controlled-NOT (CNOT)
gates together with single qubit
rotations~\cite{vidal2004universal}.
An MPS truncated to bond dimension $\chi\leq2$ can also
provide an inexpensive initial circuit through an
exact encoding with two qubit
blocks~\cite{ran2020encoding}.

\begin{figure}[!htbp]
    \centering
    \includegraphics[width=\linewidth]{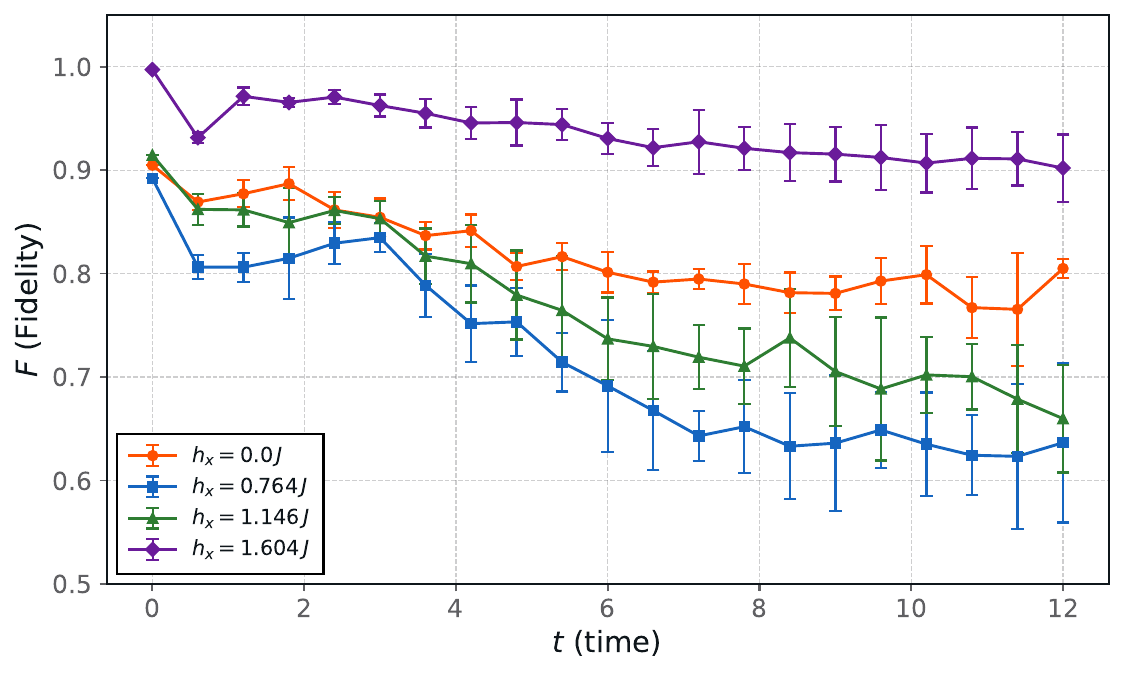}
    \caption{
    Fidelity of the compressed state preparations
    relative to their classical targets.
    At $t_k=0$, the plotted quantity is the ground
    state fidelity $F_0$ in
    Eq.~\eqref{eq:prep-fidelity_0}.
    At nonzero times $t_k=k\Delta t$, with
    $\Delta t=0.6/J$, the curves show the fidelity
    $F_k$ in Eq.~\eqref{eq:fidelity}, averaged with equal weights over the three spin channels $(xx,yy,zz)$.
    The ground state circuits use
    $(252,\ 315,\ 232,\ 77)$ two qubit gates at
    $h_x/J=(0,\ 0.764,\ 1.146,\ 1.604)$, respectively.
    Every nonzero time point uses $420$ two qubit
    gates.
    These fidelities quantify how closely the compressed circuits approximate their classical target states before hardware execution.}
    \label{fig:fidelity}
\end{figure}

For the four fields executed on quantum hardware
$h_x/J=(0,\ 0.764,\ 1.146,\ 1.604)$, we explored up to
$D=4$ layers and up to $124$ Evenbly--Vidal
optimization sweeps. Selecting circuits according to
their fidelity and gate cost gives ground state
preparations with $(252,\ 315,\ 232,\ 77)$ two qubit
gates, respectively. Their fidelities are shown at
$t_k=0$ in Figure~\ref{fig:fidelity} $(F_0=0.905, \ 0.892, \ 0.915, \ 0.997)$. The resulting state $\ket{\psi_0}$ provides the initial
state for the kick protocol.

\subsection{Kick protocol}

The local kick converts the retarded commutator into spin expectation
values that can be measured without an ancilla. This follows the
local kick protocol of Baez \emph{et
al.}~\cite{baez2020dynamical}, in which a local $\pm\pi/2$ rotation of
the source spin replaces the interferometric (Hadamard-test) measurement
of the commutator. At the source site $j_c$, we apply
\begin{equation}
    K^\beta_\pm
    =
    e^{\mp i\pi\sigma^\beta_{j_c}/4}
    =
    \frac{\mathbb{I}\mp i\sigma^\beta_{j_c}}{\sqrt{2}},
\end{equation}
corresponding to rotations by $\pm\pi/2$ about the
$\beta$ axis. Under exact evolution
$U(t)=e^{-iHt}$, the resulting states are
\begin{equation}
    \begin{aligned}
    \ket{\psi_\pm(t)}
    &=U(t)K^\beta_\pm\ket{\psi_0}
    =\frac{U(t)}{\sqrt{2}}
      \left(\ket{\psi_0}\mp i\ket{\psi_1}\right),
    \end{aligned}
    \label{eq:kicked-state}
\end{equation}
where
$\ket{\psi_1}=\sigma^\beta_{j_c}\ket{\psi_0}$.
The dependence of these states on $\beta$ is
suppressed for clarity.

Writing the measured expectation as
$\langle\sigma^\alpha_j(t)\rangle_\pm
=\bra{\psi_\pm(t)}\sigma^\alpha_j\ket{\psi_\pm(t)}$,
we obtain
\begin{equation}
    \begin{aligned}
    \langle\sigma^\alpha_j(t)\rangle_\pm
    ={}&
    \frac{1}{2}\langle\sigma^\alpha_j(t)\rangle_0
    +\frac{1}{2}\langle\sigma^\alpha_j(t)\rangle_1\\
    &\quad\pm
    \operatorname{Im}
    \braket{\psi_0|\sigma^\alpha_j(t)|\psi_1},
    \end{aligned}
    \label{eq:three-terms}
\end{equation}
with
$\langle\sigma^\alpha_j(t)\rangle_a
=\bra{\psi_a}\sigma^\alpha_j(t)\ket{\psi_a}$
for $a=0,1$.

To distinguish the prepared state from the exact
ground state, let
$G^{\alpha\beta}_{\mathrm{ret},0}$ denote the response
in Eq.~\eqref{eq:rgf-com} evaluated in
$\ket{\psi_0}$. For $t>0$,
\begin{equation}
    \begin{aligned}
    &\operatorname{Im}
    \braket{\psi_0|\sigma^\alpha_j(t)|\psi_1}\\
    &\qquad=
    -\frac{i}{2}
    \braket{
    \psi_0|
    [\sigma^\alpha_j(t),\sigma^\beta_{j_c}(0)]
    |\psi_0
    }\\
    &\qquad=
    G^{\alpha\beta}_{\mathrm{ret},0}(j,j_c,t).
    \end{aligned}
\end{equation}
The identity is exact for the chosen initial state;
its approximation to the ground state response
depends on the quality of $\ket{\psi_0}$.

Taking the difference between the two kick
measurements cancels the first two terms of
Eq.~\eqref{eq:three-terms}, giving
\begin{equation}
    G^{\alpha\beta}_{\mathrm{ret},0}(j,t)
    =
    \frac{1}{2}
    \left[
    \langle\sigma^\alpha_j(t)\rangle_+
    -
    \langle\sigma^\alpha_j(t)\rangle_-
    \right].
    \label{eq:pm}
\end{equation}
Here and below, the fixed source coordinate $j_c$
is suppressed. Thus, two kick preparations provide
the response for a general pair of spin components.

For diagonal channels, spin symmetry can reduce the
measurement requirement to a single kick.
At $h_x=0$, the Hamiltonian commutes with the global
spin rotations
$P_\gamma=\prod_i\sigma^\gamma_i$.
If $\ket{\psi_0}$ is an eigenstate of a conserved
$P_\gamma$ with $\gamma\neq\alpha$, then
$\sigma^\alpha_j(t)$ is odd under this symmetry.
Both background terms in
Eq.~\eqref{eq:three-terms} therefore vanish, yielding
\begin{equation}
    G^{\alpha\alpha}_{\mathrm{ret},0}(j,t)
    =
    \langle\sigma^\alpha_j(t)\rangle_+.
    \label{eq:single}
\end{equation}
At zero field this construction applies to all
three diagonal channels. A nonzero transverse field
preserves $P_x$, so the same reduction applies to
the $yy$ and $zz$ channels, while the $xx$ response
requires the difference in Eq.~\eqref{eq:pm}.
The single kick identities assume that the initial
state lies in the appropriate symmetry sector;
departures caused by compression or hardware noise
can leave residual background contributions.

\subsection{Time evolution}
\label{subsec: time-evo}

We construct the time evolution targets using a
second order product formula. The Hamiltonian in
Eq.~\eqref{eq:xxzhamiltonian} is separated into
contributions from even bonds, odd bonds, and the
transverse field,
$H=H_{\mathrm{e}}+H_{\mathrm{o}}+H_{\mathrm{f}}$.
Each contribution contains mutually commuting terms.
A symmetric evolution step is
\begin{equation}
    \begin{aligned}
    U_2(\Delta t)={}&
    e^{-iH_{\mathrm{o}}\Delta t/2}
    e^{-iH_{\mathrm{f}}\Delta t/2}
    e^{-iH_{\mathrm{e}}\Delta t}\\
    &\times
    e^{-iH_{\mathrm{f}}\Delta t/2}
    e^{-iH_{\mathrm{o}}\Delta t/2},
    \end{aligned}
    \label{eq:strang}
\end{equation}
with accumulated error
$\mathcal{O}(t\Delta t^2)$ at fixed Hamiltonian
parameters~\cite{suzuki1991general,childs2021trotter}.

The bond terms generate two qubit operations that
can each be synthesized with at most three CNOT
gates~\cite{vidal2004universal}, whereas the field
terms require only single qubit rotations.
Although adjacent half steps can be combined
between successive applications of
$U_2(\Delta t)$, the gate count for direct evolution
still grows with the number of time steps.

To maintain the available gate budget, we treat
each evolved state as a separate state preparation
problem. The target for a given kick and time is
\begin{equation}
    \ket{\Psi_{\pm,k}}
    =
    \left[U_2(\Delta t)\right]^k
    e^{\mp i\pi\sigma^\beta_{j_c}/4}
    \ket{\psi_\chi}.
    \label{eq:target-state}
\end{equation}
These targets are constructed classically as MPS.
For each target, we optimize a staircase circuit
$W_k$ of nearest neighbor two qubit
blocks~\cite{ran2020encoding,rudolph2024decomposition}
to maximize
\begin{equation}
    F_k=
    \left|
    \bra{\Psi_{\pm,k}}W_k\ket{0}^{\otimes L}
    \right|^2.
    \label{eq:fidelity}
\end{equation}
The dependence of $W_k$ and $F_k$ on the field,
spin component, and kick sign is implicit.
As for the ground state preparation, the
optimization proceeds through classical sweeps
over the circuit blocks.

For $L=36$, every nonzero time point is represented
by a circuit containing $420$ two qubit gates,
with a minimum channel-averaged compression fidelity of $0.62$
over the studied time window.
Each circuit prepares an approximation to the
complete state in Eq.~\eqref{eq:target-state},
incorporating the initial preparation, kick,
and evolution into a single optimized circuit.
The quantum processor therefore samples each
time point through an independent state
preparation with the same gate budget.

The fidelity $F_k$ quantifies agreement with the
classical target used in the optimization.
It is distinct from errors in the DMRG reference,
the Trotter approximation, truncation during
classical evolution, and hardware execution.
Likewise, the fixed quantum gate count over the
studied time window does not include the cost
of classical target construction and circuit
optimization. This separation provides a
controlled basis for assessing which magnetic
observables remain accessible after compression
and execution on the quantum processor.

\section{Results}\label{sec:iv}

\subsection{Classical benchmarks for the magnetic response}
\label{sec:classical-L100}

\begin{figure}[!htbp]
    \centering
    \includegraphics[width=1\linewidth]{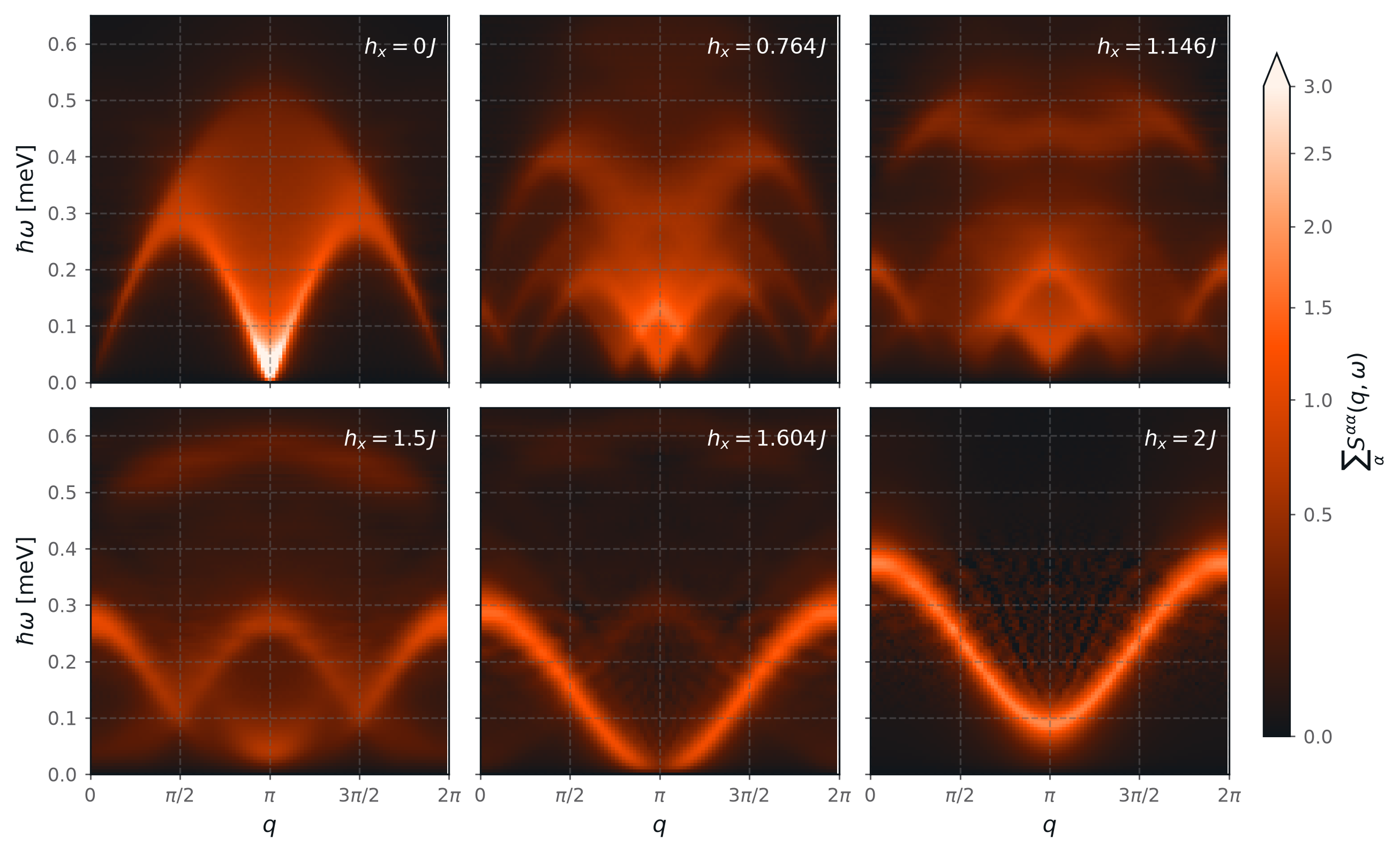}
    \caption{Total DSF of  the $L=100$ site anisotropic $(\Delta=0.25)$ open chain at $h_x/J = (0, \ 0.764, \ 1.146, \ 1.5,  \ 1.604, \ 2)$ obtained from DMRG-CV calculations with spectral broadening \(\eta=0.1J\). The spectrum evolves from a gapless two-spinon continuum at zero-field to a single gapped magnon beyond the critical point.} 
    \label{fig:dsf-60}
\end{figure}

The classical benchmarks establish both the magnetic response that the QPU should recover and the accuracy attainable under the conditions of the hardware experiment. DMRG is well suited to one-dimensional spin systems~\cite{white1992dmrg,schollwock2011dmrg}, and we use two complementary calculations. Our primary reference for the Heisenberg XXZ model is DMRG-CV, which evaluates the response directly in frequency space through the correction-vector method~\cite{Alvarez2009}. We use the DSF data of Laurell \emph{et al.}~\cite{laurell2021a} for an $L=100$ open chain with $\Delta=0.25$ and $m=1000$ retained states. This calculation avoids Trotter errors and finite real-time cutoffs and its fixed broadening $\eta=0.1\,J$ is retained throughout, without additional broadening to match the hardware observation window. Comparison with QPU+EM therefore addresses solver validation i.e. how accurately the quantum calculation reproduces the response of the same Hamiltonian~\cite{buchs2026validation}. Comparison with INS addresses model validation i.e. whether that Hamiltonian describes the magnetic response of Cs$_2$CoCl$_4$, accounting for experimental broadening, background, and statistical uncertainty.

Figure~\ref{fig:dsf-60} shows the physical evolution captured by this reference across the four hardware fields ($h_x/J=(0,\ 0.764,\ 1.146,\ 1.604)$) and the additional values $h_x/J=1.5$ and $2$, below and above the expected transition near $h_x/J=1.604$. At zero field, the gapless two-spinon continuum reflects the fractionalization of a spin excitation into propagating spinons. By $h_x=2J$, the response is dominated by a gapped magnon branch in the field-polarized regime. Following this evolution beyond the hardware field range clarifies the origin of the branch observed near the largest measured field and identifies the changes in excitation character that the QPU calculations should recover despite noise.

To assess recovery of these features under the conditions of the 36-qubit experiment, we use DMRG+Trotter as a second benchmark. We obtain the $L=36$ ground state with the DMRG implementation of TeNPy~\cite{hauschild2024tensor} at bond dimension $\chi=64$, apply the kick at $j_c=17$, and evolve through 20 uncompressed second-order Trotter steps of size $\Delta t=0.6/J$, reaching $Jt=12$. The open boundaries, reflection averaging, and spectral reconstruction match those used for the hardware data. With no circuit compression, this calculation retains the finite size, Trotter approximation, and observation window of the dynamics targeted by the hardware circuits. It therefore tests how closely QPU+EM recovers those dynamics, while DMRG-CV provides the independent frequency-domain benchmark for the model response. Because the DMRG-CV reference retains its original broadening without resolution matching, differences in spectral widths can reflect the distinct broadening and finite observation window as well as errors in the reconstructed response.

\subsection{Local spin dynamics}

\begin{figure}[!htbp]
    \centering
    \includegraphics[width=\linewidth]{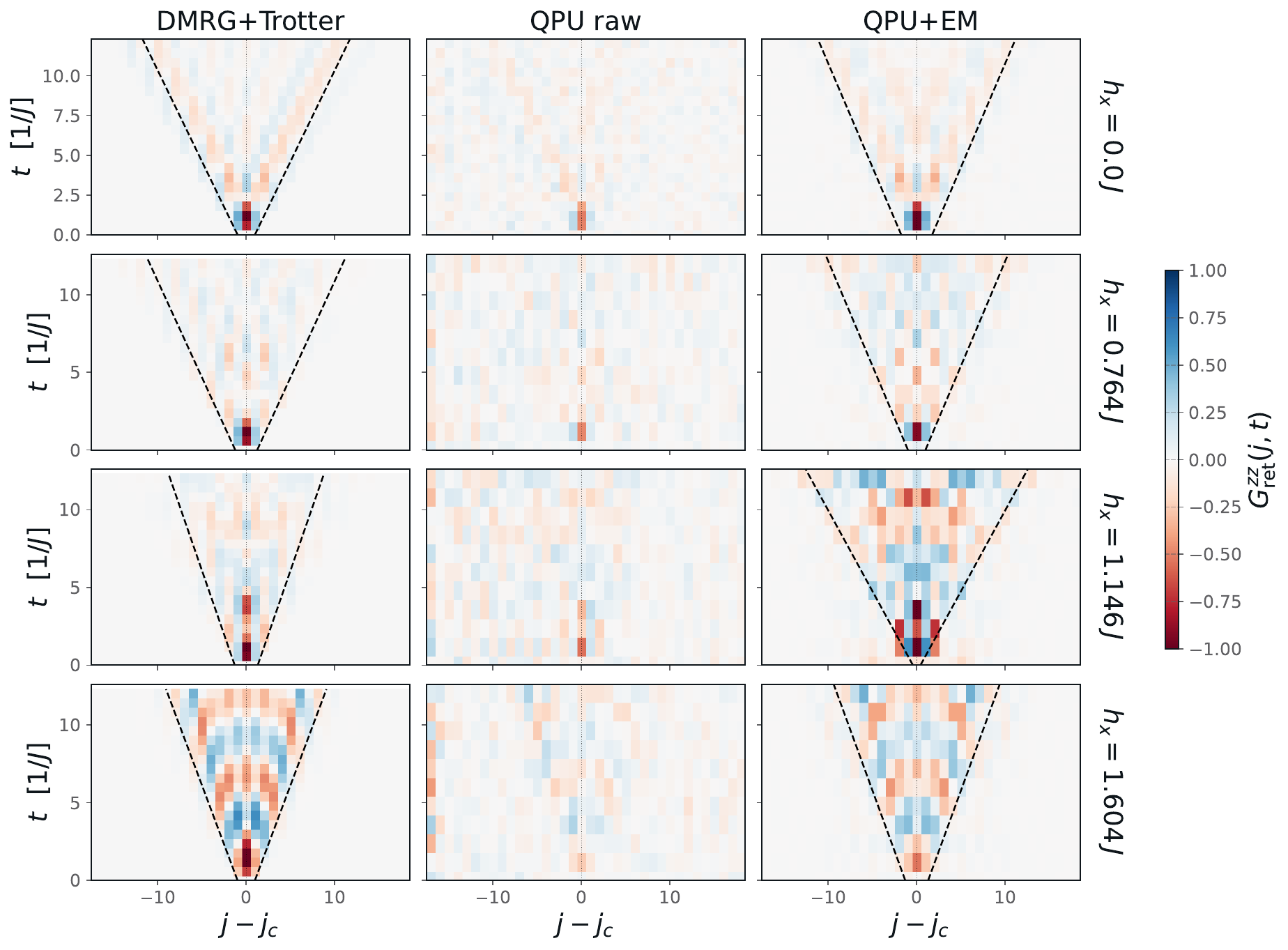}
    \caption{Retarded Green's function $G^{zz}_{ret}(j,t)$ for the $L=36$ site  anisotropic $(\Delta=0.25)$ open chain with a kick at $j_c=17$. Columns show DMRG+Trotter $(L=36)$, raw hardware, and error mitigated hardware results. }
    \label{fig:hw-rgf-zz}
\end{figure}

\begin{figure}[!htbp]
    \centering
    \includegraphics[width=\linewidth]{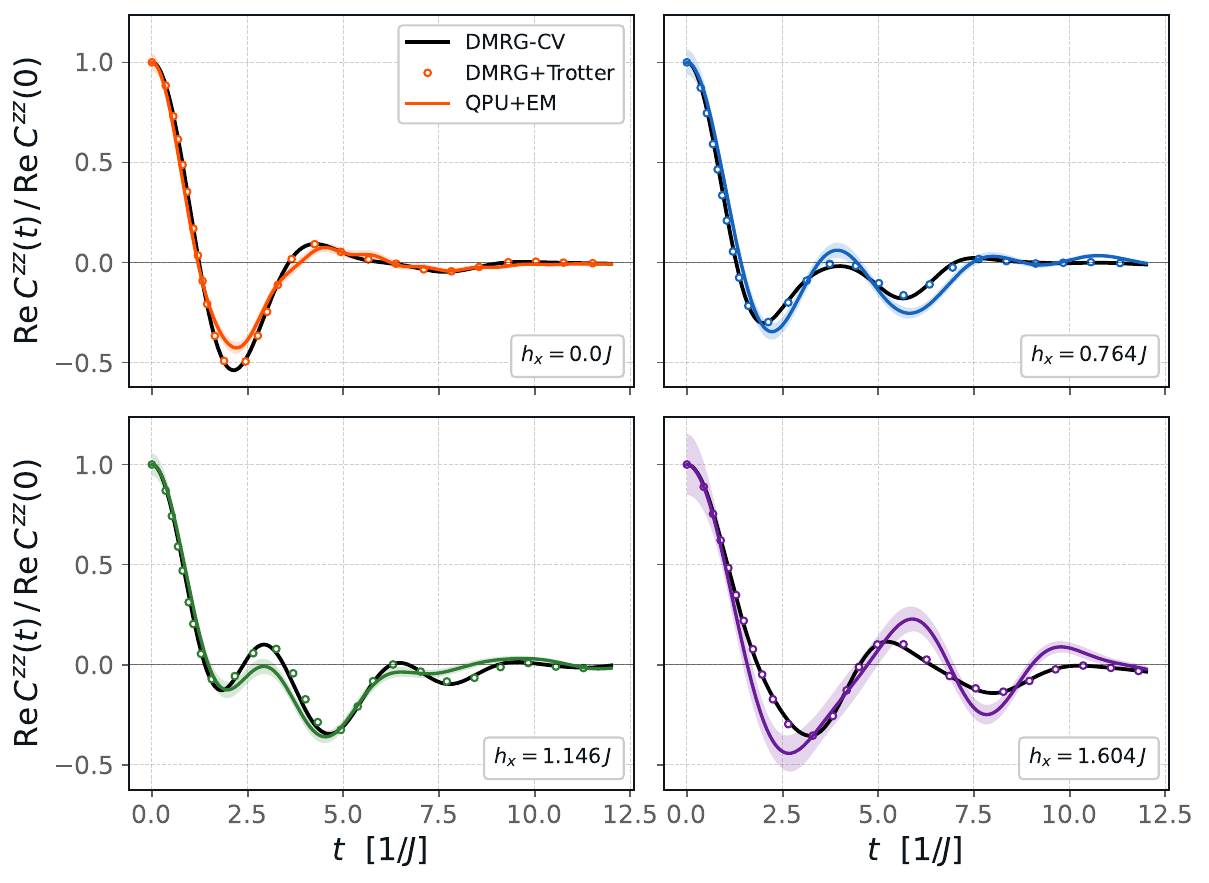}
    \caption{Real part of the on-site spin-spin autocorrelation function $\text{Re}C^{zz}(t) \equiv \text{Re}C^{z}(j_c,t) = \text{Re}\braket{\sigma^z_{j_c}(t)\sigma^z_{j_c}(0)}$  at site $j_c=17$. Each panel is normalized by the equal-time modulus $\vert C^{zz}(0) \vert = \sqrt{[\text{Re}C^{zz}(0)]^2+ [\text{Im}C^{zz}(0)]^2} = \vert \text{Re} C^{zz}(0) \vert$. Each column refers to one of the following field values: $h_x/J=(0,\ 0.764, \ 1.146, \ 1.604  )$. Solid black curves represent DMRG-CV results, solid gray curves the DMRG+Trotter results and solid colored curves represent error mitigated hardware data  with shadow bands corresponding to one standard deviation of the QPU data. }
    \label{fig:hw-acf-zz}
\end{figure}

The response to a local kick shows how magnetic excitations evolve in space and time as the transverse field changes. The retarded Green's function $G^{zz}_{\mathrm{ret}}(j,t)$ follows this response from the source site $j_c=17$. Both the light-cone maps in Figure~\ref{fig:hw-rgf-zz} and the autocorrelations in Figure~\ref{fig:hw-acf-zz} use this single-source response without momentum-space symmetrization. Figure~\ref{fig:hw-rgf-zz} compares the $L=36$ DMRG+Trotter result with hardware measurements before and after error mitigation. The changing spatial extent and oscillation pattern connect the field evolution of the spectrum to the propagation of a local spin perturbation. The principal oscillations remain visible in the hardware data, and background suppression through error mitigation makes their structure clearer.

The on-site autocorrelation $C^{zz}(t)=\langle\sigma^z_{j_c}(t)\sigma^z_{j_c}(0)\rangle$ provides a complementary view of the dynamics at the source. To construct the DMRG-CV reference for this comparison, we start from the frequency-domain spectra $S^{\alpha\alpha}(k,\omega)$ of Ref.~\cite{laurell2021a}, which contain no time record. We invert the momentum transform in Eq.~\eqref{eq:rgf} and use the zero-temperature fluctuation--dissipation relation to reconstruct the retarded Green's function of Eq.~\eqref{eq:rgf-com} on the hardware time grid. For this time-domain reconstruction, we remove the Lorentzian broadening $\eta=0.1\,J$ of the correction-vector spectra, apply a half-DPSS taper with time-bandwidth product $NW=1.5$, and extract the autocorrelation and connected correlators using Eqs.~\eqref{eq:con-acf}--\eqref{eq: con-sigma-acf}. Figure~\ref{fig:hw-acf-zz} shows the real part of the reconstructed autocorrelation, normalized by its equal-time modulus $|C^{zz}(0)|$.

At zero field, the initial decay is consistent with dephasing across the broad excitation continuum. As the field increases, changes in the decay and oscillations reflect the redistribution of spectral weight. These temporal signatures are recovered by the error mitigated hardware: across the measured fields, the curves have Pearson correlations of at least $0.952$ with DMRG-CV and MSE values no greater than $0.02891\pm0.00755$. Normalization removes a common amplitude factor, including the mirror correction, so this agreement tests the temporal shape of the response independently of that calibration. The comparisons show that error mitigation preserves access to the field-dependent magnetic dynamics despite hardware noise, allowing the spectral evolution to be understood through the underlying spin correlations. The other spin components and further details of the correlation reconstruction are presented in Appendix~\ref{sec:appendix-b}, Section~\ref{sec:supp-correlations}.

\subsection{Sign reversal of short range spin correlations}

\begin{figure}[!htbp]
    \centering
    \includegraphics[width=\linewidth]{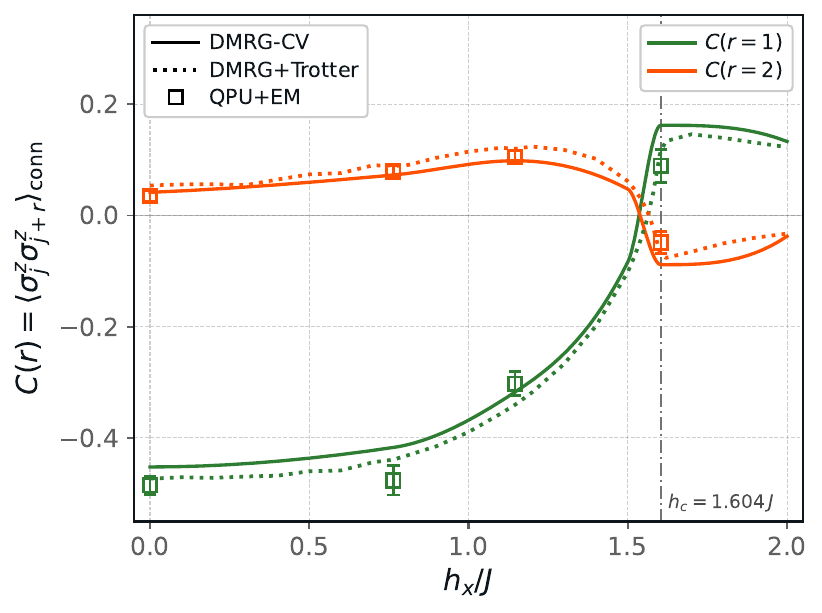}
    \caption{The connected correlator $C(r) = \langle\sigma^z_j\sigma^z_{j+r}\rangle - \langle\sigma^z_j\rangle\langle\sigma^z_{j+r}\rangle$, averaged over pairs with both sites more than three away from the kick site, from the $k = 0$ circuits run on IonQ Forte Enterprise. The error mitigated hardware measurements are compared with DMRG ground-state correlations. Note that for this plot at $t=0$, the QPU+EM only includes the mirror circuit EM without the added the lineshape/spectral EM strategies. }
    \label{fig:hw-corr-zz}
\end{figure}

Equal time ground-state correlations provide a direct test of how the transverse field reorganizes local spin fluctuations. Using the hardware data at $t=0$ ($k=0$), before any time evolution, we evaluate the connected correlation
\begin{equation}
  C(r)=\frac{1}{|\mathcal{P}_r|}
  \sum_{(j,j+r)\in\mathcal{P}_r}
  \left[
    \langle\sigma^z_j\sigma^z_{j+r}\rangle
    -\langle\sigma^z_j\rangle\langle\sigma^z_{j+r}\rangle
  \right],
  \label{eq:cr}
\end{equation}
where $\mathcal{P}_r$ contains pairs separated by $r$ sites, with both sites more than three sites away from the kick site (beyond this the correlations decay very quickly). Figure~\ref{fig:hw-corr-zz} compares these error mitigated measurements with DMRG ground-state correlations.

At the three lower measured fields, the nearest neighbor correlation $C(r=1)$ is negative, while the next nearest neighbor correlation $C(r=2)$ is positive. Both signs reverse at $h_x=1.604J$: the hardware gives $C(r=1)=+0.0889\pm0.0294$ and $C(r=2)=-0.0488\pm0.0197$ after error mitigation, with the latter following the negative value $-0.070$ obtained from DMRG-CV. The positive nearest neighbor correlation is already present in the raw data, so the observed reversal does not depend on error mitigation. Quantitative checks of this signal are given in Appendix~\ref{sec:appendix-b}, Section~\ref{sec:supp-correlations}.

Both DMRG benchmarks place the nearest neighbor zero crossing at $h_x=1.561J$, below the expected critical field $h_c\simeq1.604J$. The reversal therefore identifies a change from anticorrelated to positively correlated fluctuations of neighboring $z$ spin components that precedes the expected transition in the classical reference. Far into the phase polarized along $x$, the state approaches a product state and the connected $zz$ correlation tends to zero. The hardware thus resolves a local signature of the changing magnetic state, with the signs of the short range correlations following the classical predictions despite noise. Because these signs are extracted directly from equal time measurements, this observation is independent of the finite observation window and the reconstruction of a frequency spectrum.

\subsection{Magnetic excitation spectra and validation}\label{sec:4e}

\begin{figure*}[!htbp]
    \centering
    \includegraphics[width=0.9\linewidth]{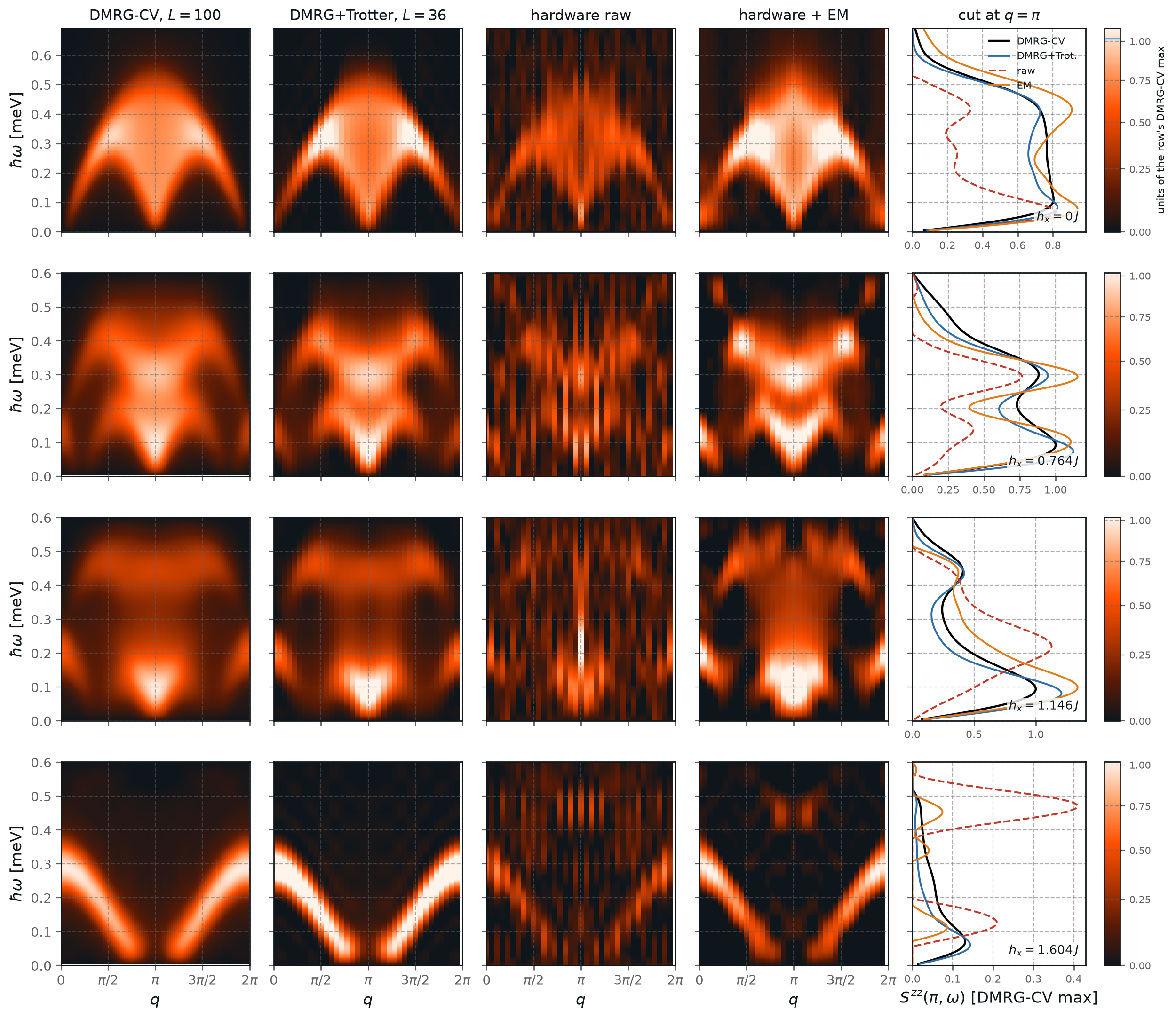}
    \caption{Longitudinal DSF $S^{zz}(q,\omega)$ of the anisotropic chain with $\Delta=0.25$. Rows correspond to $h_x/J=0$, $0.764$, $1.146$, and $1.604$, from top to bottom. Columns show, from left to right, DMRG-CV ($L=100$, $\eta=0.1\,J$), DMRG+Trotter ($L=36$), raw hardware results, error mitigated hardware results, and cuts at $q=\pi$ through the four spectra. The hardware results use $L=36$ at all four fields. The colorbars for each row are set by the DMRG-CV $L=100$ result peak value being one. }
    \label{fig:dsf_zz}
\end{figure*}

The QPU+EM spectra follow the redistribution of magnetic spectral weight with field. Figure~\ref{fig:dsf_zz} compares raw and error mitigated $S^{zz}(q,\omega)$ with both classical references at all four fields. At zero field, the easy-plane anisotropy gives dominant transverse responses $S^{xx}=S^{yy}$ by $U(1)$ symmetry, concentrated near $q=\pi$ along the lower boundary of the two spinon continuum. The $S^{zz}$ response emphasizes the upper boundary and peaks near $q=\pi/2$. A transverse field breaks this symmetry: at $h_x=0.764J$, features in the reference $S^{xx}$ spectrum move upward in energy, while those in $S^{yy}$ and $S^{zz}$ move downward. DMRG+Trotter places the $S^{zz}$ maximum near $q=\pi$ at $h_x=1.146J$ and toward $q=0$ at $h_x=1.604J\simeq h_c$, with a branch dispersing downward toward a weaker peak at $q=\pi$. This concentration of weight into a dispersive branch, accompanied by a weaker higher-energy band, is consistent with increasing magnon character near the transition~\cite{laurell2021quantifying}. QPU+EM recovers the principal redistribution and emerging branch despite compression and hardware noise. The other spin components are shown in Appendix~\ref{sec:appendix-b}.

\begin{figure*}[!htbp]
  \centering
    \includegraphics[width=\textwidth]{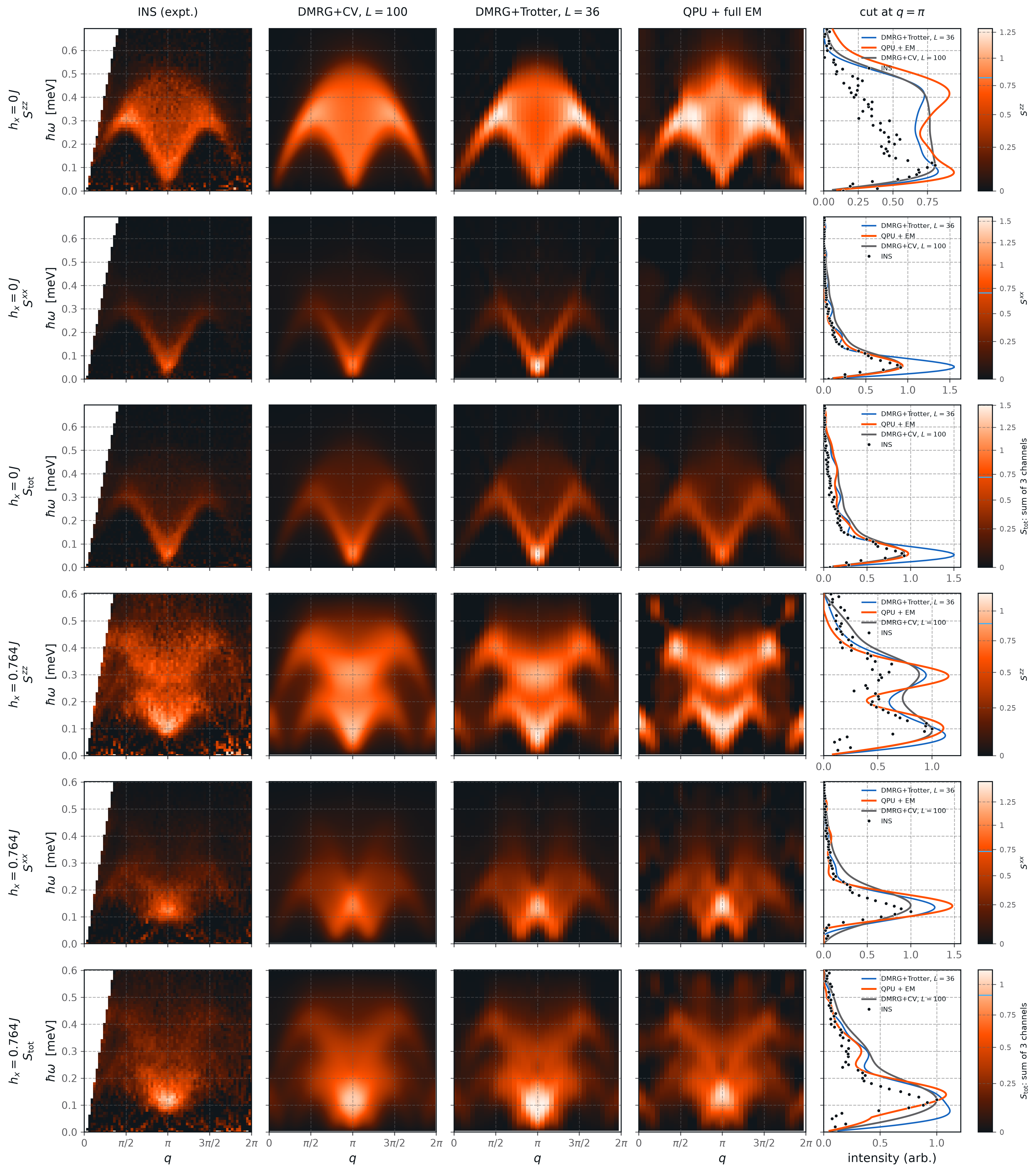}
  \caption{Total dynamical spin structure factor
    $S_{\mathrm{tot}}(q,\omega) = S^{xx} + S^{yy} + S^{zz}$ of the $L = 36$
    XXZ chain ($\Delta = 0.25$, $J = 0.23$~meV), measured on IonQ Forte Enterprise.
    Rows show the longitudinal channel $S^{zz}$, one transverse channel $S^{xx}$, and the total $S_\mathrm{tot} = S^{xx} + S^{yy} + S^{zz}$, at $h_x = 0$ (upper three) and $h_x = 0.764 J$ (lower three). The field mapping is $h_x[J] = 0.764 \times B[T]$, so these correspond to 0 T and 1 T. The colorbars for each row are set by the DMRG-CV $L=100$ result peak value being one. Columns: 1) Inelastic neutron scattering from \cite{laurell2021quantifying} ($70$ mK, inferred channel resolution) ii) DMRG-CV ($L=100$, $\eta=0.1\,J$) iii) DMRG+Trotter (L = 36) iv) QPU results with the full EM stack v) Constant-$q$ cuts at $q=\pi$ through columns (i)-(iv). }
  \label{fig:hw-total-dsf}
\end{figure*}

For quantitative validation, we use the total response $S_{\mathrm{tot}}=S^{xx}+S^{yy}+S^{zz}$ at $h_x/J=0$ and $0.764$, corresponding to $0$ and $1\,\mathrm{T}$ under the adopted rescaling (Fig.~\ref{fig:hw-total-dsf}). At zero field, $S^{yy}$ follows from $S^{xx}$ by symmetry; all three components were measured at $h_x=0.764J$. The primary benchmark is $L=100$ DMRG-CV~\cite{Alvarez2009} with fixed broadening $\eta=0.1J$, without additional broadening to match the hardware observation window. This comparison tests solver accuracy within the shared XXZ model~\cite{buchs2026validation}. The $L=36$ DMRG+Trotter baseline matches the hardware system size, Trotter approximation, and observation window, testing recovery of the targeted dynamics after compression, hardware execution, and mitigation.

Against DMRG-CV, the total spectra have Pearson correlations and structural similarity above $0.9$, normalized mean squared errors below $0.15$, and fitted amplitudes within about $20\%$ of the reference (Table~\ref{tab:metrics}). The dominant spectral pattern and overall intensity therefore survive hardware execution. At $q=\pi$, the total-spectrum peak offsets are below the nominal time-window scale of $0.06\,\mathrm{meV}$ for $t_T=12/J$; the taper gives an effective resolution near $0.10\,\mathrm{meV}$. The principal excitation energies agree within these scales, while the zero-field continuum bandwidth remains underestimated. These complementary metrics distinguish recovery of the dominant excitations from the more demanding reconstruction of the full spectral weight and boundaries. The longitudinal response follows the classical field evolution across all four fields, although intensity and shape errors grow near $h_c$ (Table~\ref{tab:metricschan}).

\begin{table*}[!htbp]
\caption{\label{tab:metrics}
Comparison of the total dynamical spin structure factor $S_{\mathrm{tot}}$
at the two fields where all three spin components were measured:
QPU+EM against $L=100$ DMRG-CV (left pair), and DMRG-CV against
the inelastic neutron scattering data of Ref.~\cite{laurell2021quantifying}
(right pair). All DMRG-CV reference spectra retain the fixed broadening
$\eta=0.1\,J$, without additional broadening to match the
observation window of the hardware data. The QPU+EM DSF values for these comparisons are unnormalized and left in absolute units. 
The \emph{Ideal} column gives the value of each metric for perfect
agreement. Entries report comparison metrics or differences,
rather than absolute values of the underlying physical observables.
The energy windows extend to $0.693$ and $0.602$~meV
at $h_x/J=0$ and $0.764$, respectively.
For the left pair, each $\Delta$ denotes QPU+EM minus DMRG-CV;
a negative value therefore means that the hardware result is smaller.
The indented uncertainties for this pair are standard deviations
from $160$ shot-noise draws propagated through the full mitigation
chain and account only for statistical noise.
An additional cross-pipeline comparison on the neutron grid,
using the comparison conventions of this table and restricted
to cells with reported INS intensities and positive uncertainties,
gives Pearson correlations of $0.878$ and $0.850$ between QPU+EM
and INS, compared with $0.891$ and $0.841$ between DMRG-CV and
INS on the same cells. }
\begin{ruledtabular}
\begin{tabular}{lccccc}
 & & \multicolumn{2}{c}{QPU $+$ EM \emph{vs} DMRG$+$CV, $L=100$}
 & \multicolumn{2}{c}{DMRG$+$CV, $L=100$ \emph{vs} INS} \\
\cline{3-4}\cline{5-6}
 & & \multicolumn{2}{c}{$h_x\,[J]$} & \multicolumn{2}{c}{$h_x\,[J]$} \\
\cline{3-4}\cline{5-6}
Quantity & Ideal & $0$ & $0.764$ & $0$ & $0.764$ \\
 & & ($0$~T) & ($1$~T) & ($0$~T) & ($1$~T) \\
\hline
$r$ & $1$ & $0.920$ & $0.962$ & $0.891$ & $0.841$ \\
\quad  &  & $\pm0.009$ & $\pm0.007$ & $\pm0.007$ & $\pm0.011$ \\
$A$ & $1$ & $0.833$ & $0.869$ & $2.536$ & $2.570$ \\
\quad  &  & $\pm0.025$ & $\pm0.029$ & $\pm0.008$ & $\pm0.011$ \\
MSE & $0$ & $0.0251$ & $0.0064$ & $0.4495$ & $0.2670$ \\
\quad  &  & $\pm0.0026$ & $\pm0.0011$ & $\pm0.0334$ & $\pm0.0403$ \\
\quad normalized, MSE$/\langle S_\mathrm{ref}^2\rangle$ & $0$ & $0.122$ & $0.049$ & $0.179$ & $0.194$ \\
\quad  &  & $\pm0.013$ & $\pm0.008$ & $\pm0.010$ & $\pm0.016$ \\
$W_1$ & $0$ & $0.0370$ & $0.0185$ & $0.0299$ & $0.0169$ \\
\quad  &  & $\pm0.0114$ & $\pm0.0044$ & $\pm0.0010$ & $\pm0.0014$ \\
SSIM & $1$ & $0.919$ & $0.914$ & $0.786$ & $0.676$ \\
\quad  &  & $\pm0.013$ & $\pm0.019$ & $\pm0.013$ & $\pm0.024$ \\
\hline
$\Delta\langle\hbar\omega\rangle_{q=\pi}$ [meV] & $0$ & $0.0298$ & $-0.0111$ & $0.0136$ & $-0.0037$ \\
\quad  &  & $\pm0.0125$ & $\pm0.0088$ & $\pm0.0029$ & $\pm0.0044$ \\
$\Delta\langle\hbar\omega\rangle_{q=\pi/2}$ [meV] & $0$ & $-0.0176$ & $0.0026$ & $-0.0107$ & $-0.0024$ \\
\quad  &  & $\pm0.0060$ & $\pm0.0141$ & $\pm0.0103$ & $\pm0.0116$ \\
$\Delta\hbar\omega_\mathrm{pk}$, $q=\pi$ [meV]$^{\dagger}$ & $0$ & $0.0402$ & $0.0125$ & $-0.0200$ & $0.0100$ \\
\quad  &  & $\pm0.0017$ & $\pm0.0032$ & $\pm0.0050$ & $\pm0.0069$ \\
$\Delta\sigma_\omega$, $q=\pi$ [meV] & $0$ & $0.0106$ & $-0.0266$ & $0.0283$ & $0.0125$ \\
\quad  &  & $\pm0.0108$ & $\pm0.0074$ & $\pm0.0025$ & $\pm0.0028$ \\
\hline
$\Delta R_{>}$ ($\hbar\omega_b=0.20$~meV) & $0$ & $0.013$ & $0.014$ & $0.052$ & $-0.016$ \\
\quad  &  & $\pm0.020$ & $\pm0.016$ & $\pm0.003$ & $\pm0.004$ \\
\hline
$\Delta$gap, $q=\pi$ [meV]$^{\dagger}$ & $0$ & $0.0000$ & $0.0025$ & $0.0000$ & $-0.0100$ \\
\quad  &  & $\pm0.0000$ & $\pm0.0018$ & $\pm0.0047$ & $\pm0.0135$ \\
\quad at a $20\%$ threshold$^{\dagger}$ & $0$ & $0.0050$ & $0.0075$ & $0.0000$ & $-0.0100$ \\
\quad  &  & $\pm0.0000$ & $\pm0.0036$ & $\pm0.0069$ & $\pm0.0215$ \\
$\Delta B$ [meV]$^{\dagger}$ & $0$ & $-0.2058$ & $0.0225$ & $-0.0200$ & $0.0000$ \\
\quad  &  & $\pm0.0562$ & $\pm0.0296$ & $\pm0.0195$ & $\pm0.0719$ \\
\end{tabular}
\end{ruledtabular}
\end{table*}

This spectral agreement persists despite appreciable errors in the compressed many-body states. Figure~\ref{fig:fidelity} measures their fidelity to the DMRG ground state at $t=0$ and the DMRG+Trotter targets at later times, before hardware execution. The fidelity for the compressed circuits averaged equally over the three spin channels falls to about $60\%$ in
the most difficult case. Nevertheless, the mitigated spectra retain the principal continuum and dispersive branch despite local distortions near $q=\pi$ (Figs.~\ref{fig:dsf_zz} and~\ref{fig:hw-total-dsf}). For these observables, substantial circuit compression remains compatible with useful spectroscopy.

The error mitigation methods we implemented to refine the raw QPU data are crucial to address both attenuation and the spurious background that obscures the excitations. Readout correction reduces measurement bias, while reflection averaging and temporal denoising reduce fluctuations. We utilized a spatial mask that restricts the response to the expected propagation region of the local perturbation, thereby suppressing background contributions to the spectrum. We also executed mirror circuit measurements on IonQ Forte Enterprise QPU for an independent hardware estimate of attenuation to determine an amplitude correction exponent for each field and spin component~\cite{Urbanek2021}. Background suppression recovers the distribution of spectral weight, and amplitude calibration compensates for attenuation. Further details about the error mitigation procedures used can be found in Appendix~\ref{sec:appendix-a}.

A positive overall rescaling leaves the Pearson correlation of an individual spin component unchanged, although different rescalings of the components can change the shape of their summed spectrum. Pearson correlation alone therefore provides an incomplete assessment of QPU accuracy. We also evaluate spectral shape, intensity, and excitation energies using complementary metrics, following the validation framework of Ref.~\cite{buchs2026validation}. The full suite of correlation measurements provide complementary checks of this magnetic response. Normalized QPU+EM autocorrelations follow DMRG-CV across the measured fields, testing the temporal shape independently of a common amplitude factor (Fig.~\ref{fig:hw-acf-zz}). The nearest neighbor $zz$ sign reversal is already visible in the raw equal time measurements and tests the field-dependent spin correlations independently of spectral reconstruction (Fig.~\ref{fig:hw-corr-zz}).

Comparison with INS tests whether the XXZ model describes the magnetic response of Cs$_2$CoCl$_4$, providing model validation~\cite{buchs2026validation}. The agreement in broad spectral features and field dependence connects the recovered QPU excitations to the neutron response (Fig.~\ref{fig:hw-total-dsf})~\cite{laurell2021quantifying}. We compare with the measured total INS spectrum while the displayed experimental spin components are inferred from it. Counting noise, other measurement uncertainties, instrumental resolution, and the isolated-chain approximation all contribute to discrepancies, so quantitative QPU accuracy is assessed against the classical references. Table~\ref{tab:metrics} makes this distinction explicit by placing QPU+EM--DMRG-CV and DMRG-CV--INS comparisons side by side. Table~\ref{tab:chi2} adds residual diagnostics after fitting an overall intensity scale and applying the stated uncertainty weights, using the conventions in Appendix~\ref{sec:appendix-b}, Section~\ref{sec:supp-validation}.

The principal continuum and dispersive features remain recognizable when the temporal taper, denoising rank, spatial mask parameter, and mirror exponent are varied (Table~\ref{tab:distortion}). Several metrics change by amounts comparable to or greater than their shot-noise uncertainties while absolute intensity is sensitive to both mirror calibration and the taper. We therefore report processing sensitivity separately from propagated shot noise. The stable excitation structure supports its physical interpretation, while the remaining amplitude and bandwidth errors identify limits of the reconstruction. Within these limits, compressed circuits with error mitigation recover magnetic behavior relevant to both classical benchmarks and neutron experiments.

\subsection{Extending the response to longer times}
\label{sec:results-continuation}

\begin{figure*}[!htbp]
\centering
\includegraphics[width=0.9\textwidth]{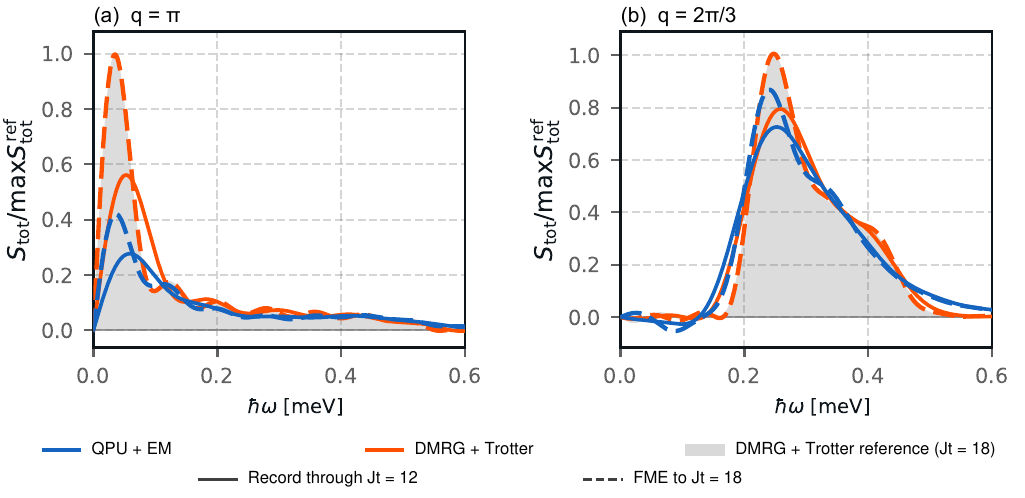}
\caption{
Finite-memory extension of the total spectrum for $L=36$ at zero field,
at (a) $q=\pi$ and (b) $q=2\pi/3$.
Blue and orange curves show QPU+EM and DMRG+Trotter results, respectively.
Solid curves use records through $Jt=12$; dashed curves include
continuation to $Jt=18$. The gray shaded spectrum is calculated
directly from the DMRG+Trotter record through $Jt=18$.
Each panel is normalized by the maximum of its reference spectrum.
The continued QPU+EM record is a prediction beyond the measured window.
See Appendix~\ref{sec:supp-finite-memory} for validation and method details.
}
\label{fig:fme-spectra-main}
\end{figure*}

Finite observation times on noisy quantum hardware can obscure low-energy excitations and closely spaced spectral features. Finite-memory extension (FME) uses the oscillatory structure of the measured response to estimate its continuation, fitting a linear model separately to each momentum and spin component and extending the record from $Jt=12$ to $Jt=18$ without additional quantum circuits. In the zero-field $L=36$ DMRG+Trotter benchmark, continuation reduces the relative $\ell_2$ error of the total spectrum from about $31\%$ to $1.5\%$ over all momenta and energies up to $0.6\,\mathrm{meV}$, compared with the directly calculated longer record. At $q=\pi$, the peak shifts from $53$ to $35\,\mu\mathrm{eV}$, matching the longer reference and recovering low-energy structure obscured by the shorter observation window (Figure~\ref{fig:fme-spectra-main}). The hardware comparisons in Table~\ref{tab:fme} also show substantial gains at zero field: continuation raises the Pearson correlation of the total spectrum from $0.89$ to $0.95$, reduces its normalized mean squared error by about $46\%$, and brings the $q=\pi$ peak into agreement with the DMRG+Trotter reference on the comparison grid, removing an offset of approximately $25\,\mu\mathrm{eV}$. The improvement is not uniform across spectral measures or fields. The zero-field rms width error increases slightly, while at $h_x=0.764J$ the total normalized mean squared error remains near $0.10$ and the Pearson correlation decreases from $0.93$ to $0.91$. These results indicate that short QPU records can retain useful information about longer-time dynamics, but the accuracy of its recovery depends on the field and observable. Further validation against longer records could make continuation useful for resolving spectral structure when hardware noise limits the accessible evolution time. Here, the extended QPU response remains a prediction because no independent hardware record is available beyond $Jt=12$. Appendix~\ref{sec:supp-finite-memory} describes the method, validation, and limitations.

\section{Conclusions and Outlook}

We have demonstrated that strongly compressed quantum circuits can recover distinct signatures of collective magnetic excitations and local spin correlations on noisy quantum hardware by simulating a 36 site transverse field XXZ chain describing Cs$_2$CoCl$_4$ on IonQ Forte Enterprise. An ancilla free kick protocol, circuit compression, and error mitigation recover the field evolution from a two spinon continuum toward a dispersive branch using only 420 two qubit gates per circuit. Comparisons against classical targets and experimental neutron scattering measurements~\cite{laurell2021quantifying} establish the robustness of reconstructed observables to circuit compression and hardware noise. With error mitigation, the total hardware spectra achieve Pearson correlations and structural similarity above $0.9$, and normalized mean squared errors below $0.15$ against DMRG-CV at both benchmarked fields (Table~\ref{tab:metrics}). Agreement across these complementary measures shows that substantial circuit compression remains compatible with quantitatively useful spectroscopy on noisy quantum hardware.

The current study offers several useful avenues for further research. One open question is whether circuit compression can be made more efficient by optimizing the dynamical observables of interest. The recovery of spectral features from imperfect state approximations suggests including errors in the relevant correlation functions in the compression objective. Testing observables and parameter regimes excluded from the optimization would establish whether this approach improves predictive accuracy at a given circuit budget. This leads to an assessment of the actual accuracy required from a quantum simulation to evaluate a specific observable of interest while accounting for experimental noise. Identifying a dispersive branch can tolerate intensity errors that would bias estimates of spectral weight or entanglement. The supplementary QFI analysis (Appendix~\ref{sec:appendix-d}) motivates testing whether sum rules and improved calibration can constrain normalization and reconstruction errors sufficiently to support entanglement bounds for the inferred ground state response. Another interesting open question is about how far the observation time can be extended at a fixed quantum circuit budget. Preparing a separately compressed state at each time limits the circuit size, but retains the classical cost of generating and optimizing the targets. A useful test to examine the extendability of observation time would be to evaluate the longest interval over which spectral accuracy can be maintained as these classical costs and compression errors increase (up to some acceptable limit). The FME results (Appendix~\ref{sec:finite-memory-extension}) also show the need to validate predictions beyond the measured interval.

Coupled chains and two dimensional magnets offer concrete systems in which to pursue these questions. Neutron studies of fractional excitations~\cite{Mourigal2013}, coupled chains in KCuF$_3$~\cite{Lake2005KCuF3}, and spinon confinement in SrCo$_2$V$_2$O$_8$~\cite{Bera2017Confinement} provide targets for studying how interchain coupling reorganizes magnetic spectra. Models of $\alpha$-RuCl$_3$~\cite{Banerjee2017RuCl3}, herbertsmithite~\cite{Han2012Herbertsmithite}, and YbMgGaO$_4$~\cite{Paddison2017YbMgGaO4} extend this scope to frustration and competing interactions. The kick protocol remains applicable, and direct connectivity between trapped ion qubits can reduce routing costs. Progress will nevertheless require circuits that represent greater entanglement, larger bond dimensions where classical compression remains practical, and preparation strategies that reduce reliance on classically generated target states. Extending the present validation framework to these problems would create a route toward distinguishing competing magnetic models and identifying fields, momenta, and energy ranges where new measurements would be most informative.

Improvements in noise control can progressively expand this scientific reach. Beyond the mitigation and calibration used here, randomized compiling can suppress the effects of coherent errors~\cite{Wallman2016RandomizedCompiling}, dynamical decoupling sequences can suppress dephasing during idle periods~\cite{Biercuk2009DynamicalDecoupling},  while virtual distillation and related purification methods provide additional possibilities, with the cost of independently prepared copies and additional measurement operations~\cite{Huggins2021VirtualDistillation,Koczor2021Purification}. Because mitigation can incur sampling costs that grow exponentially with circuit depth under relevant noise models~\cite{Takagi2022Limits}, quantum error detection is a natural complementary direction. Distance two codes can identify faults and reject flagged measurements, as demonstrated for expressive quantum circuits~\cite{Self2024Iceberg}. Dynamical spin correlations offer a concrete application for such methods: syndrome measurements, supplemented by appropriate physical symmetry checks, can diagnose errors without requiring a classical reference. Further development toward full quantum error correction would support longer observation times and more demanding dynamics. With these advances, the experimentally anchored benchmarks established here can develop into increasingly predictive simulations that help interpret magnetic spectra, discriminate between microscopic descriptions, and guide the design of future experiments.

\section{Acknowledgments}

This material is based upon work supported by the U.S. Department of Energy, Office of Science, National Quantum Information Science Research Centers, Quantum Science Center.

This manuscript has been authored by UT-Battelle, LLC, under Contract No. DE-AC0500OR22725 with the U.S. Department of Energy and LANL unclassified release designation - LA-UR-26-28437. The United States Government retains and the publisher, by accepting the article for publication, acknowledges that the United States Government retains a non-exclusive, paid-up, irrevocable, world-wide license to publish or reproduce the published form of this manuscript, or allow others to do so, for the United States Government purposes. The Department of Energy will provide public access to these results of federally sponsored research in accordance with the DOE Public Access Plan.

\bibliography{refs}
\appendix
\newpage

\section{Trapped-ion quantum hardware}
\label{sec:ion-traps-qpus}

All hardware experiments were carried out on IonQ's Forte Enterprise processor, based on the trapped-ion architecture described in Ref.~\cite{Chen2024-ik, chernyshev2026pathfinding}. The computational register comprises a chain of 36 $^{171}\mathrm{Yb}^{+}$ ions, with each qubit encoded in the ground state hyperfine levels. Laser ablation supplies the neutral atoms, which are photoionized and confined in a surface linear Paul trap housed within a vacuum package. Individual ions are addressed by beams steered independently using acousto-optic deflectors (AODs)~\cite{Kim2008-uj,Pogorelov2021-pu}. This addressing scheme reduces beam alignment errors and crosstalk, while automated calibration and control protocols maintain stable, high-fidelity operation across the ion chain.

The processor implements a universal gate set consisting of arbitrary single-qubit rotations and native $R_{ZZ}$-type entangling gates. These operations use two-photon Raman transitions driven by pulsed laser light at $355\,\mathrm{nm}$. At the time of the experiments, direct randomized benchmarking (DRB) gave a median two-qubit gate fidelity of $99.17\%$, with single-qubit fidelities near $99.98\%$. The corresponding gate durations were approximately $130\,\mu\mathrm{s}$ for single-qubit operations and $950\,\mu\mathrm{s}$ for two-qubit operations.

\section{Error Mitigation}\label{sec:appendix-a}

Here we address two sources of error: the systematic error of using a finite evolution time, using a taper, and the quantum hardware error, using various error-mitigation (EM) schemes.

\subsection{Convergence of the ground-state calculation}
\begin{figure}[h!]
    \centering
    \includegraphics[width=\linewidth]{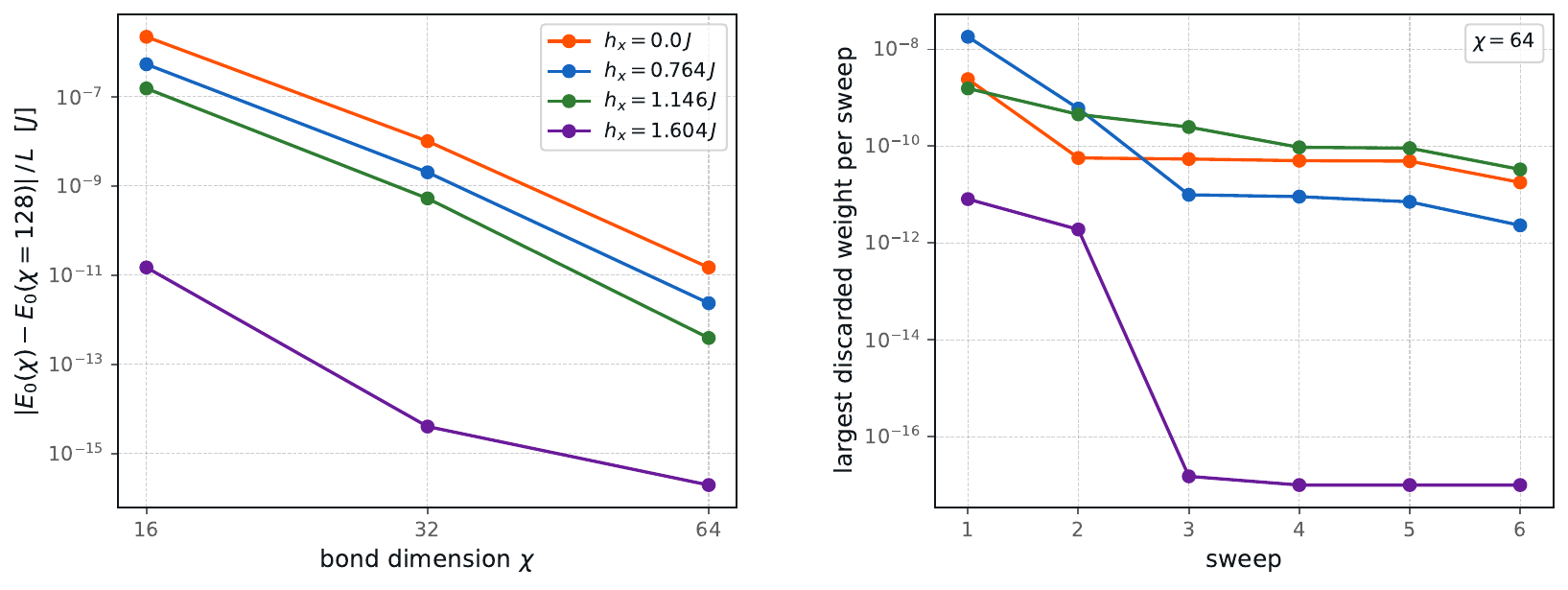}
    \caption{ Convergence of the DMRG ground state of the $L=36$, $\Delta=0.25$ chain across the four
transverse fields. Left: $|E_0(\chi)-E_0(128)|/L$ against the bond dimension $\chi$; at the
production value $\chi=64$ the energy is converged below the sweep criterion $10^{-10}|E_0|$.
Right: largest discarded Schmidt weight $\sum_{i>\chi}s_i^2$ of each DMRG sweep at $\chi=64$. The maximum energy error of $10^{-10}$ was reached by considering $6-14$ sweeps at $\chi=64$.}
    \label{fig:dmrg_con}
\end{figure}

The ground state $\ket{\psi_\chi}$ that is used as baseline for each quantum circuit is obtained by two-site density-matrix renormalization group (DMRG) on the $L=36$ chain, started
from the N\'eel product state and iterated in sweeps across the chain until the energy changes by less than $10^{-10}|E_0|$
between consecutive sweeps. In Figure~\ref{fig:dmrg_con} we monitor the convergence over all four transverse fields by studying 
the dependence of the converged energy on the bond  dimension $\chi$ through the difference $|E_0(\chi)-E_0(128)|/L$ for $\chi=16,32,64$. At $\chi=64$, we find that deviations from the $\chi=128$ energies lie bellow the $1.5\times 10^{-11} J$ threshold for every field value.For a second metric we evaluated the largest discarded weight (Schmidt weight) for fixed bond dimension $\chi=64$. Here, we find that the largest weight is reduced to less than $10^{-10}$ after 6 sweeps.

\subsection{Window Taper}
First, we address the systematic error of truncating the time evolution to a time $t_\mathrm{max}$, which convolves the true spectrum with the transform of the observation window, resulting in deterministic distortion of the dynamical spin structure factor (DSF). This is addressed using a taper window $w_k$, which redistributes this induced bias by replacing the temporal Fourier transform in the definition of the DSF to include the taper:
\begin{equation}
    \tilde{S}^{\alpha\alpha}(q,\omega) = -\frac{1}{2\pi}\,
    \mathrm{Im} \sum_{k=0}^{N_t-1} w_k\, \Delta t_k\,
    e^{i\omega t_k}\, G^{R}_{\alpha\alpha}(q, t_k). 
\end{equation}
A taper $w_k$ does not remove the truncation distortion; it damps the ringing at the cost of
resolution, as the uncertainty principle requires.
 
\begin{figure}
    \centering
    \includegraphics[width=\linewidth]{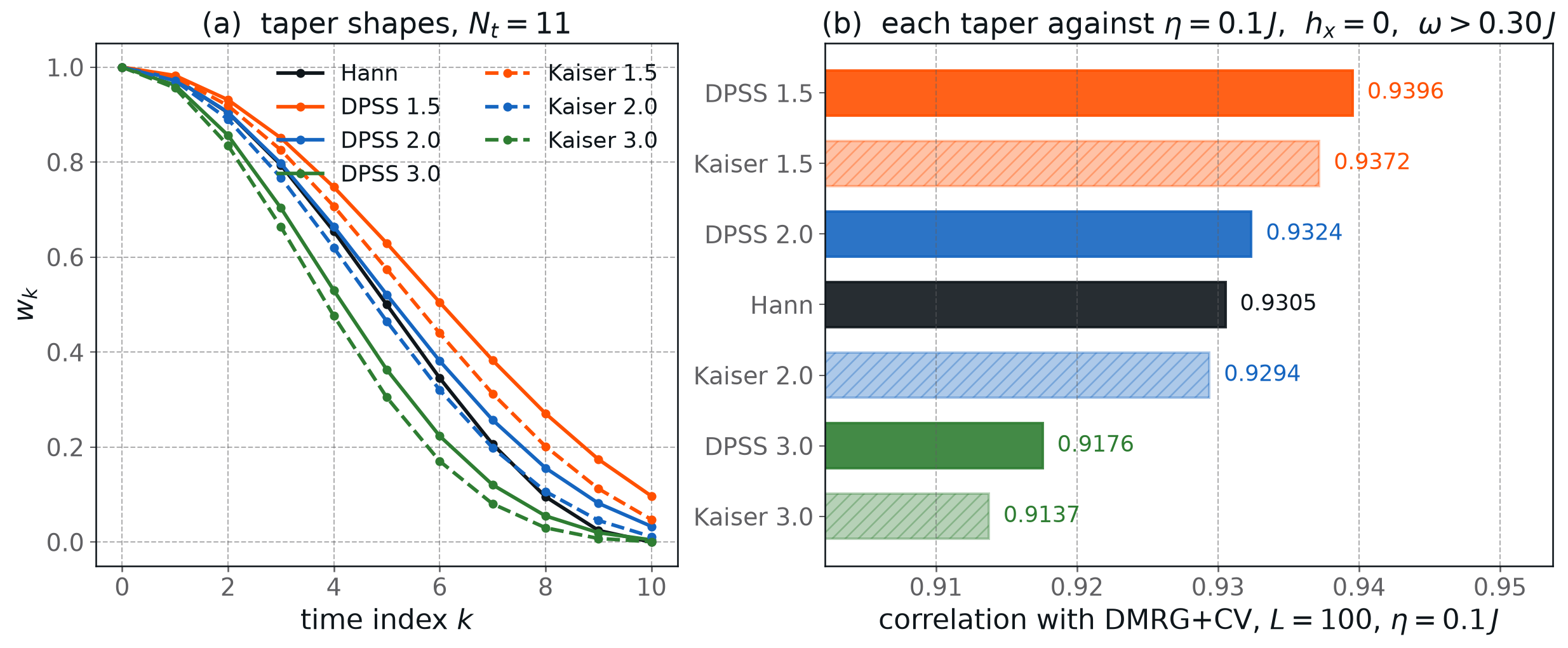}
    \caption{Left: the taper shapes in Eqs. \ref{eq:taper-dpss}-\ref{eq:taper-hann} with various $NW$ values for an example $N_t=11$ window. Right: the correlation between the QPU+EM results from the tapers for the $h_x=0$ case with DMRG-CV results. }
    \label{fig:taper}
\end{figure}

We tested 3 different tapers against the DMRG-CV result, which does not use a finite time approximation and therefore is used as the ground truth. For the tapers, for $N_t$ time steps, $M = 2N_t - 1$, and $\mathbf{v}$ a symmetric window of length
$M$; the taper applied is its descending half,
\begin{equation}
  w_k = \frac{v_{N_t-1+k}}{v_{N_t-1}}, \qquad k = 0,\dots,N_t-1,
  \label{eq:taper-half}
\end{equation}
normalized so that $w_0 = 1$.
The zeroth discrete prolate spheroidal sequence (DPSS) of time-bandwidth product $NW$ is the length-$M$ sequence maximizing the fraction of its enery inside $|f| \le W$, with $W = NW/M$. It
is the eigenvector of largest eigenvalue $\lambda_0$ of
\begin{equation}
  \sum_{n'=0}^{M-1}
    \frac{\sin\!\big[2\pi W (n-n')\big]}{\pi (n-n')}\, v_{n'}
    = \lambda_0\, v_n .
  \label{eq:taper-dpss}
\end{equation}
The Kaiser window is its closed-form Bessel approximation at matched
bandwidth, $\beta = \pi NW$,
\begin{equation}
  v_n = \frac{I_0\!\left(\beta\sqrt{1-\left(\tfrac{2n}{M-1}-1\right)^{2}}\right)}
             {I_0(\beta)},
  \label{eq:taper-kaiser}
\end{equation}
with $I_0$ the modified Bessel function of the first kind.
The Hann window is
\begin{equation}
  v_n = \tfrac{1}{2}\left[1 - \cos\!\left(\tfrac{2\pi n}{M-1}\right)\right],
  \label{eq:taper-hann}
\end{equation}
whose descending half reduces, via Eq.~\eqref{eq:taper-half}, to the
familiar $w_k = \tfrac{1}{2}\left[1 + \cos\!\left(\pi k/(N_t-1)\right)\right]$. 
The shapes and performance of these 3 taper classes with various $NW$ are given in Figure \ref{fig:taper}. The left part of the figure shows the studied taper shapes for Eqs. \ref{eq:taper-dpss}-\ref{eq:taper-hann} for various $NW$ values. The right part of the figure shows the performance of each taper, quantified as the correlation of the noiseless result with each taper with the DMRG-CV reference at fixed broadening $\eta=0.1J$, without applying a time-domain taper to that reference. Among the tested windows, the DPSS taper with $NW=1.5$ has the highest correlation $(r=0.9846)$ with the DMRG-CV result. It should be noted that high correlation doesn't imply removal of finite-time effect as peak broadening artifacts can persist. Unless explicitly identified as a taper comparison or sensitivity test, every temporal Fourier transform of a finite time record in this work uses the normalized descending half-DPSS of Eqs.~\eqref{eq:taper-half} and~\eqref{eq:taper-dpss} with $NW=1.5$. The window is recomputed for the number of samples in each record, using $M=2N_t-1$, so truncated, original, and FME-extended records follow the same prescription despite having different durations.

\subsection{Hardware EM} 
Next, we implement various low-cost error mitigation techniques that correct for errors incurred while running the circuits on the quantum processing unit (QPU). First, we apply readout error mitigation, where two calibration circuits determine the per-qubit response $\mathbb{E}[m \mid s] = a + b\,s$ with $a = p_{1|0} - p_{0|1}$ and $b = 1 - p_{1|0} - p_{0|1}$, which is then inverted~\cite{bravyi2021,nation2021}. The effect is generally small compared to the contribution from gate error, but this applies to all channels and measurements.

For certain channels and $h_x$ field values, symmetry post-selection can be applied. $S_z$ is conserved at $h_x = 0$, so shots with bitstrings measured outside the conserved sector are dropped. At
finite field $h_x\neq0$ only the global parity $P_X$ survives, which can be read out in the $xx$ channel. 

We apply a spatial mask to suppress contributions outside a chosen region around the kick site. The mask expands with time according to a fixed processing parameter and requires no additional circuits or shots. It can remove physical response as well as noise, so we assess the sensitivity of the reconstructed spectra to the mask parameter in Tables~\ref{tab:distortion} and~\ref{tab:distortionzz}.

Additionally, the matrix $G(j,t)$ is not low rank: a two-spinon continuum is intrinsically high-rank, so applying singular value shrinkage does not offer a significant improvement. Our multichannel singular spectrum analysis (MSSA) acts instead at finite momentum, where $G(q,t)$ is a sum of a small number of damped exponentials/poles of the Green's function, and a sum of $K$ exponentials has a Hankel matrix of rank $K$. Thus, rank-$K$ projection with anti-diagonal averaging \cite{cadzow1988} yields a reduction in shape error. This procedure is unique in that it directly attacks shape error, rather than just attenuation. However, the optimal $K$ is here provided explicitly.

Finally, while the previous error mitigation protocols address bias and lineshape, they do not correct for attenuation of observables due to noise, or amplitude error. Therefore, we also include an error mitigation scheme where a probe (``mirror'') circuit is used to estimate the attenuation and correct for it. We run one additional mirror circuit per channel and external field value on IonQ Forte Enterprise alongside the main experiment, matching its two-qubit gate count and shot count and using a known ideal output \cite{Urbanek2021}. Here, we use the mirror circuit $W_{[:m]}W_{[:m]}^{\dagger}$ for a 2-qubit gate count $m$. Its ideal output is $|0\cdots0\rangle$, which affords two estimators of the surviving amplitude: the single-site survival $\lambda_Z = \overline{\langle Z_j\rangle}$, which is blind to correlated error and so over-estimates it, and the full-string survival $P(0^{36})$, which counts any single flip as total loss and so under-estimates it. These
bracket the truth widely at $L = 36$, so we interpolate geometrically,
\begin{equation}   \lambda = \lambda_Z^{\,1-a}\,\bigl[P(0^{36})\bigr]^{\,a} ,  \label{eq:canary} \end{equation}
which leaves the exponent $a$ undetermined. We determine it from an independent hardware measurement: the attenuation $f$ obtained by inverting $\mathrm{Var}(\sum_j\sigma^z_j)$ on the measured $k=0$ histogram, without using a reference spectrum or neutron data. Since every circuit carries the same $420$ two-qubit gates, $f$ is a property of the circuit depth rather than of the field or the spin component, and we adopt its inverse-variance mean over the two fields where the inversion is well conditioned, $\bar f = 0.4677 \pm 0.0139$, for every record. The resulting exponents $a = \ln(\lambda_Z/\bar f)/\ln(\lambda_Z/P_0)$ are $0.401$ and $0.087$ for $xx$ and $zz$ at $h_x=0$, $0.257$, $0.050$ and $0.235$ at $0.764\,J$, $0.095$ at $1.146\,J$ and $0.180$ at $1.604\,J$. 

This EM is applied fully to the QPU results, as well as partially to the DMRG+Trotter results. There, DMRG+Trotter column passes through the record-domain stages of the mitigation chain - reflection averaging, the light-cone mask and the rank-4 MSSA - so that it is read through the same analysis as the hardware column. The EM protocols are not applied to the DMRG+CV results. 

\begin{figure}[!htbp]
    \centering
    \includegraphics[width=\linewidth]{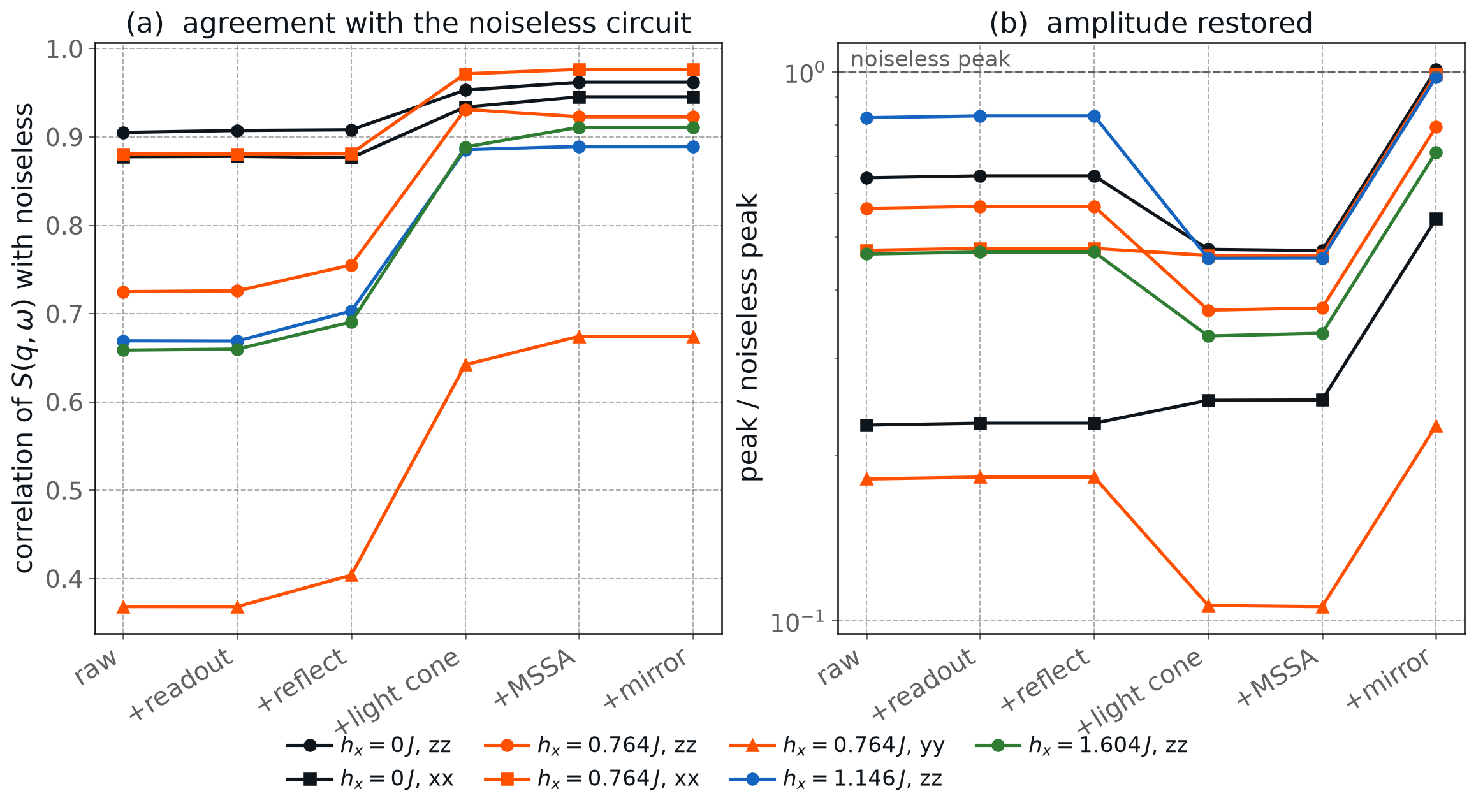}
    \caption{The effect of each error mitigation protocol on the lineshape (left), measured by correlation of hardware $S(q,\omega)$ with the noiseless result, and amplitude (right), measured as the measured hardware peak over the noiseless peak. }
    \label{fig:em_abl}
\end{figure}

The results for the various error mitigation protocols are shown in Figure \ref{fig:em_abl}. Here, the error mitigation is switched on one stack at a time, cumulatively, and compares the intermediate result to the noiseless result for two metrics: lineshape, measured as the correlation of hardware $S(q,\omega)$ with the noiseless result, and amplitude, measured as the measured hardware peak over the noiseless peak. The readout EM has a marginal effect on both the lineshape and amplitude, consistent with estimations. The reflection averaging improves the lineshape for most finite-field cases by a modest average improvement of $0.03$. The light cone mask has a significant effect on the lineshape, improving by up to $0.24$ without additional circuits or shots. The mask also decreases the amplitude by removing part of the measured response, which can include both noise and physical signal. MSSA provides a marginal affect to only the lineshape up to $0.032$. Finally, the mirror circuit leaves the lineshape unaffected but takes the correction from $0.66-0.93$ to $0.89-0.98$, the exception being the $yy$ channel at $h_x = 0.764 J$, which starts at $0.37$ and reaches only $0.67$.

Relative to the noiseless compressed circuit simulation reference, reflection averaging and the light-cone mask carry the raw hardware correlations from $0.552-0.720$ to $0.827-0.857$, and the rank-4 MSSA adds a further $0.01-0.05$. In the $zz$ channel, measured at all four fields, the mitigated values are $0.902$, $0.863$, $0.846$ and $0.850$; in the $xx$ channel, measured at 0, $0.764$ and $1.146\,J$, they are $0.892$, $0.949$ and $0.929$. There is no transverse hardware record at 1.604 J, so no xx correlation is quoted there. The readout inversion changes these correlations by less than 0.001, as expected for a near-uniform rescaling, as the gain is in the mask.

\begin{figure*}[!htbp]
    \centering
    \includegraphics[width=\linewidth]{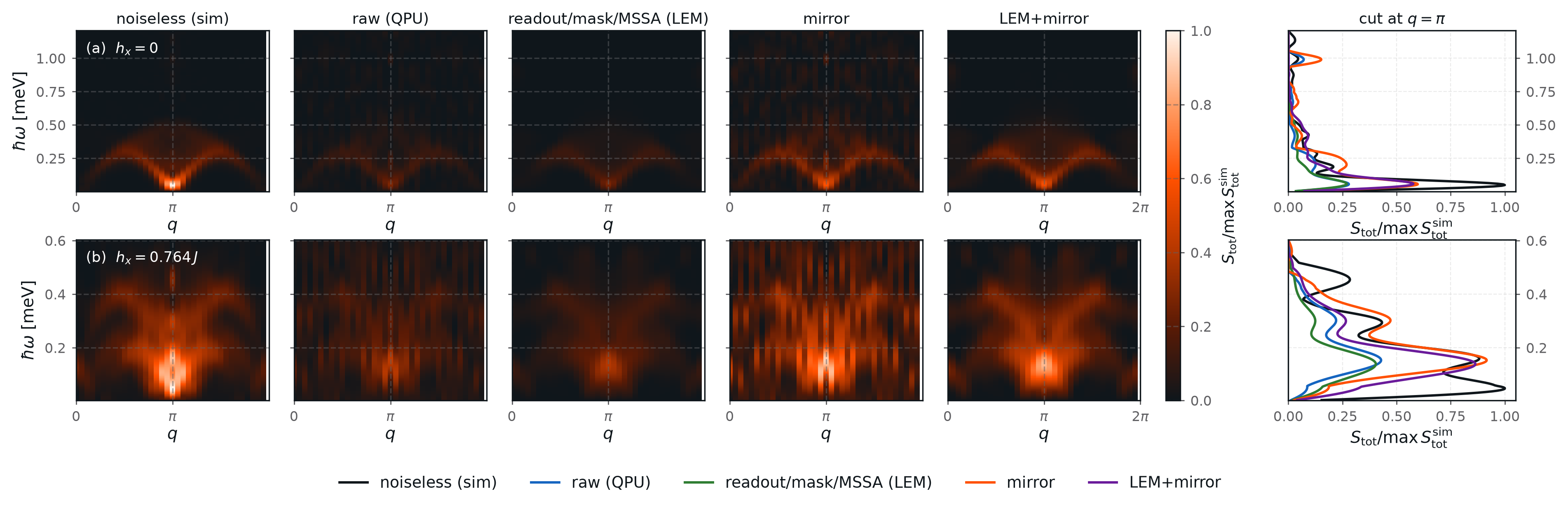}
    \caption{Total DSF of the $L=36$ anisotropic ($\Delta=0.25$) chain for the $h_x=0$ (upper) and $h_x=0.764J$ (lower) cases. The first column shows the result from running the quantum circuits on the noiseless simulator. The second column gives the raw QPU results with no EM. The third column gives the QPU results with only the lineshape error mitigation (LEM) applied (readout, light cone mask, and MSSA). The fourth column gives the QPU results with only amplitude correction applied (mirror circuit). The fifth column gives the QPU results with both the lineshape and amplitude EM applied. The sixth column then gives the line cut at $q=\pi$ for the previous five columns. The plots show the DSF normalized to the peak of the noiseless simulation of the circuits being one. }
    \label{fig:em-total-dsf_appendix}
\end{figure*}

The visual effect of each of the error mitigation strategies alone on the total DSF are shown in Figure \ref{fig:em-total-dsf_appendix}. The first column shows the noiseless result for the $h_x=0,~0.764J$ cases as reference. The second column shows the raw QPU results, which have diminished amplitude and lineshape error. The third row shows the result with the lineshape-relevant error mitigation, readout, light cone mask, and MSSA, where the spurious DSF contributions in the raw QPU results are removed. The fourth column shows the result with just the mirror circuit correction for amplitude, where the amplitude is improved, but the spurious DSF contributions remain. The fifth row has both lineshape and amplitude corrections, and has the closest match to the noiseless result. The sixth row shows the line cut at $q=\pi$ for the 5 previous columns, where the EM+mirror result matches closest to the noiseless result. 

\section{Spectral validation and spin correlations}\label{sec:appendix-b}

This appendix brings together the definitions, detailed comparisons, and correlation analysis supporting Section~\ref{sec:iv}. The validation metrics are aligned closely with the metrics defined in~\cite{buchs2026validation}. The dynamical comparisons include the DMRG ground state with bond dimension $\chi=64$ followed by second-order Trotter evolution (DMRG+Trotter), raw hardware measurements on IonQ Forte Enterprise, and the same measurements after error mitigation. DMRG-CV provides the independent spectral reference used to assess QPU+EM. The construction of the compressed circuits is described in Section~\ref{sec: qsim}.

\subsection{Definitions of spectral validation metrics}
\label{sec:spectral-metrics}

The total response is $S_{\mathrm{tot}}=S^{xx}+S^{yy}+S^{zz}$. Figure~\ref{fig:hw-total-dsf} compares the neutron data, DMRG-CV, DMRG+Trotter, and QPU+EM at $h_x=0$ and $0.764J$, where the total hardware response is available. 

The DMRG-CV reference is the published $L=100$ spectrum of Laurell \emph{et al.}~\cite{laurell2021a}, calculated using the correction-vector method~\cite{Alvarez2009} with $m=1000$ retained states and Lorentzian broadening $\eta=0.1J$, as described in Section IV A. The same fixed reference is used for comparisons with QPU+EM and INS. Its intensities are interpolated onto the appropriate comparison grid without additional convolution to match the hardware observation window or the experimental resolution. For the QPU+EM comparison, we combine the spin components with their stated weights and clip negative intensities to zero.

\begin{figure*}
    \centering
    \includegraphics[width=\linewidth]{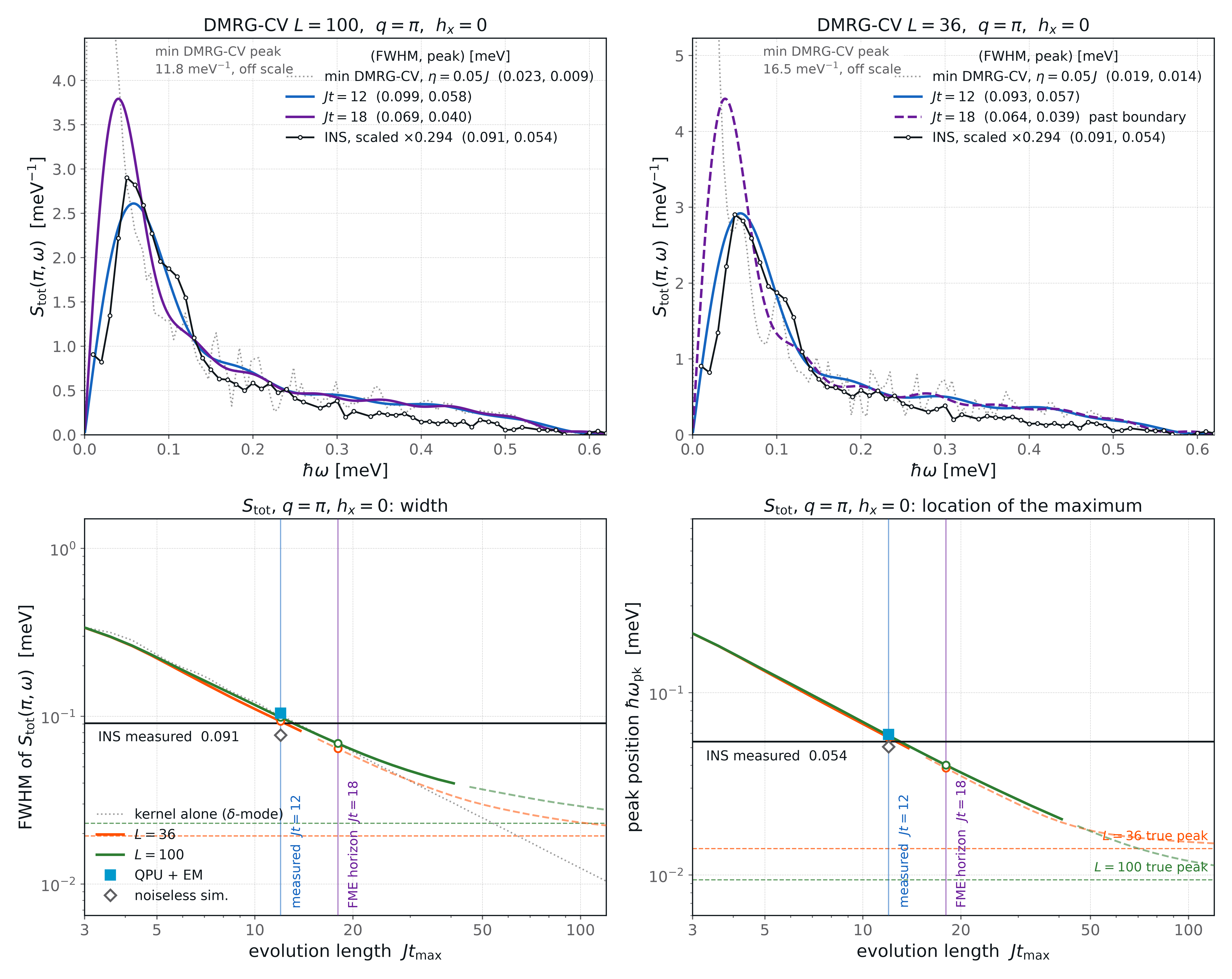}
    \caption{$S_\mathrm{tot}(\pi,\omega)$ at $h_x=0$ against the length of the time record. \emph{Top:} the $q=\pi$ cut of the DMRG-CV solve read through our estimator at $Jt=12$, the measured record, and at $Jt=18$, for $L=100$ (left) and $L=36$ (right), with the minimally broadened solve behind them as a gray dotted line; its peak runs off scale and is annotated instead. Both solves are correction vector at $\eta=0.05\,J$, so the two panels differ only in chain length. The black curve with open circles is the measured neutron cut; it is in the INS experiment's own units and so carries one least-squares scale factor, $\times0.294$, fitted against the $L=100$, $Jt=12$ curve and used unchanged in both panels. Legends give (FWHM, peak position) in meV. \emph{Bottom:} the same two quantities over the whole sweep, with the measured benchmarks marked - the error-mitigated hardware (square) and the noiseless simulation (diamond) at $Jt=12$, and the neutron values as horizontal lines. The dotted curve is the estimator's response to a single mode, with no spectrum behind it. A dashed segment marks the record outrunning the chain: solid until the front launched at $j_c$ reaches the end ($Jt=15.6$ at $L=36$, $43.3$ at $L=100$), dashed beyond. Horizontal dashed lines are the unconvolved limits the curves approach (width $0.019/0.023$~meV, peak $0.0139/0.0094$~meV for $L=36/100$). Note that these are our own correction-vector solves at $\eta=0.05\,J$, not the published results of Ref.~\cite{laurell2021a} at $\eta=0.10\,J$ used in Tables~\ref{tab:metrics}--\ref{tab:distortionzz}. The reference linewidth here is $2\eta=0.023$~meV at $L=100$ ($0.019$~meV at $L=36$, where the discrete spectrum is not a single Lorentzian), so the estimator is about $4.4$ times wider than it, against the factor $2.2$ quoted for the comparison with the published archive.}
    \label{fig:jt_scale}
\end{figure*}

\begin{figure*}
    \centering
    \includegraphics[width=\linewidth]{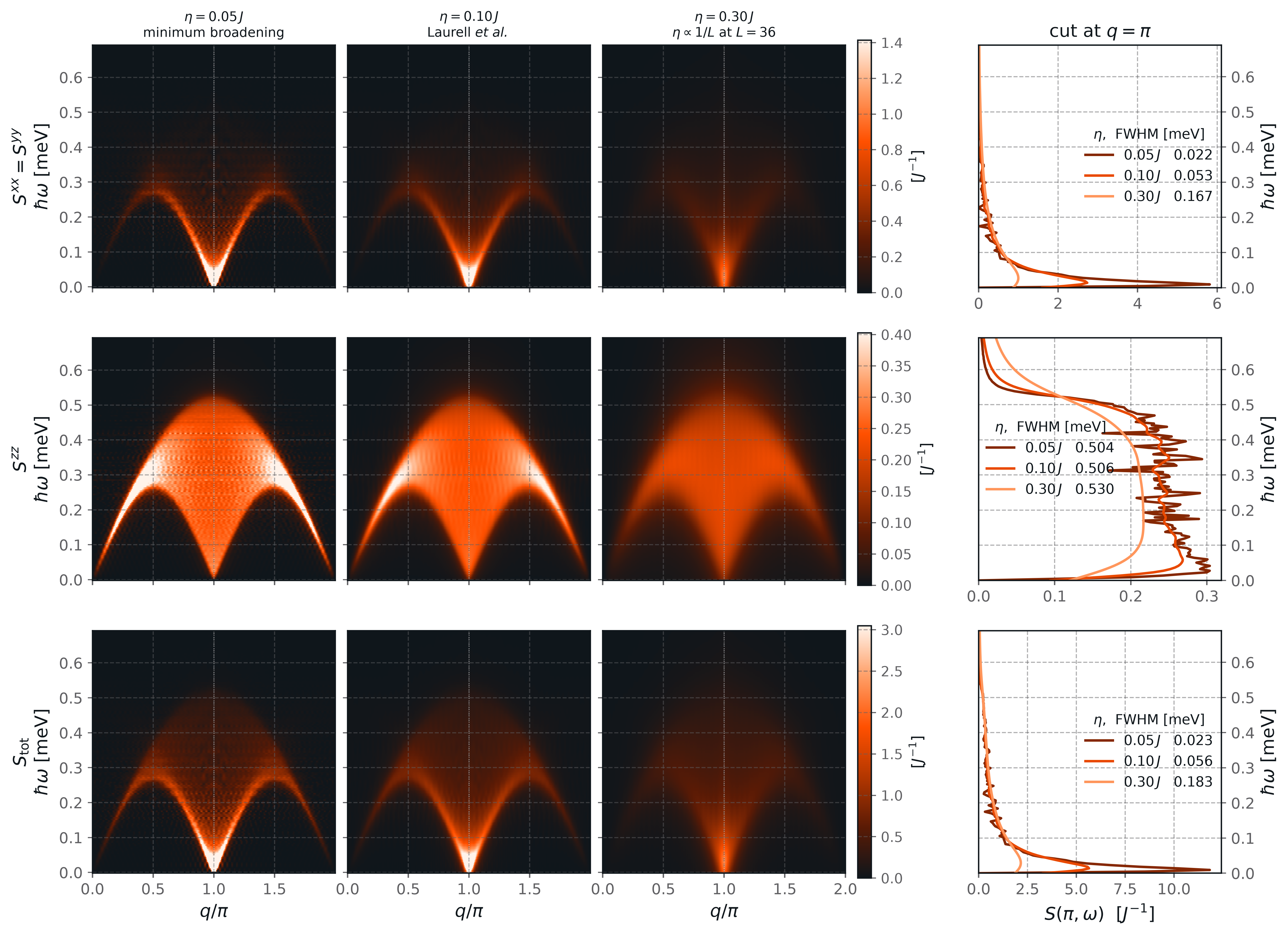}
    \caption{Effect of the correction-vector broadening on $S^{xx}=S^{yy}$, $S^{zz}$ and $S_\mathrm{tot}$ at $h_x=0$, from the $L=100$ DMRG-CV solve. Every width is generated by convolution from the single $\eta=0.05\,J$ calculation: Lorentzians compose additively, so no panel is a separate solve or a refit. The three widths are given as follows: (i) $0.05\,J$, the smallest this solve supports, (ii) $0.10\,J$, the width of the published spectra from \cite{laurell2021quantifying}, and (iii) $0.30\,J$, the $\eta\propto1/L$ prescription evaluated at the $L=36$ chain studied on hardware.}
    \label{fig:broadening}
\end{figure*}

For the INS comparison, the spectra are used at their native correction-vector resolution and interpolated onto the neutron $(q,\hbar\omega)$ grid. Then, following \cite{laurell2021quantifying}, the elastic feature is removed by subtracting at each $q$ a Lorentzian of height $S^{\alpha\alpha}(q,\omega=0)$ and of width equal to the width the feature actually carries at that point in the chain. Thus the subtraction follows any fixed broadening and uses the broadened width $\eta_\mathrm{eff}=0.1J$, not the solve width $\eta$. This value is taken from \cite{laurell2021quantifying}. The effect of various choices for $\eta_\mathrm{eff}$ is given in Figure \ref{fig:broadening}. Note that $\eta_\mathrm{eff}=0.1J$ gives a FWHM resolution of $0.046$, while the INS results have resolution $0.073-0.080$ and the QPU+EM values, incidentally, have resolution $0.099-0.101$. 
Figure \ref{fig:jt_scale} provides the total DSF at the $q=\pi$ cut for various values of $Jt$ and $L$ alongside the INS and raw DMRG-CV data. The lower plots show the FWHM of the total DSF cut at $q=\pi$ as a function of evolution time length for $L=36$ and $100$ alongside the INS resoluiton. Energy resolution at $q=\pi$ is set by the length of the time record, not by the chain. At the measured $Jt = 12$, L = 36 and L = 100 give widths of 0.093 and 0.099 meV and peaks of 0.057 and 0.058 meV - 5\% and 2\% apart - so finite size does not limit the current data. Increasing the record to $Jt=18$ reduces the widths to 0.064 and 0.069 meV for $L=36$ and $L=100$, respectively. This comparison assesses the effect of the observation window; possible boundary effects at longer times require a separate check. The measurement is already at the Jt = 12 resolution. INS gives 0.091 meV, next to QPU + EM at 0.104 and the noiseless simulation at 0.077, and far from the 0.064 meV of the Jt = 18 extension - so a finer record is beyond what the neutron data can check.

One amplitude is fitted, since the neutron intensity is in meV$^{-1}$ and the calculation is in spin units. The $\chi=64$ matrix-product state is not an exact many-body solution. It is the classical compression target against which the simulation is benchmarked, and its convergence in $\chi$ given in Figure \ref{fig:dmrg_con}. 

For the QPU comparison, let $S^Q$ and $S^D$ denote the hardware and DMRG-CV spectra on the comparison window. The QPU–DMRG-CV comparison grid is the $L=36$ plane-wave momenta $q_n=2\pi n/L$ across the full zone, and a uniform $240$-point frequency grid running to the Nyquist frequency $\pi/\Delta t$ of the measured record, restricted to $\hbar\omega\le\omega_\mathrm{max}$ - $36\times138=4968$ cells at $h_x=0$ and $36\times240=8640$ at $0.764\,J$. For the INS comparion, the spectra are evaluated on the neutron grid, which is $68$ momenta by $83$ energies, onto which the reference is interpolated while retaining its fixed broadening $\eta=0.1J$. Cells enter where the DMRG-CV results report a finite intensity and a positive uncertainty and the interpolated reference is finite, within $0<\hbar\omega\le\omega_\mathrm{max}$: $3737$ cells at $0$~T and $3323$ at $1$~T. 

The Pearson correlation is
\begin{equation}
 r=\frac{\langle(S^Q-\bar S^Q)(S^D-\bar S^D)\rangle}{\sigma_{S^Q}\sigma_{S^D}}.
 \label{eq:spectral-pearson}
\end{equation}
Here the averages and standard deviations are evaluated over the same cells. Invariance under a positive overall rescaling and a constant offset makes $r$ a measure of the pattern of spectral weight, rather than its absolute intensity. The fitted amplitude ratio is
\begin{equation}
 A=\arg\min_\alpha\lVert S^Q-\alpha S^D\rVert^2
  =\frac{\langle S^Q,S^D\rangle}{\langle S^D,S^D\rangle},
 \label{eq:spectral-amplitude}
\end{equation}
where the inner products run over the comparison window. A value near unity indicates agreement in the fitted overall scale; it does not require the two spectra to agree point by point. The earlier analysis reported that the correction vector $S^{zz}$ spectrum satisfies the single channel sum rule to $0.5\%$. Specifically, this calculation was computing $(1/N)\sum_q\!\int\!d\omega\,S^{zz}(q,\omega)=\langle S^zS^z\rangle=1/4$ for $S=1/2$, with the momentum average taken over the uniform full-zone plane-wave grid and the frequency integral by trapezoid over the full solved range $0\le\hbar\omega\le3.6\,J$, after the elastic feature has been removed. This gives a value of $0.2512$, and it does not strongly depend on these choices as restricting to the displayed window instead gives $+0.2\%$, and retaining the elastic feature gives $+0.7\%$. 

The mean squared error (MSE) is given by 
\begin{equation}
\mathrm{MSE}=\frac{1}{N}\sum_{q,\omega}\bigl[A(q,\omega)-R(q,\omega)\bigr]^2 .
\end{equation}
For the comparison between DMRG-CV and INS, the spectra are in units of meV$^{-1}$ and spin units respectively, so $A$ is first multiplied by the same fitted amplitude $\alpha$ the modified $\chi^2$ uses. The normalized form divides by the reference's mean square, $\mathrm{MSE}/\langle R^2\rangle$ with $\langle R^2\rangle=N^{-1}\sum R^2$, which is dimensionless and makes cases of different overall intensity comparable.

The Wasserstein distance $W_1$ is computed for both comparisons with the two spectra clipped to zero, coarsened in $\omega$ to $n_\omega=60$ bins by summation, and normalized to unit mass, giving distributions $p$ and $r$ on the same support. Cells are mapped to the unit square, momentum to $[0,1]$ across the full zone and frequency to $(j+\tfrac12)/n_\omega$, and the ground cost is the Euclidean distance there. $W_1$ is the minimum of $\sum_{ab}\pi_{ab}\,d_{ab}$ over couplings $\pi$ with
marginals $p$ and $r$, solved exactly by network simplex rather than by an entropic approximation. Note that $W_1$ does not involve amplitude and is scale free. 

For computing the structural similarity index measurement (SSIM), the grids are first rebinned in $\omega$ to $36$ cells by averaging, so that a cell is square in the normalized $(q,\omega)$ domain and the $h_x=0$ window ($138$ frequency cells) and the finite-field windows ($240$) are commensurate. Local means, variances and covariance are
taken with an $11\times11$ uniform window and the unbiased factor $n/(n-1)$, $n=121$, and 
\begin{equation}
\mathrm{SSIM}=\Bigl\langle\frac{(2\mu_A\mu_R+c_1)(2\sigma_{AR}+c_2)}
{(\mu_A^2+\mu_R^2+c_1)(\sigma_A^2+\sigma_R^2+c_2)}\Bigr\rangle ,
\end{equation}
with $c_1=(0.01\,D)^2$, $c_2=(0.03\,D)^2$ and $D$ the dynamic range, taken as the larger of the two spectra' ranges. The average runs over the interior, cropping five cells on each side so that no window overhangs the boundary, and since the neutron footprint is not rectangular (there is a kinematic wedge with no data), for that comparison the local statistics are computed on zero-filled arrays and the resulting SSIM value is averaged only over cells where neutron data exist.

Energy moments complement these cell-by-cell comparisons. Negative intensities are clipped to zero before evaluating the moments. The centroid,
\begin{equation}
 \langle\hbar\omega\rangle_q=
 \frac{\int\hbar\omega\,S(q,\omega)\,\mathrm{d}(\hbar\omega)}
 {\int S(q,\omega)\,\mathrm{d}(\hbar\omega)},
 \label{eq:spectral-centroid}
\end{equation}
is evaluated at $q=\pi$ and $q=\pi/2$. It uses the full lineshape and can vary continuously even when the frequency grid is discrete. The root mean square width at $q=\pi$ is
\begin{equation}
 \sigma_\omega(q)=\left[
 \frac{\int(\hbar\omega-\langle\hbar\omega\rangle_q)^2S(q,\omega)\,\mathrm{d}(\hbar\omega)}
 {\int S(q,\omega)\,\mathrm{d}(\hbar\omega)}\right]^{1/2}.
 \label{eq:spectral-width}
\end{equation}
These integrals use the accessible spectral window. Unlike a fitted peak width, the second moment remains defined for a flat or bimodal spectrum with nonzero integrated weight.

The fraction of spectral weight above a fixed energy is
\begin{equation}
 R_{>}=
 \frac{\int\mathrm{d}q\int_{E_b}^{E_{\max}}S(q,E/\hbar)\,\mathrm{d}E}
 {\int\mathrm{d}q\int_0^{E_{\max}}S(q,E/\hbar)\,\mathrm{d}E},
 \label{eq:spectral-partial-weight}
\end{equation}
where $E=\hbar\omega$, $E_b=0.20\,\mathrm{meV}$, and $E_{\max}$ is the upper energy cutoff. Holding $E_b$ fixed allows changes in the distribution of weight to be compared across fields, even when peak positions agree. The upper cutoffs remain those specified for each field in Table~\ref{tab:metrics}.

We also report the peak energy $\hbar\omega_{\mathrm{pk}}(q)$, the energy at which $S(q,\omega)$ is maximal, including the cut at $q=\pi$. It requires no lineshape model but is quantized by the evaluation grid and can be unstable for flat or bimodal spectra. The threshold estimate of the gap is
\begin{equation}
 \Delta_{\mathrm{gap}}=\min\left\{\hbar\omega:
 S(\pi,\omega)\ge f\max_{\omega'}S(\pi,\omega')\right\},
 \label{eq:spectral-gap}
\end{equation}
with $f=0.10$; repeating the estimate at $f=0.20$ tests its threshold sensitivity. The bandwidth estimator is
\begin{equation}
 B=\max_q\hbar\omega_{\mathrm{pk}}(q)-\min_q\hbar\omega_{\mathrm{pk}}(q).
 \label{eq:spectral-bandwidth}
\end{equation}
We report the difference $\Delta B$ between the mitigated hardware and DMRG-CV values, in the convention of the preceding paragraph. Both inherit the peak grid discretization. The bandwidth additionally takes extrema over 36 independently located peaks, making it particularly sensitive to peak selection.
Note that the evaluation grid is 240 points running from $0.02\,J$ to the Nyquist frequency $\pi/\Delta$ of the record being analyzed. Because the $h_x=0$ records are sampled at $\Delta t=0.6/J$
and the finite-field records at $\Delta t=1.2/J$, the grid step is $5.02\,\mu$eV at zero field and $2.50\,\mu$eV at finite field; the quoted $0.0025$~meV gap shifts are one step of the latter. This grid is a choice of evaluation points, not an FFT bin structure: the estimator evaluates the transform directly at each requested frequency from the tapered time record, so no zero padding is involved and none would add information. The reference is linearly interpolated onto the same grid. The nominal scale $\pi/t_\mathrm{max}$ is $0.060$~meV at every field, since all records share $t_\mathrm{max}=12/J$. The effective resolution, the full width at half maximum of the estimator's response to a single mode, is $0.101$~meV at $h_x=0$ and $0.099$~meV at finite field; the excess over $\pi/t_\mathrm{max}$ is the half-DPSS taper.  
Tables~\ref{tab:metrics}--\ref{tab:distortionzz} compare against the published DMRG-CV spectra at fixed broadening $\eta=0.1J$, corresponding to a Lorentzian full width at half maximum of $0.046\,\mathrm{meV}$. The reference is not convolved with the response of the measured time record. The effective QPU reconstruction resolution is approximately $2.2$ times wider, so differences in spectral widths, centroids, threshold gaps, and peak-derived bandwidths can reflect the distinct resolution functions as well as errors in the reconstructed response.

The distinct resolution functions affect the validation metrics differently. Spectral widths are directly sensitive to broadening, while peak positions can shift when nearby features overlap or their relative intensities change. The bandwidth estimator $\Delta B$ takes extrema over peaks identified independently at 36 momenta and is therefore sensitive to these changes in peak selection. Width and bandwidth discrepancies must consequently be interpreted together with the reconstruction uncertainties and the fixed resolution difference between the two calculations.

Finally, we define how we treat uncertainty quantification, following Ref. \cite{buchs2026validation}. Every uncertainty starts from the native observable of its pipeline
For the hardware the native observable is the measured histogram itself, which we retain, so the covariance of the record is constructed from it rather than modeled: at a fixed time slice $t_n$ all $L$ site expectations are estimated from one histogram of $N_n$ shots, and writing $x_s(j)=\pm1$ for the eigenvalue of $\sigma^z_j$ in bitstring $s$ and $p_s$ for its measured frequency, the estimator $\hat e(j)=\sum_s p_s x_s(j)$ has the multinomial covariance
\begin{equation}
  \mathrm{Cov}\!\left[\hat e(i),\hat e(j)\right]
  = \frac{1}{N_n}\Big[\sum_s p_s\,x_s(i)x_s(j)
  - \hat e(i)\,\hat e(j)\Big].
  \label{eq:multinom}
\end{equation}
The readout inversion is affine, $G=w\,[\hat e-(p_{10}-p_{01})]/b$ with $b=1-p_{10}-p_{01}$ and $w=\tfrac12$ for the records measured as an opposite-kick pair, so the covariance of the corrected record at slice $n$ is $\Sigma_n=(w/b)^2\,\mathrm{Cov}[\hat e]$, summed over the two kick signs where both are present; separate slices were separate jobs and are independent, so the full covariance is block diagonal in time. The diagonal of Eq.~\eqref{eq:multinom} is $(1-\hat e(j)^2)/N_n$, which recovers the per-site form $\sigma=A_\mathrm{ro}\sqrt{(1-G^2)/ N_n}$ used previously; retaining the off-diagonal block, the correlation between sites read out in the same shots, changes the propagated uncertainty on $S(q,\omega)$ by $1$--$3\%$ in the median and by up to $20\%$ pixel by pixel. We then propagate this covariance through the composed map $\mathcal{M}=f\circ\mathcal{E}$ of the metric $f$ after the analysis chain $\mathcal{E}$ --- readout inversion, reflection averaging, the spatial mask parameter, rank-$4$ MSSA, the mirror amplitude, the tapered transform and the fluctuation--dissipation factor --- by linearizing at the measured record,
\begin{equation}
  \sigma_M^2 = \sum_n
  \left(\frac{\partial \mathcal{M}}{\partial G_n}\right)^{\!\top}
  \Sigma_n
  \left(\frac{\partial \mathcal{M}}{\partial G_n}\right),
  \label{eq:prop}
\end{equation}
with the derivatives taken by central differences of the analysis code itself rather than from an independent analytic expression, so that the object differentiated is the same chain that produces the central value. Inputs annihilated by the light-cone mask contribute exactly zero and are omitted: reflection averaging pairs site $j$ with $2j_c-j$ and therefore preserves $|j-j_c|$, so it commutes with the mask and no weight enters from outside the cone. 

Three of the shape diagnostics return an element of the frequency grid - the peak energy is $\omega$ at the maximum of a column, the gap is the lowest $\omega$ at which a column exceeds a fixed fraction of its own maximum, and the bandwidth is the spread of the peak energy across $q$ - so each is piecewise constant in the data, its derivative vanishes almost everywhere, and Eq.~\eqref{eq:prop} would return $\sigma_M=0$ rather than an inaccurate $\sigma_M$. For those rows, and only those, we retain resampling: $160$ multinomial resamples of the measured histograms, each pushed through the unlinearized chain with every metric recomputed, so that the quoted $\sigma_M$ is again the spread of the metric itself. Where every resample falls in the same frequency bin the quoted value is an upper bound set by the grid step rather than a vanishing spread. 

The linearization is checked against the same resampling on the rows where both apply: stratified by intensity, the ratio of the resampled standard deviation to Eq.~\eqref{eq:prop} on $S_\mathrm{tot}(q,\omega)$ is $1.01$ in the median with a $5$--$95\%$ range of $0.98$--$1.07$ for pixels above half the peak, and stays within $8\%$ in the median down to $5\%$ of the peak, degrading only on the faint background where the positivity constraint of the estimator is active and the spectrum is consequently one-sided.

For the neutron comparison there is no counting statistics to propagate: the reported experimental uncertainties are used as given and the map is resampled within them over the same $160$ draws. No spectrum is renormalized before a metric is computed; all comparisons are made in absolute units on a common $\omega$ window, which is what allows the amplitude ratio $A$ to be reported as a result rather than fitted away. 

\begin{figure}
    \centering
    \includegraphics[width=\linewidth]{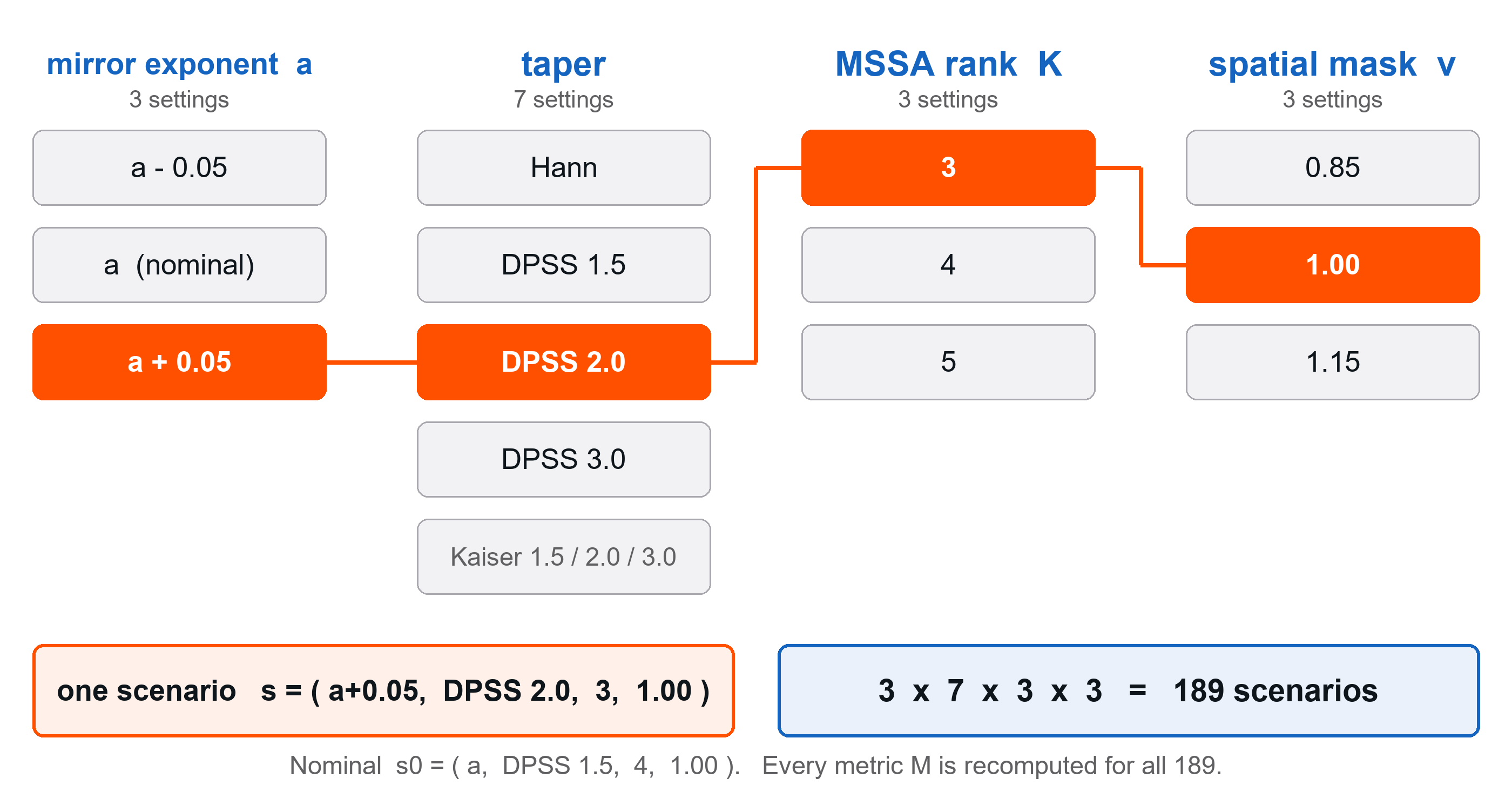}
    \caption{Schematic showing how the distortion scenarios are generated. The analysis contains four choices that are not measured quantities: the mirror exponent $a$, the temporal taper, the rank $K$ of the MSSA, and the spatial mask parameter $v$. Each is assigned a set of admissible settings - $a$ displaced by $\pm0.05$ about its measured value, the taper over a half-Hann window together with half-DPSS and half-Kaiser windows at time-bandwidth $NW=1.5$, $2.0$ and $3.0$, $K\in\{3,4,5\}$, and $v\in\{0.85,1.00,1.15\}$ - and a scenario fixes one setting from each, so the admissible ensemble is the Cartesian product of the four sets, $3\times7\times3\times3=189$ settings~\cite{buchs2026validation}. }
    \label{fig:scenario_knobs}
\end{figure}

These uncertainties give the statistical band $\sigma_M$ alone; the distortion band $\delta_M$ reported beside it in Tables~\ref{tab:distortion} and~\ref{tab:distortionzz} is a different object, obtained by re-running the identical analysis over the Cartesian product of the admissible processing choices - the mirror exponent, the taper, the MSSA rank and the spatial mask parameter ($3\times7\times3\times3=189$ settings) - and taking the two-sided spread of each metric about its nominal value, so it measures the latitude left in the analysis rather than counting statistics (Fig.~\ref{fig:scenario_knobs}). It is deliberately the spread over the joint ensemble and not the quadrature sum of the four single-knob columns. 

\subsection{Spectra across the field sweep}
\label{sec:supp-spectra}

\begin{figure*}[!htbp]
    \centering
    \includegraphics[width=0.9\linewidth]{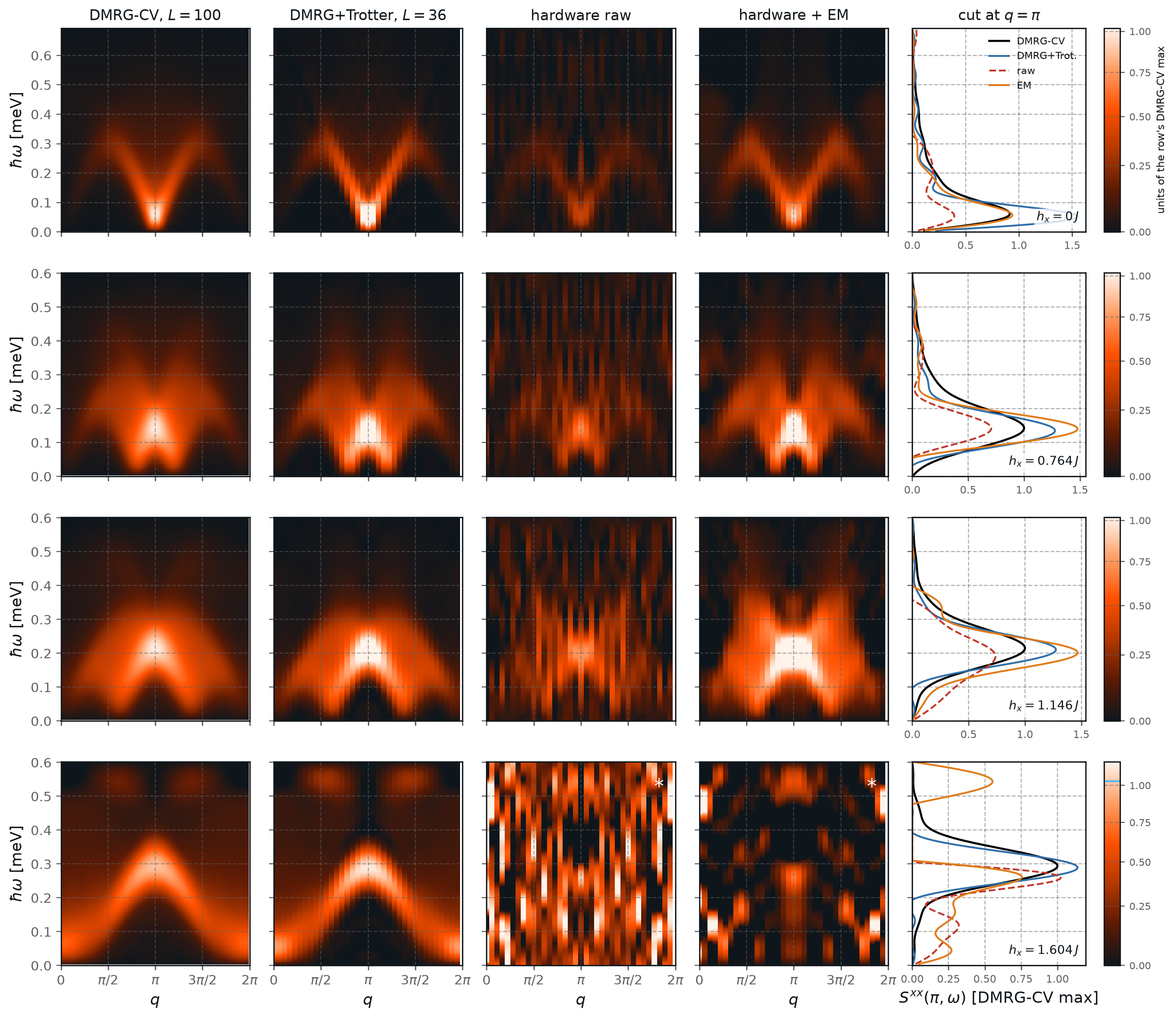}
    \caption{Transverse DSF $S^{xx}(q,\omega)$ of the $L=36$ anisotropic $(\Delta=0.25)$ chain, with each panel  normalized to the DMRG-CV maximum. Rows: plot the heatmap for at fields $h_x/J=(0,\ 0.764, \ 1.146, \ 1.604)$. Columns: correspond to DMRG-CV, DMRG+Trotter, raw hardware and error mitigated hardware results. ${}^{*}$For data-points without hardware runs $(h_x=1.604 J)$, we replace the plots with noisy simulation data of the corresponding QPU circuits with the following noise parameters: readout bit-flip $p_{ro}=5.9\times10^{-3}$, one-qubit depolarizing $p_1=3\times 10^{-4}$, two-qubit depolarizing $p_2=9.375\times 10^{-3}$, and $2496$ shots for each circuit. }
    \label{fig:dsf_xx}
\end{figure*}

\begin{figure*}[!htbp]
    \centering
    \includegraphics[width=0.9\linewidth]{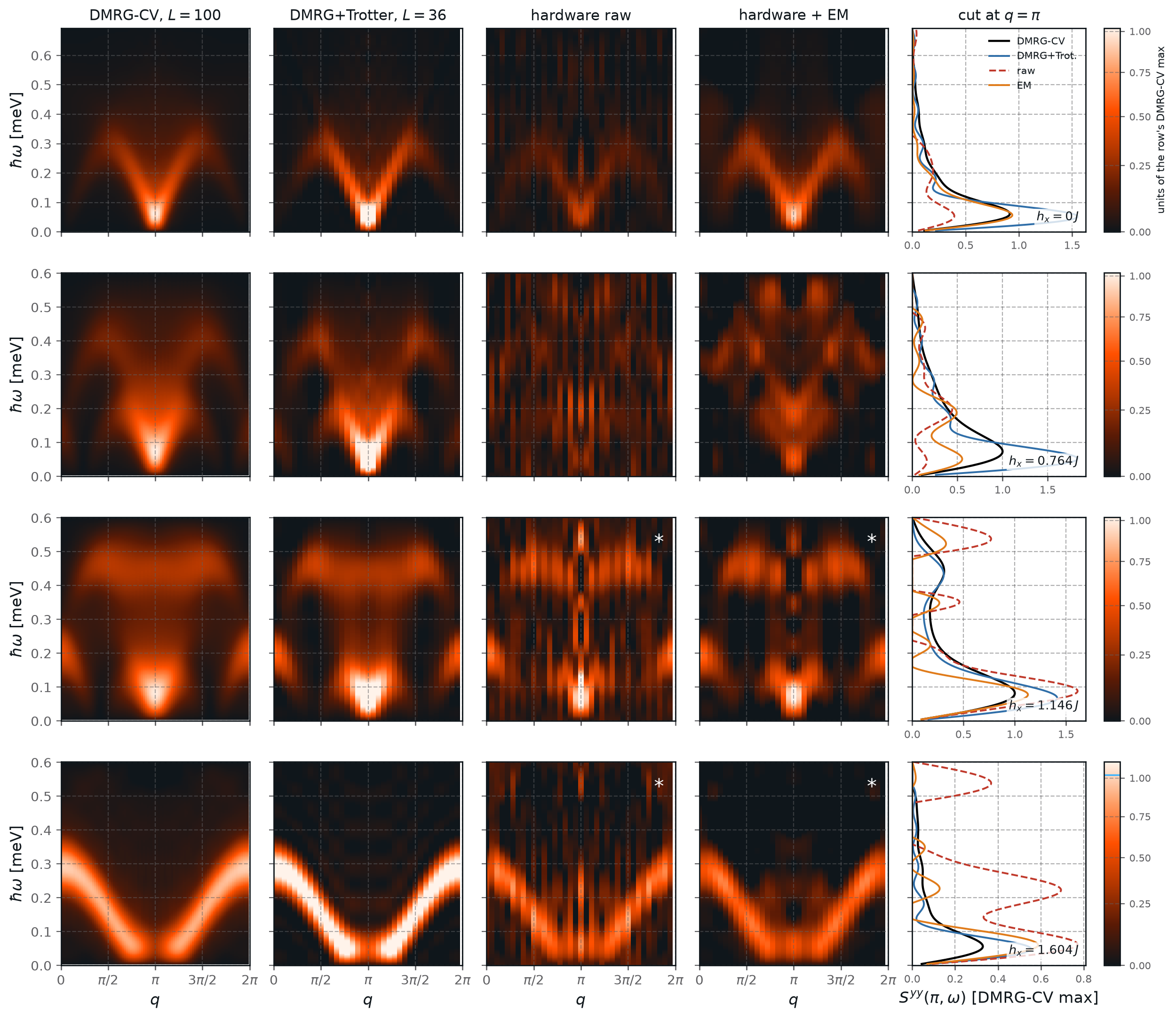}
    \caption{Transverse DSF $S^{yy}(q,\omega)$ of the $L=36$ anisotropic $(\Delta=0.25)$ chain, with each panel  normalized to the DMRG-CV maximum. Rows: plot the heatmap for at fields $h_x/J=(0,\ 0.764, \ 1.146, \ 1.604)$. Columns: correspond to DMRG-CV, DMRG+Trotter, raw hardware and error mitigated hardware results. ${}^{*}$For data-points without hardware runs $(h_x=1.146J, \ 1.604 J)$, we replace the plots with noisy simulation data of the corresponding QPU circuits with the following noise parameters: readout bit-flip $p_{ro}=5.9\times10^{-3}$, one-qubit depolarizing $p_1=3\times 10^{-4}$, two-qubit depolarizing $p_2=9.375\times 10^{-3}$, and $2496$ shots for each circuit. }
    \label{fig:dsf_yy}
\end{figure*}

\begin{figure*}[!htbp]
    \centering
    \includegraphics[width=0.9\linewidth]{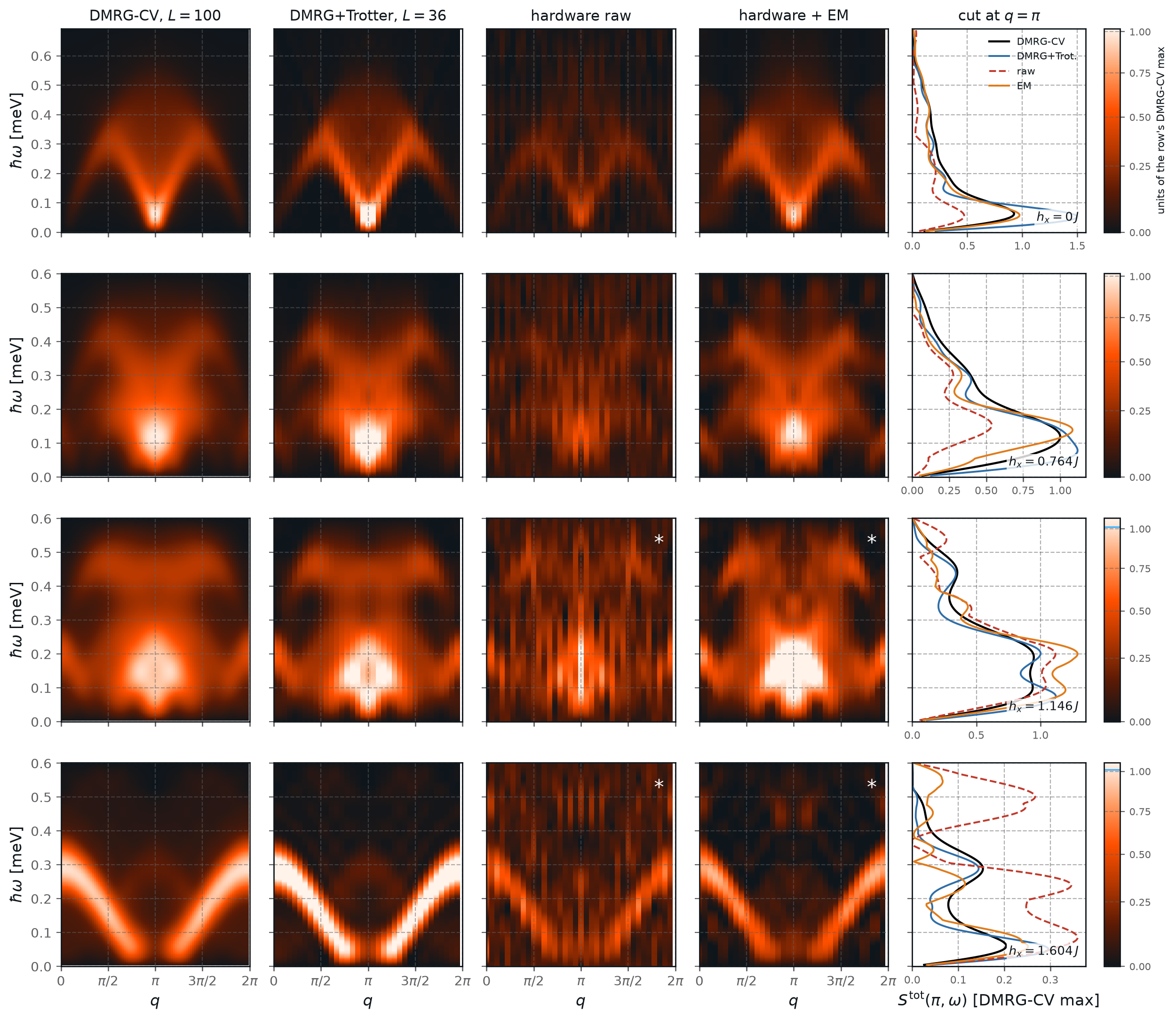}
    \caption{Total DSF $S_{\text{tot}}(q,\omega)=S^{xx}(q,\omega)+S^{yy}(q,\omega)+S^{zz}(q,\omega)$ of the $L=36$ anisotropic $(\Delta=0.25)$ chain, with each panel  normalized to its own maximum. Rows: plot the heatmap for at fields $h_x/J=(0,\ 0.764, \ 1.146, \ 1.604)$. Columns: correspond to DMRG-CV, DMRG+Trotter, raw hardware and error mitigated hardware results. ${}^{*}$For data-points without hardware runs $(h_x=1.146J, \ 1.604 J)$, we replace the plots with noisy simulation data of the corresponding QPU circuits with the following noise parameters: readout bit-flip $p_{ro}=5.9\times10^{-3}$, one-qubit depolarizing $p_1=3\times 10^{-4}$, two-qubit depolarizing $p_2=9.375\times 10^{-3}$, and $2496$ shots for each circuit. For the transverse field values $(h_x=1.604 J)$, we used both QPU and noisy simulations to calculate $S_{\text{tot}}(q, \omega)$.  }
    \label{fig:dsf_total}
\end{figure*}

Figure~\ref{fig:dsf_zz} and Figures.~\ref{fig:dsf_xx}--\ref{fig:dsf_total} show the spin components and total DSF at $h_x/J=(0,\,0.764,\,1.146,\,1.604)$. At zero field, the lower boundary of the two spinon continuum is visible in $S^{xx}$ and $S^{yy}$ and follows $\omega_L(q)=v_s|\sin q|$, with $v_s=1.154J$ for $\Delta=\cos\gamma=0.25$. The upper boundary, $\omega_U(q)=2v_s|\sin(q/2)|$, is more visible in $S^{zz}$ and the total response. These two boundaries were experimentally resolved by INS for a near-ideal Heinsenberg chain KCuF$_3$ in Ref. \cite{lake2013multispinon}. The field separates the spin channels and concentrates weight into a prominent dispersive branch, with a weaker second band at intermediate fields. The peak locations and their interpretation are given in Section~\ref{sec:iv}.

The hardware spectra retain the principal continuum and branch features seen in DMRG+Trotter, although localized discrepancies remain near $q=\pi$. The minimum compression channel-averaged fidelity over the studied targets is $0.62$ (this is an observed minimum over $124$ EV sweeps during the optimization of the compressed state). Raw hardware spectra retain these structures but develop a noise floor of approximately $5\%$ to $10\%$ in regions with little reference intensity. Error mitigation makes the boundaries and branch positions clearer across the measured fields, although defects remain, including double peaks near $q=\pi$ and $\hbar\omega\simeq0.5\,\mathrm{meV}$ at $h_x=1.604J$.

\subsection{Quantitative comparisons and uncertainty analysis}
\label{sec:supp-validation}

Table~\ref{tab:metrics} summarizes the total spectra, while Table~\ref{tab:metricschan} gives the selected component comparisons: $xx/yy$ and $zz$ at zero field, $xx$ and $zz$ at $h_x=0.764J$, and $zz$ at $h_x=1.146J$ and $1.604J$. All three spin components were measured at $h_x=0.764J$ and contribute to the total response; only the $yy$ column at that field has been omitted from the component table. Every difference is (QPU $+$ EM) minus DMRG-CV, evaluated on the
common estimator grid over $0<\hbar\omega\le\omega_\mathrm{max}$. The Pearson and amplitude comparisons in Section~\ref{sec:4e} distinguish agreement in spectral pattern from the residual overall intensity offsets. The energy moments, widths, and partial weights test changes within the lineshape that a single gain factor cannot describe. A narrower line can have a larger peak without having a larger fitted overall amplitude.

At zero field the measured response is broader than the reference, with an rms width at $q=\pi$ of $0.156$ against $0.145$~meV. At $h_x = 0.764\,J$ it is narrower, $0.104$ against $0.131$~meV, and the fitted amplitude $A = 0.869$ lies $13\%$ below unity. Where the text calls the finite-field response strong, it refers to peak height, not to $A$.

The width differences are $+0.0106\pm0.0108$~meV at zero field and $-0.0266\pm0.0074$~meV at $0.764\,J$. These correspond to approximately 1.0 and 3.6 statistical standard deviations, respectively. Their distortion bands for these metrics are ($-0.0059,+0.0097$) and ($-0.0046,+0.0052$)~meV (Table~\ref{tab:distortion}). The finite-field spectrum is therefore narrower than the reference throughout the tested processing variations. The centroid differences at $q=\pi$ are $+0.0298\pm0.0125$ and $-0.0111\pm0.0088$~meV. Their processing sensitivity ranges are $(-0.0072,+0.0386)$ and $(-0.0034,+0.0120)$ meV, demonstrating that the inferred offsets also depend on reconstruction choices. The peak positions differ by $+0.0402\pm0.0017$ and $+0.0125\pm0.0032$ meV, corresponding to approximately eight and five evaluation-grid steps. Both offsets are below the nominal finite-time scale of $0.060$ meV; the finer evaluation grid does not increase the physical spectral resolution.

The partial weight above $0.20\,\mathrm{meV}$ tests whether spectral weight is redistributed even when the peak energies agree. Measured over upper cutoffs of $0.693$ and $0.602$~meV, the DMRG-CV differences are $+0.013\pm0.020$ and $+0.014\pm0.016$ at $h_x/J=0$ and $0.764$, respectively. Both are consistent with zero within the quoted statistical uncertainties. The processing sensitivity ranges about these nominal differences are $(-0.012,+0.022)$ and $(-0.017,+0.030)$, respectively (Table \ref{tab:distortion}. These results do not establish a field-dependent discrepancy in the partial spectral weight.

The threshold gap difference is zero at zero field and $+0.0025\pm0.0018\,\mathrm{meV}$ at $h_x=0.764J$ at a 10\% threshold. Raising the threshold to 20\% changes these differences to \(+0.0050\) and \(+0.0075\pm0.0036\) meV (Table~\ref{tab:metrics}). These threshold-dependent estimates should not be interpreted as resolving gaps below the finite-time spectral resolution.The bandwidth is less stable because it depends on extremal peak positions throughout the zone. It is also the one clear discrepancy: $\Delta B = -0.2058 \pm 0.0562$ meV at zero field, $3.7$ standard deviations from zero and above the effective resolution of $0.101$ meV, against $+0.0225 \pm 0.0296$ meV at $0.764\,J$, which is consistent with zero. Because the DMRG-CV reference retains $\eta=0.1J$ without matching the hardware observation window, the difference in resolution can contribute to this bandwidth deficit alongside errors in the reconstructed response. The zero-field distortion band reaches $0.090$ meV; it is the quantitative form of the underestimated continuum extent described in Section \ref{sec:4e}.

\begin{table*}[t]
\caption{\label{tab:metricschan}Error-mitigated hardware against the published $L=100$ DMRG-CV reference resolved by spin component. $S^{zz}$ was measured on hardware at every field and all three components at $h_x=0.764\,J$; at $h_x=0$ the $U(1)$ symmetry about $z$ makes $S^{yy}$ identical to $S^{xx}$, and that column stands for both. The second column gives the value each quantity would take for a perfect reconstruction, and every entry is an error or a ratio, not an absolute value. Indented rows are the propagated shot-noise standard deviation over $160$ draws through the full mitigation chain, and are statistical only. The QPU+EM DSF values for these comparisons are unnormalized and left in absolute units. }
\begin{ruledtabular}
\begin{tabular}{lccccccc}
 & & \multicolumn{2}{c}{$h_x=0$} & \multicolumn{2}{c}{$h_x=0.764\,J$} & $h_x=1.146\,J$ & $h_x=1.604\,J$ \\
\cline{3-4}\cline{5-6}\cline{7-7}\cline{8-8}
Quantity & Ideal & $S^{xx}\!=\!S^{yy}$ & $S^{zz}$ & $S^{xx}$ & $S^{zz}$ & $S^{zz}$ & $S^{zz}$ \\
\hline
$r$ & $1$ & $0.896$ & $0.955$ & $0.973$ & $0.896$ & $0.896$ & $0.862$ \\
\quad & & $\pm0.014$ & $\pm0.005$ & $\pm0.004$ & $\pm0.020$ & $\pm0.012$ & $\pm0.015$ \\
$A$ & $1$ & $0.712$ & $1.095$ & $1.005$ & $0.969$ & $1.077$ & $0.522$ \\
\quad & & $\pm0.031$ & $\pm0.025$ & $\pm0.025$ & $\pm0.045$ & $\pm0.041$ & $\pm0.024$ \\
\hline
MSE & $0$ & $0.0056$ & $0.0018$ & $0.0010$ & $0.0021$ & $0.0026$ & $0.0108$ \\
\quad & & $\pm0.0006$ & $\pm0.0002$ & $\pm0.0001$ & $\pm0.0004$ & $\pm0.0004$ & $\pm0.0007$ \\
\quad normalized, MSE$/\langle S_\mathrm{ref}^2\rangle$ & $0$ & $0.184$ & $0.085$ & $0.038$ & $0.119$ & $0.157$ & $0.299$ \\
\quad & & $\pm0.019$ & $\pm0.010$ & $\pm0.006$ & $\pm0.021$ & $\pm0.024$ & $\pm0.021$ \\
$W_1$ & $0$ & $0.0488$ & $0.0261$ & $0.0125$ & $0.0422$ & $0.0360$ & $0.0594$ \\
\quad & & $\pm0.0171$ & $\pm0.0070$ & $\pm0.0031$ & $\pm0.0118$ & $\pm0.0089$ & $\pm0.0116$ \\
SSIM & $1$ & $0.861$ & $0.880$ & $0.923$ & $0.840$ & $0.706$ & $0.414$ \\
\quad & & $\pm0.058$ & $\pm0.017$ & $\pm0.013$ & $\pm0.025$ & $\pm0.037$ & $\pm0.044$ \\
\hline
$\Delta\langle\hbar\omega\rangle_{q=\pi}$ [meV] & $0$ & $-0.0008$ & $0.0249$ & $-0.0126$ & $-0.0279$ & $-0.0251$ & $0.0960$ \\
\quad & & $\pm0.0238$ & $\pm0.0079$ & $\pm0.0070$ & $\pm0.0103$ & $\pm0.0111$ & $\pm0.1127$ \\
$\Delta\langle\hbar\omega\rangle_{q=\pi/2}$ [meV] & $0$ & $-0.0316$ & $-0.0009$ & $0.0043$ & $-0.0036$ & $0.1123$ & $-0.0200$ \\
\quad & & $\pm0.0085$ & $\pm0.0079$ & $\pm0.0160$ & $\pm0.0275$ & $\pm0.0208$ & $\pm0.0276$ \\
$\Delta\hbar\omega_\mathrm{pk}$, $q=\pi$ [meV]$^{\dagger}$ & $0$ & $0.0351$ & $0.0151$ & $0.0075$ & $0.2300$ & $0.0250$ & $0.0725$ \\
\quad & & $\pm0.0025$ & $\pm0.1641$ & $\pm0.0021$ & $\pm0.0926$ & $\pm0.0065$ & $\pm0.1645$ \\
$\Delta\sigma_\omega$, $q=\pi$ [meV] & $0$ & $-0.0230$ & $0.0126$ & $-0.0119$ & $-0.0319$ & $-0.0318$ & $-0.0083$ \\
\quad & & $\pm0.0485$ & $\pm0.0064$ & $\pm0.0075$ & $\pm0.0067$ & $\pm0.0077$ & $\pm0.0257$ \\
\hline
$\Delta R_{>}$ ($\hbar\omega_b=0.20$~meV) & $0$ & $0.003$ & $-0.016$ & $-0.002$ & $-0.039$ & $-0.036$ & $0.074$ \\
\quad & & $\pm0.038$ & $\pm0.015$ & $\pm0.018$ & $\pm0.029$ & $\pm0.021$ & $\pm0.027$ \\
\hline
$\Delta$gap, $q=\pi$ [meV]$^{\dagger}$ & $0$ & $0.0000$ & $0.0000$ & $0.0075$ & $0.0000$ & $-0.0050$ & $0.0700$ \\
\quad & & $\pm0.0000$ & $\pm0.0025$ & $\pm0.0036$ & $\pm0.0019$ & $\pm0.0012$ & $\pm0.0387$ \\
\quad at a $20\%$ threshold$^{\dagger}$ & $0$ & $0.0050$ & $0.0000$ & $-0.0100$ & $0.0025$ & $-0.0050$ & $0.0700$ \\
\quad & & $\pm0.0000$ & $\pm0.0024$ & $\pm0.0030$ & $\pm0.0038$ & $\pm0.0019$ & $\pm0.0613$ \\
$\Delta B$ [meV]$^{\dagger}$ & $0$ & $0.0201$ & $0.0502$ & $0.0750$ & $0.1375$ & $-0.0025$ & $0.1500$ \\
\quad & & $\pm0.1078$ & $\pm0.0411$ & $\pm0.0634$ & $\pm0.0657$ & $\pm0.0136$ & $\pm0.0401$ \\
\end{tabular}
\end{ruledtabular}
\end{table*}

The neutron comparison supplies an experimental context for validation against DMRG-CV. As discussed in Section~\ref{sec:4e}, QPU+EM versus INS should be interpreted alongside DMRG-CV versus INS on the same comparison window, rather than against perfect correlation. A separate consistency check compares the measured total neutron spectrum with the sum of its three inferred spin components, giving correlations of $0.979$ and $0.960$ at $h_x=0$ and $0.764J$.

Reflection of the neutron spectrum, $S(q,\omega)\mapsto S(2\pi-q,\omega)$, provides an additional check on experimental consistency. Writing the measured map as $x=s+n$, its reflection is $\tilde x=s+\tilde n$ provided the true spectrum is inversion symmetric, so that $\rho=\mathrm{corr}(x,\tilde x)=V_s/(V_s+V_n)$ is the reliability of the map. Under these assumptions, the underlying signal $s$ correlates with the measured map as $\mathrm{corr}(s,x)=\sqrt{\rho}$, setting the upper limit for a prediction independent of the experimental noise. Noise enters both maps in the reflection comparison, but only $x$ in the comparison with $s$, which accounts for the square root. We find $\rho=0.919$ and $0.860$ at $0$ and $1$~T, giving $\sqrt{\rho}=0.959$ and $0.927$; the pooled value $0.902$ obscures a real difference in data quality between the two fields. This is a lower bound on the attainable correlation rather than the attainable correlation itself, since genuine asymmetry in $s$ would depress $\rho$ alongside the noise. It assumes inversion symmetry of the true spectrum, independence of the noise at reflected cells, and equal noise variance there.

The residual measures in Table~\ref{tab:chi2} complement Pearson correlation and the unweighted amplitude fit. Let $I_i$ be the reference, $M_i$ the map fitted to it, and $i$ index the $N$ cells in the comparison window. The fitted scale is
\begin{equation}
 \alpha=\frac{\sum_i w_iM_iI_i}{\sum_i w_iM_i^2},
 \label{eq:alpha}
\end{equation}
with $w_i=\sigma_{\mathrm{INS},i}^{-2}$ when the neutron uncertainties alone are used, $w_i=\sigma_{\mathrm{QPU},i}^{-2}$ when the hardware uncertainties alone are used, and $w_i=(\sigma_{\mathrm{INS},i}^2+\alpha^2\sigma_{\mathrm{QPU},i}^2)^{-1}$ when both are included. The modified reduced chi squared statistic and profile residual are
\begin{equation}
 \chi^2_{\mathrm{red}}=\frac{1}{N-p}\sum_iw_i(\alpha M_i-I_i)^2,
 \label{eq:chi2}
\end{equation}
\begin{equation}
 R_p=\frac{\sum_i|\alpha M_i-I_i|}{\sum_i|I_i|}.
 \label{eq:rp}
\end{equation}
The first measures squared residuals relative to their assigned uncertainties, while the second measures their total absolute magnitude relative to the reference intensity. The weighted profile residual, expected residual, and goodness of fit are
\begin{equation}
 R_{wp}=\left[\frac{\sum_iw_i(\alpha M_i-I_i)^2}{\sum_iw_iI_i^2}\right]^{1/2},
 \label{eq:rwp}
\end{equation}
\begin{equation}
 R_{\mathrm{exp}}=\left[\frac{N-p}{\sum_iw_iI_i^2}\right]^{1/2},
 \label{eq:rexp}
\end{equation}
\begin{equation}
 \mathrm{GoF}=\frac{R_{wp}}{R_{\mathrm{exp}}}=\sqrt{\chi^2_{\mathrm{red}}}.
 \label{eq:gof}
\end{equation}
A GoF near unity indicates residuals commensurate with the assigned errors under the weighting model. 
One parameter is fitted in every row, $p=1$, namely the scale $\alpha$; no other quantity is adjusted. The fit is not symmetric in its two arguments, since the denominators of $R_p$ and $R_{wp}$ are formed from the data side, so we state the direction explicitly: in the three neutron rows the neutron map is the data and the calculation or the hardware result is the model, while in the last row the hardware map is the data and DMRG-CV is the model. This is why $\alpha\simeq2.6$ in the former, where it carries the conversion from meV$^{-1}$ to spin units, and $\alpha\simeq0.9$ in the latter, where both sides already share units.

When both uncertainties are included the weight $w_i=(\sigma_{\mathrm{INS},i}^2+\alpha^2\sigma_{\mathrm{QPU},i}^2)^{-1}$ depends on the scale being fitted, so Eq.~\ref{eq:alpha} and the weights are solved together by fixed-point iteration from $\alpha=1$; the scale is stable to five decimal places within five iterations. The weight matrix is diagonal, and correlations between map cells are not modeled but are present; the correlated residuals leave the expectation of $\chi^2_\mathrm{red}$ unchanged when the assigned variances are correct, but they inflate its sampling variance, so differences between rows at the level of $0.1$ in GoF should not be read as meaningful. Finally, $R_\mathrm{exp}$ is the value $R_{wp}$ - not $R_p$ - would take for a perfect model with the assigned errors. 

\begin{table*}[t]
\caption{\label{tab:chi2} Modified reduced chi-square for the QPU+EM, DMRG-CV, and INS $S_\mathrm{tot}$ comparisons, weighted by measurement uncertainty. $\chi^2_\mathrm{red}$ is then the mean squared residual in units of the error bar, and its square root is the typical miss per cell. The QPU+EM DSF values for these comparisons are unnormalized and left in absolute units. }
\begin{ruledtabular}
\begin{tabular}{llcccccc}
 & & \multicolumn{3}{c}{$h_x=0$} & \multicolumn{3}{c}{$h_x=0.764\,J$} \\
\cline{3-5}\cline{6-8}
Comparison & Weight & $R_p$ [\%] & $\chi^2_\mathrm{red}$ & GoF & $R_p$ [\%] & $\chi^2_\mathrm{red}$ & GoF \\
\hline
INS \textit{vs.} DMRG-CV & $\sigma_\mathrm{INS}$ & $42.2$ & $5.64$ & $2.38$ & $39.5$ & $2.66$ & $1.63$ \\
INS \textit{vs.} QPU $+$ EM & $\sigma_\mathrm{INS}$ & $52.5$ & $9.59$ & $3.10$ & $42.2$ & $3.35$ & $1.83$ \\
\quad same, both errors & $\sigma_\mathrm{INS}\oplus\sigma_\mathrm{QPU}$ & $51.8$ & $5.40$ & $2.32$ & $42.0$ & $2.16$ & $1.47$ \\
\textit{floor} $R_\mathrm{exp}$ & & \textit{16.8} & --- & \textit{1} & \textit{21.8} & --- & \textit{1} \\
\hline
DMRG-CV \textit{vs.} QPU $+$ EM & $\sigma_\mathrm{QPU}$ & $33.3$ & $9.17$ & $3.03$ & $48.4$ & $21.52$ & $4.64$ \\
\textit{floor} $R_\mathrm{exp}$ & & \textit{14.7} & --- & \textit{1} & \textit{14.5} & --- & \textit{1} \\
\end{tabular}
\end{ruledtabular}
\end{table*}

With neutron uncertainties used for both comparisons, Table~\ref{tab:chi2} gives $R_p=(42.2\%,\,39.5\%)$ for DMRG-CV and $(52.5\%, 42.2\%)$ for QPU+EM at the two fields. With neutron uncertainties used for both fits, the hardware profile residual exceeds the classical profile residual by $10.3$ percentage points at zero field and $2.7$ percentage points at $h_x=0.764\,J$. That comparison is not like for like, however: it charges the hardware with its own shot noise while assigning it no uncertainty. Including both neutron and hardware uncertainties reduces the hardware $\chi^2_\mathrm{red}$ from $9.59$ to $5.40$ at zero field and from $3.35$ to $2.16$ at $0.764\,J$, giving $\mathrm{GoF}=(2.32,\,1.47)$ against $(2.38,\,1.63)$ for DMRG-CV against INS. The difference in GoF is $0.06$ and $0.16$ at zero field and $h_x=0.764\,J$, respectively. These values are similar in magnitude, but the comparisons use different uncertainty models and neglect correlations between spectral cells; they therefore do not establish equivalent agreement with the neutron data. The direct comparison of DMRG-CV with QPU $+$ EM gives $R_p=(33.3\%,\,48.4\%)$ with $\mathrm{GoF}=(3.03,\,4.64)$. The residuals therefore exceed the scale accounted for by the assigned hardware statistical uncertainties. The tabulated $R_\mathrm{exp}$ values, $(16.8\%,\,21.8\%)$ for the neutron rows and $(14.7\%,\,14.5\%)$ for the direct comparison, are the values $R_{wp}$ would take for a perfect model carrying the assigned errors. They are floors for $R_{wp}$, not for $R_p$, and the two are not ordered in general.

Tables~\ref{tab:distortion} and~\ref{tab:distortionzz} separate propagated shot noise from sensitivity to the chosen processing parameters. The former uses the total spectra at the two fields with a complete hardware response; the latter follows $S^{zz}$ across all four fields. The temporal taper, denoising rank, spatial mask parameter, and mirror exponent are varied around their nominal settings, with the reference recalculated whenever the taper changes. The resulting sensitivity bands measure dependence on these analysis choices and are reported separately from statistical uncertainty.

Absolute intensity is particularly sensitive to the mirror calibration. 
For the total response the amplitude carries $A=0.833$ with $\sigma_A=0.025$ and $\delta_A= (-0.213,+0.116)$ at zero field, and $A=0.869$ with $\sigma_A=0.029$ and $\delta_A=(-0.232,+0.167)$ at $0.764\,J$. The mirror exponent alone contributes $(-0.094, +0.110)$ and $(-0.112, +0.133)$. The systematic band exceeds the statistical one by a factor $4.6$-$8.5$ at zero field and $5.8$-$8.0$ at $0.764\,J$. 
A common amplitude rescaling leaves Pearson correlation unchanged.
 
Tables \ref{tab:distortion} and \ref{tab:distortionzz} show that sensitivity to processing choices can be comparable to or larger than the propagated shot-noise uncertainty. For example, the largest absolute processing-induced change in the total-spectrum Pearson correlation is $0.061$ at zero field and $0.014$ at $h_x=0.764J$, compared with statistical uncertainties of $0.009$ and $0.007$, respectively. For the zero-field centroid at $q=\pi$, the largest change is $0.0386$ meV, approximately $3.1$ times the statistical uncertainty of $0.0125$ meV. Statistical error bars alone therefore do not describe the full dependence of the results on the analysis. We report the statistical uncertainties and processing sensitivity ranges separately because the latter are not probabilistic confidence intervals.
A small sensitivity band certifies that a number is stable under reprocessing; it does not make the number small. For example, the zero-field total
bandwidth has a value of
$-0.206$~meV, above the effective resolution, while its distortion band reaches $0.090$~meV - substantial, but still smaller than the deficit it qualifies. The discrepancy is robust to processing and larger
than the resolution at the same time.

Table~\ref{tab:distortion} reports a nonzero change in the Pearson coefficient when only the mirror exponent is varied, although that exponent acts as a multiplicative rescaling. The two are consistent because the rescaling is per channel: each component carries its own fitted exponent, so varying them together reweights the mixture that forms $S_\mathrm{tot}$ and alters its shape. The single-channel tables confirm this - in every $S^{zz}$ row the mirror-exponent entry for $r$ is identically zero, as a pure rescaling requires, while for the total spectra,
Table \ref{tab:distortion} reports changes of $(-0.002,+0.001)$ at
$h_x=0$ and zero at the reported precision at
$h_x=0.764J$.

\begin{table*}[t]
\caption{\label{tab:distortion}Stochastic and distortion bands on each metric of Table~\ref{tab:metrics}, for the total dynamical spin structure factor $S^{\mathrm{tot}}$. $\sigma_M$ is the statistical band already quoted there: shot noise resampled $160$ times through the full mitigation chain. $\delta_M$ is the distortion band, the sensitivity of the same metric to the four choices made in the analysis, none of which is a measurement. Each is varied about its nominal setting - the mirror exponent by $\pm0.05$, the taper over half-Hann, half-DPSS $2.0$ and half-Kaiser $1.5$ against the nominal half-DPSS $1.5$, the MSSA rank over $3$ and $5$ against $4$, and the spatial mask parameter over $0.85$ and $1.15$ against $1$ - and the tabulated value is the largest deviation from nominal over that knob's settings, with $\delta_M$ joint Cartesian band over $3\times7\times3\times3 = 189$ settings. The two bands are kept separate because they do not behave alike: $\sigma_M$ falls as more shots are taken and $\delta_M$ does not.  }
\scriptsize
\setlength{\tabcolsep}{1.5pt}%
\renewcommand{\arraystretch}{1.15}%
\begin{ruledtabular}
\begin{tabular}{lccccccc}
 & & & \multicolumn{4}{c}{deviation per knob} & \\
\cline{4-7}
Quantity & $M$ & $\sigma_M$ & mirror exp.\ $a$ & taper & MSSA rank & spatial mask $v$ & $\delta_M$\\
\hline
\multicolumn{8}{c}{$h_x = 0\,J$ (0~T)}\\
\hline
$r$ & $0.920$ & $0.009$ & $(-0.002,\,+0.001)$ & $(-0.053,\,+0.000)$ & $0$ & $(-0.007,\,+0.002)$ & $(-0.061,\,+0.003)$ \\
$A$ & $0.833$ & $0.025$ & $(-0.094,\,+0.110)$ & $(-0.121,\,+0.000)$ & $(-0.011,\,+0.002)$ & $(-0.010,\,+0.003)$ & $(-0.213,\,+0.116)$ \\
\hline
MSE & $0.0251$ & $0.0026$ & $(-0.0000,\,+0.0038)$ & $(-0.0000,\,+0.0171)$ & $(-0.0000,\,+0.0004)$ & $(-0.0007,\,+0.0019)$ & $(-0.0007,\,+0.0238)$ \\
\quad normalized, MSE$/\langle S^2_\mathrm{ref}\rangle$ & $0.122$ & $0.013$ & $(-0.000,\,+0.018)$ & $(-0.000,\,+0.083)$ & $(-0.000,\,+0.002)$ & $(-0.003,\,+0.009)$ & $(-0.003,\,+0.116)$ \\
$W_1$ & $0.0370$ & $0.0114$ & $(-0.0002,\,+0.0009)$ & $(-0.0000,\,+0.0052)$ & $(-0.0001,\,+0.0022)$ & $(-0.0014,\,+0.0000)$ & $(-0.0017,\,+0.0133)$ \\
SSIM & $0.919$ & $0.013$ & $(-0.017,\,+0.000)$ & $(-0.034,\,+0.000)$ & $(-0.008,\,+0.000)$ & $(-0.005,\,+0.001)$ & $(-0.068,\,+0.001)$ \\
\hline
$\Delta\langle\hbar\omega\rangle_{q=\pi}$ [meV] & $0.0298$ & $0.0125$ & $(-0.0058,\,+0.0060)$ & $(-0.0000,\,+0.0235)$ & $(-0.0007,\,+0.0038)$ & $(-0.0017,\,+0.0010)$ & $(-0.0072,\,+0.0386)$ \\
$\Delta\langle\hbar\omega\rangle_{q=\pi/2}$ [meV] & $-0.0176$ & $0.0060$ & $(-0.0013,\,+0.0013)$ & $(-0.0001,\,+0.0001)$ & $(-0.0003,\,+0.0061)$ & $(-0.0023,\,+0.0012)$ & $(-0.0038,\,+0.0103)$ \\
$\Delta\hbar\omega_\mathrm{pk}$, $q=\pi$ [meV]$^{\dagger}$ & $0.0402$ & $0.0017$ & $0$ & $(-0.0000,\,+0.0201)$ & $0$ & $0$ & $(-0.0000,\,+0.0251)$ \\
$\Delta\sigma_\omega$, $q=\pi$ [meV] & $0.0106$ & $0.0108$ & $(-0.0027,\,+0.0025)$ & $(-0.0005,\,+0.0000)$ & $(-0.0000,\,+0.0039)$ & $(-0.0030,\,+0.0000)$ & $(-0.0059,\,+0.0097)$ \\
\hline
$\Delta R_{>}$ & $0.013$ & $0.020$ & $(-0.008,\,+0.008)$ & $(-0.000,\,+0.011)$ & $(-0.004,\,+0.003)$ & $(-0.003,\,+0.000)$ & $(-0.012,\,+0.022)$ \\
\hline
$\Delta$gap, $q=\pi$ [meV]$^{\dagger}$ & $0.0000$ & $0.0000$ & $0$ & $(-0.0000,\,+0.0050)$ & $0$ & $0$ & $(-0.0000,\,+0.0050)$ \\
$\Delta B$ [meV]$^{\dagger}$ & $-0.2058$ & $0.0562$ & $0$ & $(-0.0201,\,+0.0000)$ & $(-0.0000,\,+0.0050)$ & $(-0.0050,\,+0.0000)$ & $(-0.0201,\,+0.0904)$ \\
\hline
\multicolumn{8}{c}{$h_x = 0.764\,J$ (1~T)}\\
\hline
$r$ & $0.962$ & $0.007$ & $0$ & $(-0.011,\,+0.000)$ & $(-0.004,\,+0.000)$ & $(-0.000,\,+0.001)$ & $(-0.014,\,+0.001)$ \\
$A$ & $0.869$ & $0.029$ & $(-0.112,\,+0.133)$ & $(-0.071,\,+0.000)$ & $(-0.069,\,+0.027)$ & $(-0.005,\,+0.002)$ & $(-0.232,\,+0.167)$ \\
\hline
MSE & $0.0064$ & $0.0011$ & $(-0.0009,\,+0.0046)$ & $(-0.0000,\,+0.0035)$ & $(-0.0000,\,+0.0027)$ & $(-0.0000,\,+0.0001)$ & $(-0.0011,\,+0.0140)$ \\
\quad normalized, MSE$/\langle S^2_\mathrm{ref}\rangle$ & $0.049$ & $0.008$ & $(-0.007,\,+0.035)$ & $(-0.000,\,+0.027)$ & $(-0.000,\,+0.020)$ & $(-0.000,\,+0.001)$ & $(-0.008,\,+0.106)$ \\
$W_1$ & $0.0185$ & $0.0044$ & $(-0.0000,\,+0.0003)$ & $(-0.0000,\,+0.0018)$ & $(-0.0031,\,+0.0043)$ & $(-0.0012,\,+0.0005)$ & $(-0.0037,\,+0.0054)$ \\
SSIM & $0.914$ & $0.019$ & $(-0.030,\,+0.000)$ & $(-0.042,\,+0.000)$ & $(-0.036,\,+0.000)$ & $(-0.004,\,+0.000)$ & $(-0.149,\,+0.000)$ \\
\hline
$\Delta\langle\hbar\omega\rangle_{q=\pi}$ [meV] & $-0.0111$ & $0.0088$ & $(-0.0008,\,+0.0007)$ & $(-0.0000,\,+0.0070)$ & $(-0.0018,\,+0.0031)$ & $(-0.0000,\,+0.0021)$ & $(-0.0034,\,+0.0120)$ \\
$\Delta\langle\hbar\omega\rangle_{q=\pi/2}$ [meV] & $0.0026$ & $0.0141$ & $(-0.0017,\,+0.0019)$ & $(-0.0027,\,+0.0000)$ & $(-0.0000,\,+0.0060)$ & $(-0.0030,\,+0.0000)$ & $(-0.0052,\,+0.0076)$ \\
$\Delta\hbar\omega_\mathrm{pk}$, $q=\pi$ [meV]$^{\dagger}$ & $0.0125$ & $0.0032$ & $(-0.0025,\,+0.0000)$ & $(-0.0050,\,+0.0000)$ & $0$ & $(-0.0025,\,+0.0025)$ & $(-0.0050,\,+0.0050)$ \\
$\Delta\sigma_\omega$, $q=\pi$ [meV] & $-0.0266$ & $0.0074$ & $(-0.0005,\,+0.0005)$ & $(-0.0007,\,+0.0012)$ & $(-0.0031,\,+0.0033)$ & $(-0.0000,\,+0.0012)$ & $(-0.0046,\,+0.0052)$ \\
\hline
$\Delta R_{>}$ & $0.014$ & $0.016$ & $(-0.005,\,+0.005)$ & $(-0.000,\,+0.021)$ & $(-0.008,\,+0.004)$ & $(-0.000,\,+0.002)$ & $(-0.017,\,+0.030)$ \\
\hline
$\Delta$gap, $q=\pi$ [meV]$^{\dagger}$ & $0.0025$ & $0.0018$ & $(-0.0025,\,+0.0000)$ & $0$ & $(-0.0025,\,+0.0000)$ & $(-0.0025,\,+0.0000)$ & $(-0.0025,\,+0.0025)$ \\
$\Delta B$ [meV]$^{\dagger}$ & $0.0225$ & $0.0296$ & $(-0.0025,\,+0.0025)$ & $(-0.0225,\,+0.0000)$ & $(-0.0075,\,+0.0000)$ & $(-0.0025,\,+0.0025)$ & $(-0.0225,\,+0.0075)$ \\
\hline
\end{tabular}
\end{ruledtabular}
\end{table*}

\begin{table*}[t]
\caption{\label{tab:distortionzz}Stochastic and distortion bands on the $S^{zz}$ metrics across the full field sweep. $S^{zz}$ is the only component measured on hardware at every field, so it is the only one whose analysis systematics can be followed through the transition. The construction is that of Table~\ref{tab:distortion}. }
\scriptsize
\setlength{\tabcolsep}{1.5pt}%
\renewcommand{\arraystretch}{1.00}%
\begin{ruledtabular}
\begin{tabular}{lccccccc}
 & & & \multicolumn{4}{c}{deviation per knob} & \\
\cline{4-7}
Quantity & $M$ & $\sigma_M$ & mirror exp.\ $a$ & taper & MSSA rank & spatial mask $v$ & $\delta_M$\\
\hline
\multicolumn{8}{c}{$h_x = 0\,J$ (0~T)}\\
\hline
$r$ & $0.955$ & $0.005$ & $0$ & $(-0.015,\,+0.000)$ & $(-0.002,\,+0.000)$ & $(-0.008,\,+0.009)$ & $(-0.023,\,+0.009)$ \\
$A$ & $1.095$ & $0.025$ & $(-0.207,\,+0.255)$ & $(-0.082,\,+0.000)$ & $(-0.019,\,+0.004)$ & $(-0.010,\,+0.013)$ & $(-0.285,\,+0.275)$ \\
\hline
MSE & $0.0018$ & $0.0002$ & $(-0.0005,\,+0.0033)$ & $(-0.0000,\,+0.0002)$ & $0$ & $(-0.0002,\,+0.0002)$ & $(-0.0007,\,+0.0035)$ \\
\quad normalized, MSE$/\langle S^2_\mathrm{ref}\rangle$ & $0.085$ & $0.010$ & $(-0.022,\,+0.153)$ & $(-0.002,\,+0.007)$ & $(-0.002,\,+0.002)$ & $(-0.011,\,+0.010)$ & $(-0.034,\,+0.165)$ \\
$W_1$ & $0.0261$ & $0.0070$ & $0$ & $(-0.0001,\,+0.0020)$ & $(-0.0001,\,+0.0021)$ & $(-0.0021,\,+0.0010)$ & $(-0.0025,\,+0.0050)$ \\
SSIM & $0.880$ & $0.017$ & $(-0.079,\,+0.013)$ & $(-0.031,\,+0.000)$ & $(-0.005,\,+0.000)$ & $(-0.002,\,+0.010)$ & $(-0.094,\,+0.024)$ \\
\hline
$\Delta\langle\hbar\omega\rangle_{q=\pi}$ [meV] & $0.0249$ & $0.0079$ & $0$ & $(-0.0000,\,+0.0108)$ & $(-0.0053,\,+0.0000)$ & $(-0.0093,\,+0.0017)$ & $(-0.0123,\,+0.0108)$ \\
$\Delta\langle\hbar\omega\rangle_{q=\pi/2}$ [meV] & $-0.0009$ & $0.0079$ & $0$ & $(-0.0000,\,+0.0002)$ & $(-0.0007,\,+0.0087)$ & $(-0.0000,\,+0.0013)$ & $(-0.0008,\,+0.0125)$ \\
$\Delta\hbar\omega_\mathrm{pk}$, $q=\pi$ [meV]$^{\dagger}$ & $0.0151$ & $0.1641$ & $0$ & $(-0.0000,\,+0.3263)$ & $0$ & $(-0.0050,\,+0.3162)$ & $(-0.0050,\,+0.3363)$ \\
$\Delta\sigma_\omega$, $q=\pi$ [meV] & $0.0126$ & $0.0064$ & $0$ & $(-0.0025,\,+0.0000)$ & $(-0.0014,\,+0.0014)$ & $(-0.0038,\,+0.0000)$ & $(-0.0069,\,+0.0014)$ \\
\hline
$\Delta R_{>}$ & $-0.016$ & $0.015$ & $0$ & $(-0.003,\,+0.000)$ & $(-0.006,\,+0.007)$ & $(-0.004,\,+0.000)$ & $(-0.009,\,+0.007)$ \\
\hline
$\Delta$gap, $q=\pi$ [meV]$^{\dagger}$ & $0.0000$ & $0.0025$ & $0$ & $(-0.0000,\,+0.0050)$ & $0$ & $(-0.0000,\,+0.0050)$ & $(-0.0000,\,+0.0050)$ \\
$\Delta B$ [meV]$^{\dagger}$ & $0.0502$ & $0.0411$ & $0$ & $(-0.0251,\,+0.0000)$ & $(-0.0602,\,+0.0050)$ & $(-0.0602,\,+0.0000)$ & $(-0.1104,\,+0.0050)$ \\
\hline
\multicolumn{8}{c}{$h_x = 0.764\,J$ (1~T)}\\
\hline
$r$ & $0.896$ & $0.020$ & $0$ & $(-0.002,\,+0.005)$ & $(-0.010,\,+0.000)$ & $(-0.001,\,+0.003)$ & $(-0.022,\,+0.010)$ \\
$A$ & $0.969$ & $0.045$ & $(-0.106,\,+0.119)$ & $(-0.097,\,+0.000)$ & $(-0.076,\,+0.029)$ & $(-0.004,\,+0.002)$ & $(-0.249,\,+0.155)$ \\
\hline
MSE & $0.0021$ & $0.0004$ & $(-0.0001,\,+0.0007)$ & $(-0.0004,\,+0.0000)$ & $(-0.0000,\,+0.0001)$ & $0$ & $(-0.0005,\,+0.0008)$ \\
\quad normalized, MSE$/\langle S^2_\mathrm{ref}\rangle$ & $0.119$ & $0.021$ & $(-0.007,\,+0.038)$ & $(-0.020,\,+0.000)$ & $(-0.001,\,+0.008)$ & $(-0.002,\,+0.000)$ & $(-0.030,\,+0.046)$ \\
$W_1$ & $0.0422$ & $0.0118$ & $0$ & $(-0.0050,\,+0.0000)$ & $(-0.0000,\,+0.0033)$ & $(-0.0019,\,+0.0015)$ & $(-0.0064,\,+0.0079)$ \\
SSIM & $0.840$ & $0.025$ & $(-0.018,\,+0.000)$ & $(-0.039,\,+0.006)$ & $(-0.037,\,+0.000)$ & $(-0.004,\,+0.000)$ & $(-0.143,\,+0.013)$ \\
\hline
$\Delta\langle\hbar\omega\rangle_{q=\pi}$ [meV] & $-0.0279$ & $0.0103$ & $0$ & $(-0.0000,\,+0.0112)$ & $(-0.0119,\,+0.0020)$ & $(-0.0023,\,+0.0024)$ & $(-0.0119,\,+0.0139)$ \\
$\Delta\langle\hbar\omega\rangle_{q=\pi/2}$ [meV] & $-0.0036$ & $0.0275$ & $0$ & $(-0.0013,\,+0.0000)$ & $(-0.0000,\,+0.0104)$ & $(-0.0000,\,+0.0034)$ & $(-0.0013,\,+0.0172)$ \\
$\Delta\hbar\omega_\mathrm{pk}$, $q=\pi$ [meV]$^{\dagger}$ & $0.2300$ & $0.0926$ & $0$ & $(-0.1950,\,+0.0025)$ & $(-0.0025,\,+0.0000)$ & $0$ & $(-0.2025,\,+0.0050)$ \\
$\Delta\sigma_\omega$, $q=\pi$ [meV] & $-0.0319$ & $0.0067$ & $0$ & $(-0.0001,\,+0.0008)$ & $(-0.0059,\,+0.0000)$ & $(-0.0010,\,+0.0013)$ & $(-0.0062,\,+0.0017)$ \\
\hline
$\Delta R_{>}$ & $-0.039$ & $0.029$ & $0$ & $(-0.000,\,+0.020)$ & $(-0.005,\,+0.003)$ & $(-0.003,\,+0.006)$ & $(-0.006,\,+0.027)$ \\
\hline
$\Delta$gap, $q=\pi$ [meV]$^{\dagger}$ & $0.0000$ & $0.0019$ & $0$ & $0$ & $0$ & $0$ & $(-0.0000,\,+0.0025)$ \\
$\Delta B$ [meV]$^{\dagger}$ & $0.1375$ & $0.0657$ & $0$ & $(-0.0125,\,+0.0000)$ & $(-0.0025,\,+0.0000)$ & $(-0.0050,\,+0.0025)$ & $(-0.1650,\,+0.0050)$ \\
\hline
\multicolumn{8}{c}{$h_x = 1.146\,J$ (1.5~T)}\\
\hline
$r$ & $0.896$ & $0.012$ & $0$ & $(-0.048,\,+0.000)$ & $(-0.004,\,+0.000)$ & $(-0.017,\,+0.001)$ & $(-0.069,\,+0.001)$ \\
$A$ & $1.077$ & $0.041$ & $(-0.195,\,+0.238)$ & $(-0.142,\,+0.000)$ & $(-0.035,\,+0.031)$ & $(-0.012,\,+0.000)$ & $(-0.341,\,+0.275)$ \\
\hline
MSE & $0.0026$ & $0.0004$ & $(-0.0007,\,+0.0028)$ & $(-0.0002,\,+0.0000)$ & $(-0.0000,\,+0.0001)$ & $(-0.0000,\,+0.0004)$ & $(-0.0008,\,+0.0038)$ \\
\quad normalized, MSE$/\langle S^2_\mathrm{ref}\rangle$ & $0.157$ & $0.024$ & $(-0.042,\,+0.167)$ & $(-0.012,\,+0.000)$ & $(-0.001,\,+0.005)$ & $(-0.002,\,+0.026)$ & $(-0.048,\,+0.226)$ \\
$W_1$ & $0.0360$ & $0.0089$ & $0$ & $(-0.0000,\,+0.0017)$ & $(-0.0032,\,+0.0086)$ & $(-0.0000,\,+0.0031)$ & $(-0.0042,\,+0.0109)$ \\
SSIM & $0.706$ & $0.037$ & $(-0.044,\,+0.003)$ & $(-0.053,\,+0.000)$ & $(-0.018,\,+0.008)$ & $(-0.047,\,+0.001)$ & $(-0.131,\,+0.021)$ \\
\hline
$\Delta\langle\hbar\omega\rangle_{q=\pi}$ [meV] & $-0.0251$ & $0.0111$ & $0$ & $(-0.0000,\,+0.0076)$ & $(-0.0080,\,+0.0028)$ & $(-0.0010,\,+0.0003)$ & $(-0.0080,\,+0.0165)$ \\
$\Delta\langle\hbar\omega\rangle_{q=\pi/2}$ [meV] & $0.1123$ & $0.0208$ & $0$ & $(-0.0000,\,+0.0024)$ & $(-0.0119,\,+0.0077)$ & $(-0.0151,\,+0.0000)$ & $(-0.0346,\,+0.0151)$ \\
$\Delta\hbar\omega_\mathrm{pk}$, $q=\pi$ [meV]$^{\dagger}$ & $0.0250$ & $0.0065$ & $0$ & $(-0.0000,\,+0.0150)$ & $(-0.0025,\,+0.0000)$ & $(-0.0025,\,+0.0050)$ & $(-0.0025,\,+0.0200)$ \\
$\Delta\sigma_\omega$, $q=\pi$ [meV] & $-0.0318$ & $0.0077$ & $0$ & $(-0.0031,\,+0.0000)$ & $(-0.0050,\,+0.0024)$ & $(-0.0053,\,+0.0000)$ & $(-0.0054,\,+0.0033)$ \\
\hline
$\Delta R_{>}$ & $-0.036$ & $0.021$ & $0$ & $(-0.000,\,+0.028)$ & $(-0.009,\,+0.003)$ & $(-0.004,\,+0.000)$ & $(-0.010,\,+0.037)$ \\
\hline
$\Delta$gap, $q=\pi$ [meV]$^{\dagger}$ & $-0.0050$ & $0.0012$ & $0$ & $0$ & $0$ & $0$ & $(-0.0000,\,+0.0025)$ \\
$\Delta B$ [meV]$^{\dagger}$ & $-0.0025$ & $0.0136$ & $0$ & $(-0.0175,\,+0.0000)$ & $(-0.0100,\,+0.0025)$ & $(-0.0000,\,+0.0075)$ & $(-0.0275,\,+0.0075)$ \\
\hline
\multicolumn{8}{c}{$h_x = 1.604\,J$ (2.1~T)}\\
\hline
$r$ & $0.862$ & $0.015$ & $0$ & $(-0.066,\,+0.000)$ & $(-0.007,\,+0.000)$ & $(-0.001,\,+0.001)$ & $(-0.077,\,+0.002)$ \\
$A$ & $0.522$ & $0.024$ & $(-0.066,\,+0.076)$ & $(-0.155,\,+0.000)$ & $(-0.012,\,+0.002)$ & $0$ & $(-0.208,\,+0.080)$ \\
\hline
MSE & $0.0108$ & $0.0007$ & $(-0.0016,\,+0.0018)$ & $(-0.0000,\,+0.0056)$ & $(-0.0000,\,+0.0003)$ & $0$ & $(-0.0016,\,+0.0077)$ \\
\quad normalized, MSE$/\langle S^2_\mathrm{ref}\rangle$ & $0.299$ & $0.021$ & $(-0.044,\,+0.051)$ & $(-0.000,\,+0.155)$ & $(-0.000,\,+0.010)$ & $0$ & $(-0.045,\,+0.213)$ \\
$W_1$ & $0.0594$ & $0.0116$ & $0$ & $(-0.0000,\,+0.0125)$ & $(-0.0000,\,+0.0109)$ & $(-0.0014,\,+0.0000)$ & $(-0.0014,\,+0.0259)$ \\
SSIM & $0.414$ & $0.044$ & $(-0.031,\,+0.032)$ & $(-0.035,\,+0.000)$ & $(-0.019,\,+0.000)$ & $(-0.000,\,+0.005)$ & $(-0.061,\,+0.037)$ \\
\hline
$\Delta\langle\hbar\omega\rangle_{q=\pi}$ [meV] & $0.0960$ & $0.1127$ & $0$ & $(-0.0000,\,+0.0480)$ & $(-0.0885,\,+0.0279)$ & $(-0.0298,\,+0.0000)$ & $(-0.0952,\,+0.0761)$ \\
$\Delta\langle\hbar\omega\rangle_{q=\pi/2}$ [meV] & $-0.0200$ & $0.0276$ & $0$ & $(-0.0276,\,+0.0000)$ & $(-0.0276,\,+0.0425)$ & $(-0.0077,\,+0.0112)$ & $(-0.0368,\,+0.0475)$ \\
$\Delta\hbar\omega_\mathrm{pk}$, $q=\pi$ [meV]$^{\dagger}$ & $0.0725$ & $0.1645$ & $0$ & $(-0.0000,\,+0.3300)$ & $(-0.0000,\,+0.3425)$ & $(-0.0000,\,+0.0025)$ & $(-0.0050,\,+0.3450)$ \\
$\Delta\sigma_\omega$, $q=\pi$ [meV] & $-0.0083$ & $0.0257$ & $0$ & $(-0.0087,\,+0.0000)$ & $(-0.0123,\,+0.0000)$ & $(-0.0000,\,+0.0036)$ & $(-0.0451,\,+0.0096)$ \\
\hline
$\Delta R_{>}$ & $0.074$ & $0.027$ & $0$ & $(-0.001,\,+0.008)$ & $(-0.000,\,+0.017)$ & $(-0.003,\,+0.000)$ & $(-0.004,\,+0.034)$ \\
\hline
$\Delta$gap, $q=\pi$ [meV]$^{\dagger}$ & $0.0700$ & $0.0387$ & $0$ & $(-0.0050,\,+0.0050)$ & $(-0.0100,\,+0.0075)$ & $(-0.0025,\,+0.0000)$ & $(-0.0600,\,+0.0350)$ \\
$\Delta B$ [meV]$^{\dagger}$ & $0.1500$ & $0.0401$ & $0$ & $(-0.0175,\,+0.0000)$ & $(-0.0000,\,+0.0050)$ & $(-0.0025,\,+0.0000)$ & $(-0.0200,\,+0.0050)$ \\
\hline
\end{tabular}
\end{ruledtabular}
\end{table*}

\subsection{Local spin response and correlation functions}
\label{sec:supp-correlations}

\begin{figure*}[!htbp]
    \centering    \includegraphics[width=0.75\linewidth]{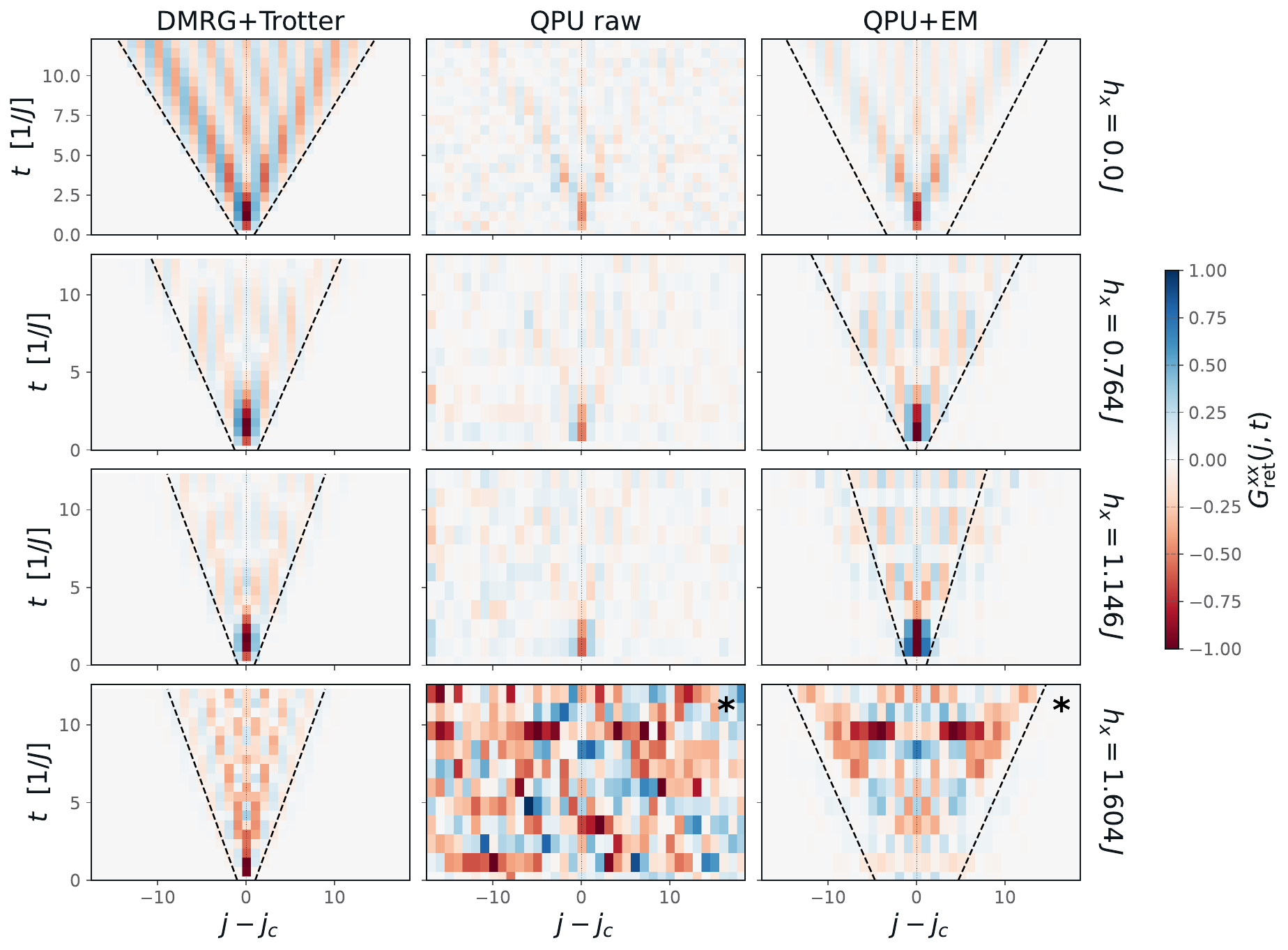}
    \caption{Retarded Green's function $G^{xx}_{ret}(j,t)$ for the $L=36$ site  anisotropic $(\Delta=0.25)$ open chain with a kick at $j_c=17$. Columns show DMRG+Trotter $(L=36)$, raw hardware, and error mitigated hardware results.  ${}^{*}$For data-points without hardware runs $(h_x=1.604 J)$, we replace the plots with noisy simulation results obtained by evaluating the corresponding compressed circuits with the following noise parameters: readout bit-flip $p_{ro}=5.9\times10^{-3}$, one-qubit depolarizing $p_1=3\times 10^{-4}$, two-qubit depolarizing $p_2=9.375\times 10^{-3}$, and $2496$ shots for each circuit.}
    \label{fig:rgf_xx}
\end{figure*}

\begin{figure*}[!htbp]
    \centering \includegraphics[width=0.75\linewidth]{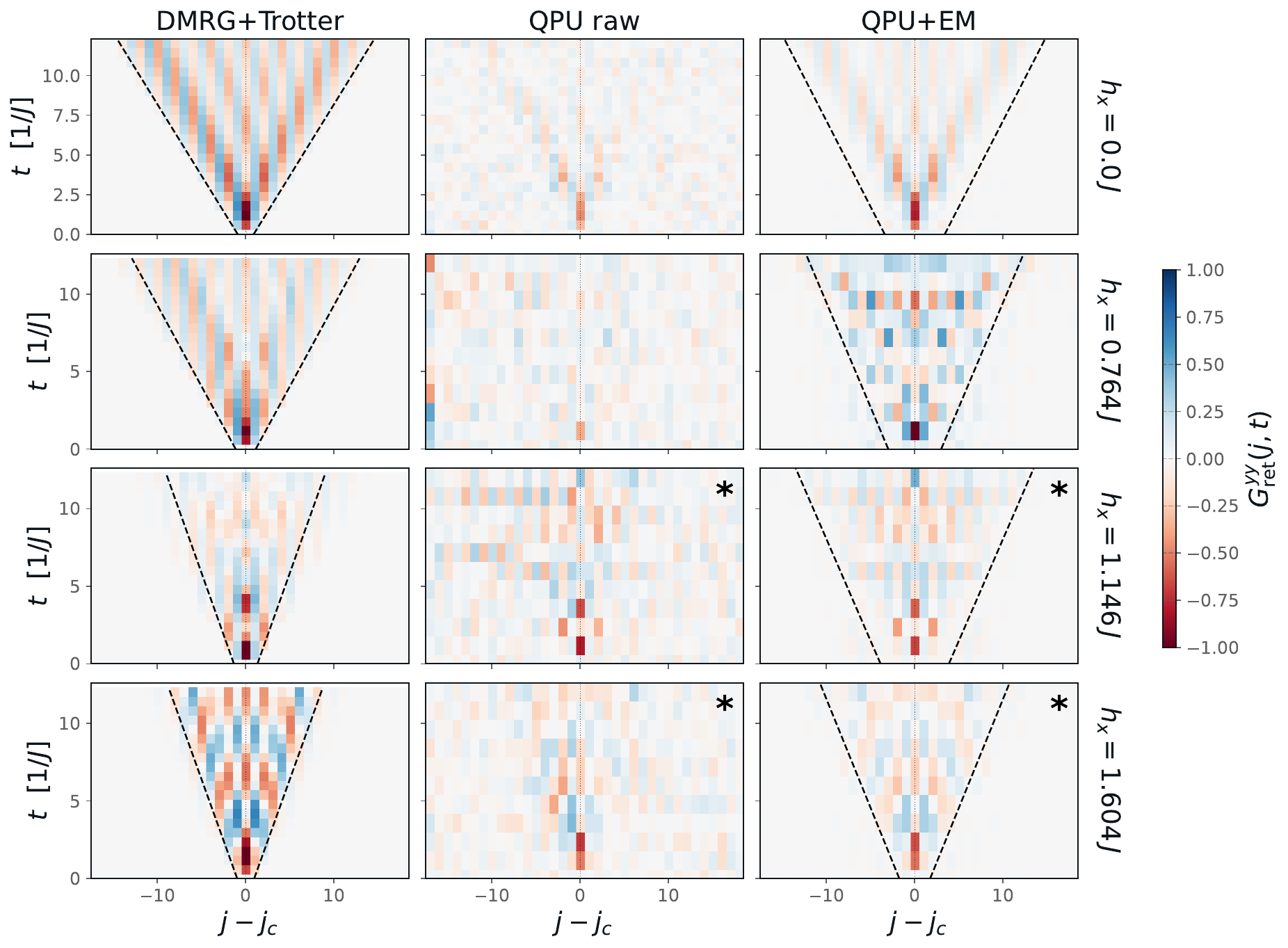}
    \caption{Retarded Green's function $G^{yy}_{ret}(j,t)$ for the $L=36$ site  anisotropic $(\Delta=0.25)$ open chain with a kick at $j_c=17$. Columns show DMRG+Trotter $(L=36)$, raw hardware, and error mitigated hardware results.  ${}^{*}$For data-points without hardware runs $(h_x=1.146J, \ 1.604 J)$, we replace the plots with noisy simulation results obtained by evaluating the corresponding compressed circuits with the following noise parameters: readout bit-flip $p_{ro}=5.9\times10^{-3}$, one-qubit depolarizing $p_1=3\times 10^{-4}$, two-qubit depolarizing $p_2=9.375\times 10^{-3}$, and $2496$ shots for each circuit.}
    \label{fig:rgf_yy}
\end{figure*}

Figures~\ref{fig:hw-rgf-zz},~\ref{fig:rgf_xx}, and~\ref{fig:rgf_yy}, show the three components retarded response at $h_x/J=(0,\,0.764,\,1.146,\,1.604)$, with each panel normalized to its own maximum. DMRG+Trotter provides the classical reference for every channel and field.. The reported hardware coverage is $G^{xx}$, $G^{yy}$, and $G^{zz}$ at the first two fields, $G^{xx}$ and $G^{zz}$ at $h_x=1.146J$, and $G^{zz}$ at $h_x=1.604J$; asterisks mark noisy simulations used for combinations without hardware data. Figure~\ref{fig:hw-rgf-zz} presents the corresponding $zz$ comparison in the main text. The perturbation is applied at $j_c=17$, and the spatial extent and oscillation pattern of the response change with the field. The earlier qualitative description of the interior was slow and extended oscillations at zero field, with progressively faster alternation as the field increases.

The connected correlation function in the Pauli basis is reconstructed from the response at a fixed source site $j_c$ through
\begin{equation}\label{eq:con-acf}
C_{\text{con}}^{\alpha \beta}(j,j_c,t)  \simeq \frac{-2\Delta \omega}{ \pi} \sum_{m=0}^{M-1} e^{-i \omega_m t} \mathrm{Im}\tilde{G}^{\alpha \beta}_{\text{ret}}(j,j_c, \omega_m) \ ,
\end{equation}
with
\begin{equation}
\tilde{G}^{\alpha \beta}_{\text{ret}}(j,j_c, \omega_m)  =\Delta t \sum_{k=0}^{N_t-1} e^{i \omega_m t_k} e^{-\eta t_k}  G^{\alpha \beta}_{\text{ret}} (j,j_c,t_k) \ ,
\end{equation}
Here $\eta$ specifies exponential damping of the finite time series and the sum uses $\omega_m\ge0$. The original construction retains the imaginary part of the frequency response to recover a connected correlation function. Its interpretation is expressed by
\begin{equation} \label{eq: con-sigma-acf}
C_{\text{con}}^{\alpha \beta}(j,j_c,t)  =  \braket{\sigma^\alpha_j (t)\sigma^\beta_{j_c}(0)}   - \braket{\sigma^\alpha_j (t)}  \braket{\sigma^\beta_{j_c}(0)}\ ,
\end{equation}
with the full correlator
\begin{equation} \label{eq: sigma-acf}
C^{\alpha \beta}(j,j_c,t)  =  \braket{\sigma^\alpha_j (t)\sigma^\beta_{j_c}(0)} \ .
\end{equation}
The real part is associated with symmetrized fluctuations, and the imaginary part with the commutator response, subject to the normalization and connected subtraction in these equations.

\begin{figure*}[!t]
    \centering
    \includegraphics[width=0.8\linewidth]{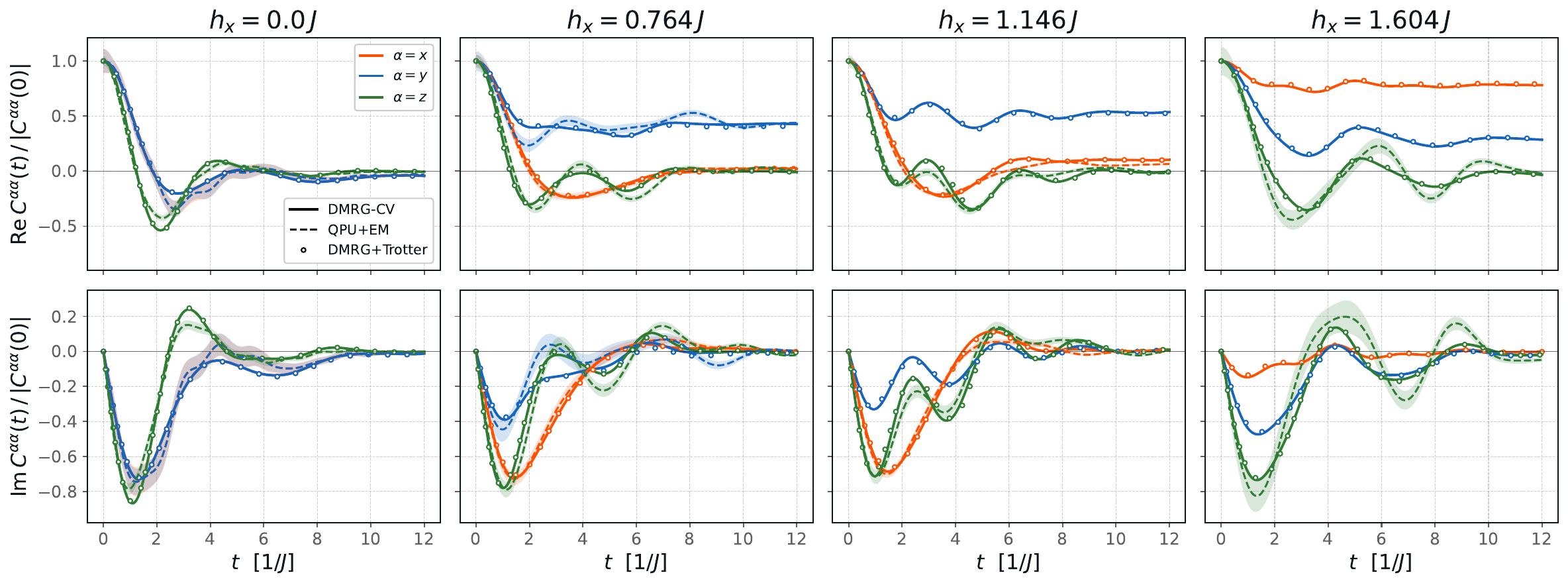}
    \caption{On-site spin-spin autocorrelation function $C^{\alpha \alpha}(t) \equiv C^{\alpha \alpha}(j_c,t) = \braket{\sigma^\alpha_{j_c}(t)\sigma^\alpha_{j_c}(0)}$  at site $j_c=17$, real (top) and imaginary (bottom) components, each normalized by the equal-time modulus $\vert C^{\alpha \alpha}(0) \vert = \sqrt{[\text{Re}C^{\alpha \alpha}(0)]^2+ [\text{Im}C^{\alpha \alpha}(0)]^2}$ . Each column refers to one of the following field values $h_x/J=(0,\ 0.764, \ 1.146, \ 1.604  )$. Solid and dotted curves represent DMRG-CV and DMRG+Trotter results, whereas dashed lines correspond error mitigated hardware results with shadow bands corresponding to one standard deviation of the QPU data.}
    \label{fig:acf-onsite}
\end{figure*}

At the source site the full autocorrelation describes the overlap of
$\sigma^{\alpha}_{j_c}(t)|\psi_0\rangle$ with
$\sigma^{\alpha}_{j_c}(0)|\psi_0\rangle$, and its decay and
oscillation complement the redistribution of spectral weight. Section IV B reports the normalized $zz$ results and their four field correlations; normalization by $|C^{zz}(0)|$ removes a common mirror amplitude factor. Figure~\ref{fig:acf-onsite} extends the comparison to the other available channels and to the real and imaginary components. The reconstructed zero time value was reported as $0.948$ to $0.978$ of unity. This quantity is reconstructed through the frequency domain and not read off the time record. The retarded function vanishes identically at $t=0$ at the source site, since the equal-time commutator of an operator with itself is zero, so the autocorrelation is not directly measured. It is therefore independent
of the plane-wave transform and no spatial transform entering at $j=j_c$, but it is not independent of the temporal processing, as the taper, the finite window and the positive-frequency projection all act on it. 

The reconstructed zero-time value is $0.948$ to $0.978$ of unity. The target is exactly unity for this channel, and the two possible conventions coincide here: the onsite Pauli product $\sigma^z\sigma^z$ is the identity, and the connected value subtracts $\langle\sigma^z\rangle^2$, which vanishes at every field because the applied field lies along $x$ and $|\langle S^z\rangle|\le7\times 10^{-7}$ throughout. The $2$-$5\%$ shortfall is therefore spectral weight lost outside the reconstruction window, not a difference of convention. The distinction does matter for the transverse channel, where $1-\langle\sigma^x\rangle^2$ falls to $0.941$, $0.857$ and $0.215$ at $h_x=0.8$, $1.1$ and $1.604\,J$; any autocorrelation check applied there must state which product it normalizes against.

For the averaging convention in Eq.~\eqref{eq:cr}, the ratios of the error mitigated hardware $C(1)$ to the DMRG values are $1.025$, $1.085$, $0.888$, and $0.767$ at $h_x/J=0$, $0.764$, $1.146$, and $1.604$, respectively. The corresponding ratios for $C(2)$ range from $0.64$ to $0.88$. At the largest field, the raw hardware value is $C(1)=+0.0849\pm0.0039$, approximately 22 quoted statistical standard deviations above zero. This significance uses the statistical uncertainty alone. The positive sign therefore precedes error mitigation. A positive multiplicative attenuation of the correlator, as in a simple depolarizing description, preserves its sign.

\begin{figure*}[!htbp]
    \centering
    \includegraphics[width=0.8\linewidth]{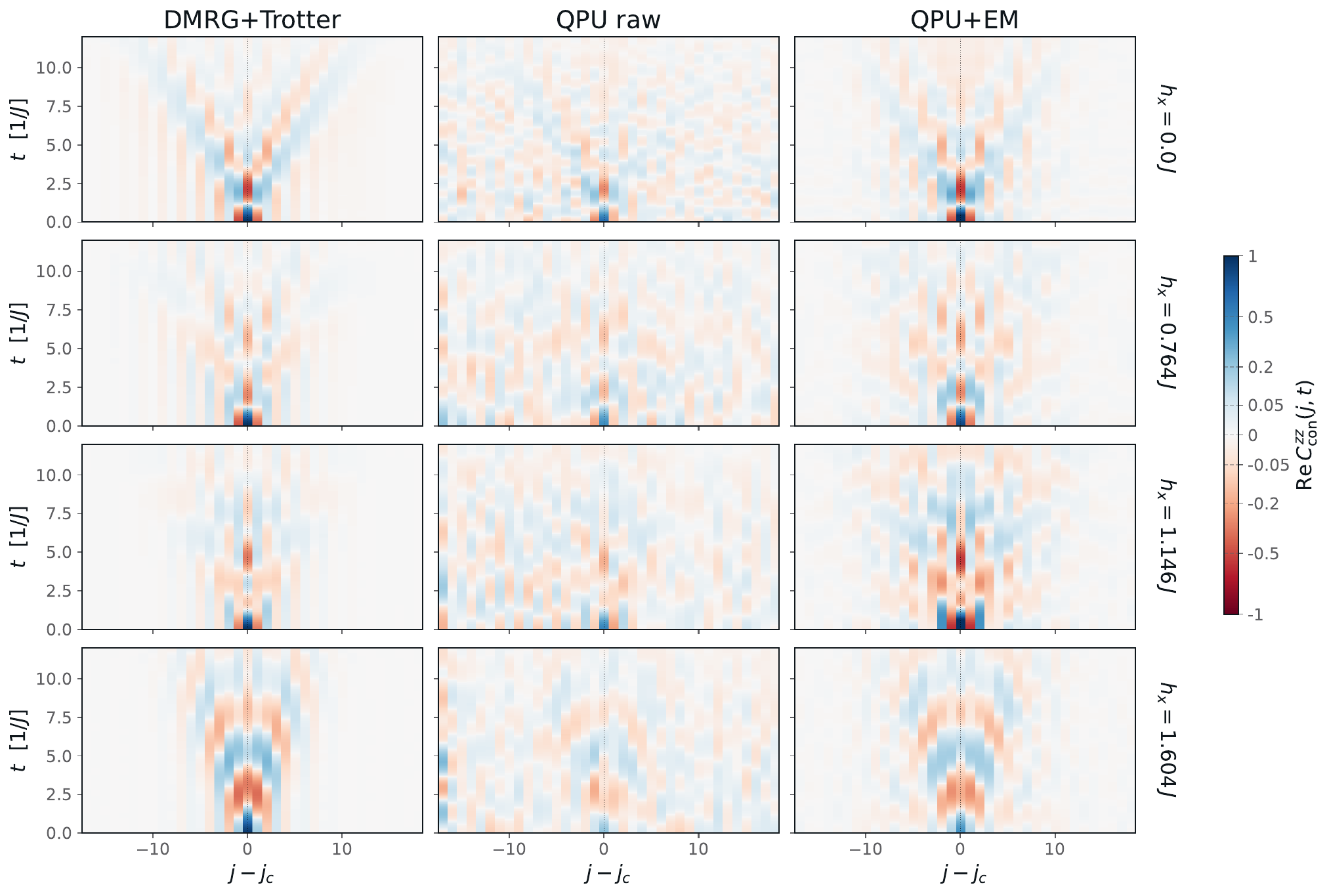}
    \caption{Real part of the connected ACF $C^{zz}_{\text{con}}(j,t)$ displaying the light cone spread for the $L=36$ site  anisotropic $(\Delta=0.25)$ open chain with a kick at $j_c=17$. Columns show DMRG+Trotter $(L=36)$, raw hardware, and error mitigated hardware results in the same order.}
    \label{fig:acf-lc-zz}
\end{figure*}

Figure~\ref{fig:acf-lc-zz} shows the spatially resolved connected correlation reconstructed from the full kick time series. Its $t=0$ row provides a comparison with the direct measurements in Figure~\ref{fig:hw-corr-zz}, but the two use disjoint pairs: the reconstruction is referenced to the source, so its nearest-neighbour entry is the single pair $(j_c, j_c+1)$, while Eq. \ref{eq:cr} averages over pairs more than three sites from the kick and excludes exactly that neighborhood. Across the four fields, the reconstructed nearest neighbor values are $(-0.484,\,-0.334,\,-0.360,\,+0.123)$, compared with $(-0.485,\,-0.476,\,-0.302,\,+0.089)$ from direct measurements - agreeing to 0.2\% at zero field, differing by $-30\%$, +19\% and +38\% at the finite fields. The reconstruction is larger at two of them, so the difference is not lost weight above $\hbar\omega_\mathrm{max}$.The two procedures give the same sign at each measured field. Interpreting their differences in magnitude requires reconciling the spatial averaging conventions and accounting for the spectral window used in the reconstruction.

\section{Finite-memory extension and its limitations}
\label{sec:supp-finite-memory}
\label{sec:finite-memory-extension}

\subsection{Method and fitting procedure}

The finite observation window broadens spectral features and limits the separation of nearby excitation energies. Extending the directly calculated or measured time series generally requires deeper circuits or a loss of compression fidelity (Figure~\ref{fig:fidelity}). We therefore examine a finite-memory extension (FME) as a way to estimate a longer response from existing data. The FME introduces a fitted model for the unobserved dynamics. We first form the momentum-dependent response
\begin{equation}
  \widetilde{G}^{\alpha\alpha}_{\mathrm{ret}}(q,t)
  =\sum_{j=0}^{L-1}e^{-iq(j-j_c)}G^{\alpha\alpha}_{\mathrm{ret}}(j,j_c,t),
  \label{eq:momentum-time-response}
\end{equation}
and approximate each diagonal spin channel and momentum time series, $x_n^\alpha(q)=\widetilde{G}^{\alpha\alpha}_{\mathrm{ret}}(q,t_n)$,
by the linear recurrence relation
\begin{equation}
  x_n^\alpha(q)\simeq
  \sum_{k=1}^{K}a_k^\alpha(q)x_{n-k}^\alpha(q).
  \label{eq:finite-memory-recurrence}
\end{equation}

For the $L=36$, $h_x=0$ study, the fitting record contains $N=20$ samples after the kick at $Jt_n=0.6,1.2,\ldots,12$. Suppressing the channel and momentum indices, we define $X_{n-K,k}=x_{n-k}$ and $y_{n-K}=x_n$, where $n=K+1,\ldots,N$ and $k=1,\ldots,K$, and fit
\begin{equation}
  \widehat{\mathbf a}
  ={\Bigl(X^\dagger X+\lambda s_X I\Bigr)}^{-1}X^\dagger\mathbf y,
  \qquad
  s_X=\frac{1}{K}\operatorname{tr}(X^\dagger X).
  \label{eq:supp-finite-memory-fit}
\end{equation}
The normalization $s_X$ makes $\lambda$ invariant under an overall rescaling of the response. Within the oscillatory model defined below, we scan $K=1,\ldots,8$ and $\lambda\in\{10^{-6},10^{-5},\ldots,10\}$ using global chronological validation over all 36 momenta. The first 17 post-kick samples are used for fitting and the final three samples within $Jt\leq12$ are held out for validation. For each candidate pair, the objective is the relative $\ell_2$ error of the real total momentum-space response over these held-out points, aggregated over all momenta. This procedure selects $K=7$ and $\lambda=10^{-4}$, rather than tuning the hyperparameters at a single momentum. The pair is fixed before the DMRG$+$Trotter validation and QPU$+$EM application. A separate coefficient vector is then refitted to each time series using all 20 post-kick samples, and FME supplies the response through $Jt=18$. The value at $t=0$ is retained for the Fourier transform but excluded from the homogeneous fit.

For a finite, closed system, the response is a sum of terms $\exp[-i(E_m-E_n)t]$ and therefore has discrete-time poles of unit modulus. We impose this oscillatory structure on the empirical recurrence. For every nonzero root of the companion polynomial
\begin{equation}
  r^K-a_1r^{K-1}-\cdots-a_K=0
\end{equation}
we write $r_\ell=\rho_\ell e^{i\theta_\ell}$ and replace it by
\begin{equation}
  \widetilde r_\ell=e^{i\theta_\ell}
  =\frac{r_\ell}{|r_\ell|}.
  \label{eq:finite-memory-unit-circle}
\end{equation}
The FME coefficients are reconstructed from the projected roots before propagation. This retains the learned oscillation frequencies while removing artificial exponential growth or decay.

\subsection{Validation against longer classical dynamics}

We evaluate the $L=36$, $h_x=0$ FME against the directly calculated DMRG$+$Trotter trajectory over $12<Jt\leq18$. The FME coefficients are fitted only to the DMRG$+$Trotter prefix through $Jt=12$, and no sample in the displayed tail enters either the coefficient fits or hyperparameter selection. At $q=\pi$, the FME total response and DMRG$+$Trotter reference have centered correlation $r=0.990$, relative $\ell_2$ error $0.015$, and standard-deviation ratio $1.121$. At $q=2\pi/3$, the corresponding values are $0.987$, $0.266$, and $1.122$. The individual $zz$ channel is more difficult to extend: its correlation, relative error, and standard-deviation ratio are $0.532$, $0.595$, and $0.362$ at $q=\pi$, and $0.954$, $0.567$, and $1.373$ at $q=2\pi/3$, respectively.

\begin{figure*}[!htbp]
    \centering
    \includegraphics[width=\textwidth]{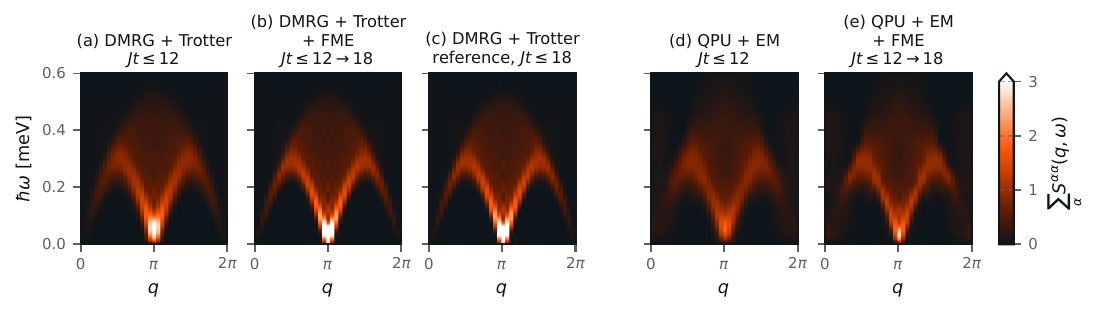}
    \caption{Momentum-resolved validation of the finite-memory extension (FME) for the total dynamical spin structure factor, $S_{\mathrm{tot}}(q,\omega)=\sum_{\alpha=x,y,z}S^{\alpha\alpha}(q,\omega)$, for $L=36$, $h_x=0$, and $\Delta t=0.6/J$. (a) DMRG$+$Trotter spectrum calculated through $Jt=12$. (b) DMRG$+$Trotter spectrum after applying FME with the fixed global parameters $K=7$ and $\lambda=10^{-4}$ to each diagonal spin channel and momentum over $0.6\leq Jt\leq12$, projecting every nonzero companion root to unit modulus, and extending the response to $Jt=18$. (c) DMRG$+$Trotter reference calculated directly through $Jt=18$; samples after $Jt=12$ are not used in the FME coefficient fits or hyperparameter selection. (d) QPU $+$ EM spectrum through $Jt=12$. (e) QPU $+$ EM $+$ FME spectrum through $Jt=18$, using the same fixed model and hyperparameters but coefficients refitted to each QPU time series. Panel (e) has no independent long-time QPU reference and is therefore a prediction of unmeasured dynamics.}
    \label{fig:finite-memory-dsf-maps}
\end{figure*}

For the corresponding spectral comparison over $0\leq\hbar\omega\leq0.4\,\mathrm{meV}$, the relative error at $q=\pi$ decreases from $0.481$ for the truncated DMRG$+$Trotter record to $0.005$ after FME. The FME and reference peak energies are both $0.035\,\mathrm{meV}$, compared with $0.053\,\mathrm{meV}$ for the short record. At $q=2\pi/3$, the relative error decreases from $0.177$ to $0.023$; the reference, FME, and short-record peaks occur at $0.249$, $0.248$, and $0.258\,\mathrm{meV}$, respectively. Figure~\ref{fig:finite-memory-dsf-maps} extends this comparison to all 36 momenta. Over $0\leq\hbar\omega\leq0.6\,\mathrm{meV}$, the relative $\ell_2$ error in the total spectrum decreases from $0.307$ to $0.015$; these errors are evaluated before display clipping. These comparisons validate the total-spectrum FME for this DMRG$+$Trotter benchmark but do not imply uniform improvement of every individual channel and momentum. The additional frequency information supplied by FME remains a model prediction rather than an independent measurement.

\begin{figure*}[!htbp]
  \centering
  \includegraphics[width=0.9\textwidth]{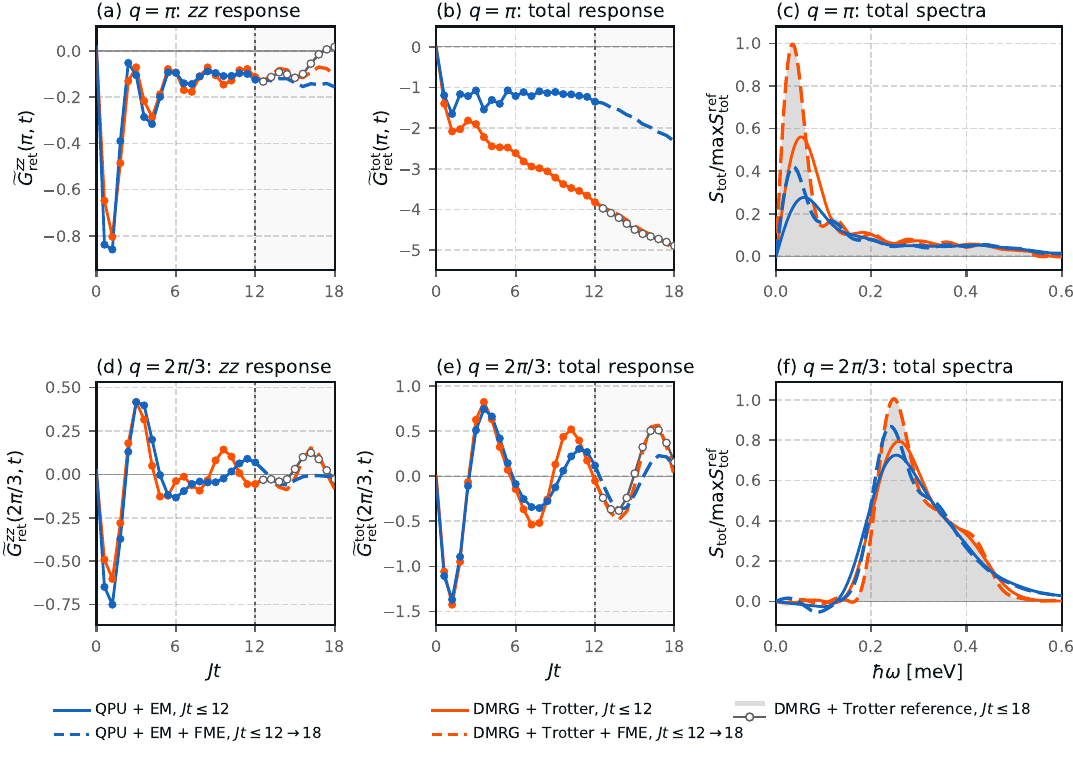}
  \caption{Finite-memory extension (FME) of the momentum-space retarded response at $q=\pi$ (top row) and $q=2\pi/3$ (bottom row) for $L=36$, $h_x=0$, and $\Delta t=0.6/J$. Panels (a) and (d) show the $zz$ component, panels (b) and (e) the total response, and panels (c) and (f) the corresponding total spectra. Solid orange lines and filled circles show DMRG$+$Trotter samples through $Jt=12$, while solid blue lines and filled circles show QPU $+$ EM samples over the same interval. A separate FME model is fitted to each momentum-resolved diagonal spin channel using the 20 post-kick samples over $0.6\leq Jt\leq12$, with the fixed global pair $K=7$ and $\lambda=10^{-4}$. Dashed lines show the FME results through $Jt=18$, with every nonzero companion root projected to unit modulus. The total response is obtained by summing the independently extended spin channels. Gray lines and open circles show the directly calculated DMRG$+$Trotter reference, which is not used in the coefficient fits or hyperparameter selection. In panels (c) and (f), the gray shaded region is the DMRG$+$Trotter reference evaluated directly through $Jt=18$; each row is normalized by the maximum of that reference spectrum.}
  \label{fig:finite-memory-extension}
\end{figure*}

\subsection{Application to the measured QPU response}

We apply the same fixed $K=7$, $\lambda=10^{-4}$ unit-circle FME model to the fully error-mitigated QPU record for $L=36$ and $h_x=0$, fitting new coefficients to each hardware time series. FME supplies the predicted response between $Jt=12$ and $18$ without additional quantum circuits. Panels (c) and (f) of Figure~\ref{fig:finite-memory-extension} compare the truncated and FME-extended DMRG$+$Trotter and QPU records at $q=\pi$ and $2\pi/3$ with the directly calculated $Jt\leq18$ DMRG$+$Trotter references; each row is normalized by the maximum of its corresponding reference. The DMRG$+$Trotter tail provides an independent check of the classical FME result, whereas the QPU$+$EM prediction cannot be independently validated over $12<Jt\leq18$ because no QPU trajectory is available in that interval.

Table~\ref{tab:fme} compares the total, $xx$, and $zz$ spectra reconstructed from QPU+EM records through $Jt=12$ and their FME extensions through $Jt=18$ against the directly calculated $L=36$ DMRG+Trotter reference at the longer time. This complementary assessment uses the order-$7$ recurrence with $\lambda=10^{-4}$ specified in the table caption. At zero field, continuation improves the total and $xx$ spectra: their normalized mean squared errors decrease by approximately $46\%$ and $45\%$, respectively, while Pearson correlations increase to approximately $0.95$ and $0.94$. SSIM increases and the Wasserstein distance decreases in both channels, and their peak energies at $q=\pi$ agree with the reference on the evaluation grid. The gains are less uniform in $S^{zz}$, where Pearson correlation and SSIM improve but the normalized mean squared error changes little and the width discrepancy increases. At $h_x=0.764J$, Pearson correlation and SSIM decrease in all three spectra, and the normalized mean squared error for $S^{zz}$ more than doubles. Some peak and centroid offsets nevertheless decrease, showing that better recovery of individual spectral features can coexist with poorer agreement across the full spectrum. The zero-field results demonstrate that continuation can improve spectral estimates from a short hardware record, while the field dependence makes validation against longer classical calculations essential.
\begin{table*}[t]
\caption{\label{tab:fme}Effect of the finite-memory extension on the metrics of Table~\ref{tab:metrics}, for the $xx/yy$ and $zz$ channels as well as the total DSF. Within each group the left entry is the hardware record at $Jt=12$ after the full mitigation chain and the right entry is the same record extended to $Jt=18$ by the finite-memory model (order $5$, ridge $10^{-2}$, every nonzero recurrence root projected onto the unit circle, $t=0$ excluded from the fit, causal descending half-DPSS $1.5$ taper). Both are scored against DMRG+Trotter at L=36 at the $Jt=18$ time scale. Metrics are evaluated over $\hbar\omega\le0.693$~meV at $h_x=0$ and $\le0.602$~meV at $h_x=0.764\,J$. The finite-field limit is the Nyquist frequency of that record; the zero-field limit is a fixed analysis cutoff, below that record's Nyquist frequency of $1.204$~meV. Neither is changed by the extension, which lengthens the record without altering its sampling interval. The entry closer to the ideal is set in bold: the larger value for $r$ and SSIM, the smaller for MSE, its normalized form and $W_1$, the one nearer $1$ for $A$, and the smaller in magnitude for every $\Delta$ row, all of which are model-minus-reference differences with target $0$. Ties are unbolded. The extrapolated samples are predictions of the rank-$5$ recurrence fitted to the measured record. Changes in the metrics quantify how this continuation affects agreement with the longer DMRG+Trotter reference. The added samples are not independent hardware measurements. } 
\scriptsize
\setlength{\tabcolsep}{2.0pt}%
\begin{ruledtabular}
\begin{tabular}{lccccccc}
 & & \multicolumn{2}{c}{$S_\mathrm{tot}$} & \multicolumn{2}{c}{$S^{xx}$} & \multicolumn{2}{c}{$S^{zz}$}\\
\cline{3-4}\cline{5-6}\cline{7-8}
Quantity & Ideal & QPU$+$EM & $+$FME & QPU$+$EM & $+$FME & QPU$+$EM & $+$FME\\
\hline
\multicolumn{8}{c}{$h_x = 0$ (0~T) }\\
\hline
$r$ & $1$ & $0.887$ & $\mathbf{0.945}$ & $0.864$ & $\mathbf{0.939}$ & $0.949$ & $\mathbf{0.956}$ \\
$A$ & $1$ & $0.688$ & $\mathbf{0.793}$ & $0.557$ & $\mathbf{0.681}$ & $\mathbf{1.032}$ & $1.091$ \\
\hline
MSE & $0$ & $0.0342$ & $\mathbf{0.0185}$ & $0.0078$ & $\mathbf{0.0043}$ & $\mathbf{0.0014}$ & $0.0014$ \\
\quad normalized, MSE$/\langle S^2_\mathrm{ref}\rangle$ & $0$ & $0.214$ & $\mathbf{0.116}$ & $0.295$ & $\mathbf{0.162}$ & $\mathbf{0.102}$ & $0.104$ \\
$W_1$ & $0$ & $0.0579$ & $\mathbf{0.0549}$ & $0.0631$ & $\mathbf{0.0587}$ & $\mathbf{0.0405}$ & $0.0410$ \\
SSIM & $1$ & $0.932$ & $\mathbf{0.943}$ & $0.951$ & $\mathbf{0.959}$ & $0.699$ & $\mathbf{0.712}$ \\
\hline
$\Delta\langle\hbar\omega\rangle_{q=\pi}$ [meV] & $0$ & $0.0564$ & $\mathbf{0.0374}$ & $0.0139$ & $\mathbf{-0.0001}$ & $0.0525$ & $\mathbf{0.0441}$ \\
$\Delta\langle\hbar\omega\rangle_{q=\pi/2}$ [meV] & $0$ & $\mathbf{-0.0081}$ & $-0.0082$ & $-0.0249$ & $\mathbf{-0.0246}$ & $0.0116$ & $\mathbf{0.0111}$ \\
$\Delta\hbar\omega_\mathrm{pk}$, $q=\pi$ [meV] & $0$ & $0.0251$ & $\mathbf{0.0000}$ & $0.0201$ & $\mathbf{0.0000}$ & $0.0301$ & $\mathbf{-0.0050}$ \\
$\Delta\sigma_\omega$, $q=\pi$ [meV] & $0$ & $\mathbf{0.0152}$ & $0.0166$ & $-0.0366$ & $\mathbf{-0.0308}$ & $\mathbf{0.0335}$ & $0.0361$ \\
\hline
$\Delta R_{>}$ & $0$ & $0.066$ & $\mathbf{0.059}$ & $0.072$ & $\mathbf{0.063}$ & $0.004$ & $\mathbf{0.001}$ \\
\hline
$\Delta$gap, $q=\pi$ [meV] & $0$ & $0.0000$ & $0.0000$ & $0.0000$ & $0.0000$ & $0.0000$ & $0.0000$ \\
$\Delta B$ [meV] & $0$ & $0.3815$ & $\mathbf{-0.2058}$ & $\mathbf{0.3815}$ & $0.4166$ & $\mathbf{0.0502}$ & $0.0753$ \\
\hline
\multicolumn{8}{c}{$h_x = 0.764\,J$ (1~T) }\\
\hline
$r$ & $1$ & $\mathbf{0.925}$ & $0.914$ & $\mathbf{0.967}$ & $0.965$ & $\mathbf{0.918}$ & $0.860$ \\
$A$ & $1$ & $0.802$ & $\mathbf{0.856}$ & $0.936$ & $\mathbf{1.035}$ & $\mathbf{0.971}$ & $1.063$ \\
\hline
MSE & $0$ & $0.0150$ & $\mathbf{0.0149}$ & $\mathbf{0.0014}$ & $0.0016$ & $\mathbf{0.0017}$ & $0.0039$ \\
\quad normalized, MSE$/\langle S^2_\mathrm{ref}\rangle$ & $0$ & $0.102$ & $\mathbf{0.101}$ & $\mathbf{0.049}$ & $0.057$ & $\mathbf{0.093}$ & $0.218$ \\
$W_1$ & $0$ & $\mathbf{0.0432}$ & $0.0462$ & $0.0323$ & $\mathbf{0.0315}$ & $\mathbf{0.0255}$ & $0.0340$ \\
SSIM & $1$ & $\mathbf{0.906}$ & $0.896$ & $\mathbf{0.907}$ & $0.895$ & $\mathbf{0.890}$ & $0.826$ \\
\hline
$\Delta\langle\hbar\omega\rangle_{q=\pi}$ [meV] & $0$ & $0.0260$ & $\mathbf{0.0236}$ & $\mathbf{0.0009}$ & $-0.0010$ & $0.0029$ & $\mathbf{0.0010}$ \\
$\Delta\langle\hbar\omega\rangle_{q=\pi/2}$ [meV] & $0$ & $0.0117$ & $\mathbf{0.0116}$ & $\mathbf{0.0137}$ & $0.0160$ & $0.0017$ & $\mathbf{-0.0006}$ \\
$\Delta\hbar\omega_\mathrm{pk}$, $q=\pi$ [meV] & $0$ & $0.1050$ & $\mathbf{0.1025}$ & $0.0075$ & $\mathbf{0.0025}$ & $0.2400$ & $\mathbf{0.2325}$ \\
$\Delta\sigma_\omega$, $q=\pi$ [meV] & $0$ & $-0.0136$ & $\mathbf{-0.0112}$ & $0.0022$ & $\mathbf{-0.0007}$ & $\mathbf{-0.0183}$ & $-0.0186$ \\
\hline
$\Delta R_{>}$ & $0$ & $0.070$ & $\mathbf{0.055}$ & $0.055$ & $\mathbf{0.045}$ & $\mathbf{-0.014}$ & $-0.041$ \\
\hline
$\Delta$gap, $q=\pi$ [meV] & $0$ & $0.0075$ & $\mathbf{0.0050}$ & $\mathbf{-0.0100}$ & $0.0125$ & $\mathbf{0.0025}$ & $0.0125$ \\
$\Delta B$ [meV] & $0$ & $-0.0425$ & $-0.0425$ & $0.0425$ & $0.0425$ & $\mathbf{0.1300}$ & $0.1775$ \\
\hline
\end{tabular}
\end{ruledtabular}
\end{table*}

\section{Additional Observables}\label{sec:appendix-d}
\subsection*{Magnetization and susceptibility}
The field-induced phase transition changes the magnetic order of the system and the properties of the ground state. Here, we study this by calculating the magnetization $m^\alpha$ and the static susceptibility $\chi^\alpha(q)$. The first tells us how polarized the spins of the system have become, and the latter how easy it is to further magnetize the system.

For the finite open chain we define the magnetization across each channel by
\begin{equation}\label{eq:mag}
m^\alpha    = \frac{1}{L} \sum_j \braket{S^\alpha_j} \ .
\end{equation}

\begin{figure}[h!]
    \centering
    \includegraphics[width=0.95\linewidth]{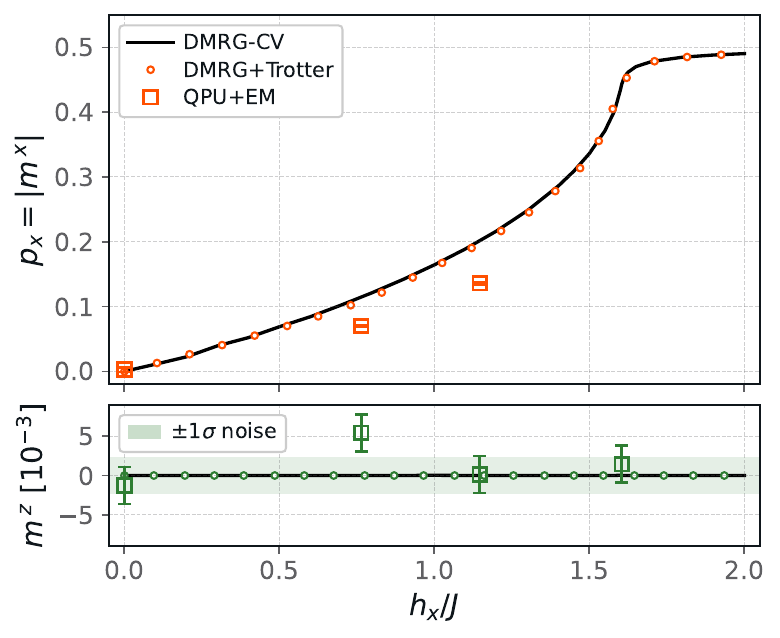}
    \caption{Uniform magnetization $m^\alpha=\frac{1}{18}\sum_{j=9}^{26} \braket{S_j^\alpha}$ per site averaged over $18$ interior sites $j \in \{9, ..., 26\}$ to avoid edges effects, evaluated at $t=0$. The solid black curve
    uses the DMRG-CV ($L=100$ sites) results and the solid gray curve uses the DMRG+Trotter ($L=36$ sites) results; the error mitigated hardware data are shown at the four fields $h_x/J=(0,\ 0.764, \ 1.146, \ 1.604)$ with  hollow squares. Top:
    $p^x=|m^x|$ vanishes at $h_x=0$, grows with the field, rises steeply through
    $h_c=1.604J$ and saturates towards the fully polarised value $0.5$ ($p^x=0.49$ at $h_x=2J$), with error mitigated values $p^x=(0.0027 \pm 0.0037, \ 0.07\pm 0.0017,\ 0.1359\pm0.0016)$ for the three first $h_x$ values. These errors bars are not explicitly visible in the figure due to the $y$ axis scale. Bottom: The magnetization $m^z$ in units of $10^{-3}$. Both 
    DMRG-CV and DMRG+Trotter results sit on zero, and the hardware values
$(-1.3,\,+5.4,\,+0.1,\,+1.4)\times10^{-3}$ at $h_x/J=(0,\ 0.764, \ 1.146, \ 1.604)$ are consistent within $\pm1\sigma=\pm2.4\times10^{-3}$ (except from the $0.764J$ point with $2.3\sigma$).}
    \label{fig:magnetization}
\end{figure}

With the field convention in Eq.~\eqref{eq:xxzhamiltonian}, positive $h_x$ favors negative signed magnetization $m^x$. In Fig.~\ref{fig:magnetization} and the discussion below, we report its magnitude, $p_x=|m^x|$, which approaches $1/2$ in the large-field limit. The $z$ component retains its signed definition.
To calculate the magnetization along the $\alpha=(x,y,z)$ axis we notice that at time $t=0$, the positive and negative kicks, as defined in Eq. \eqref{eq:three-terms}, yield the same expectation value  and we can write
\begin{equation}\label{eq:rgf-mag}
\braket{\sigma^\alpha_j(0)}_+=\braket{\sigma^\alpha_j(0)}_-  = \braket{\sigma^\alpha_j (0)} \ ,
\end{equation}
where the last term in \eqref{eq:rgf-mag} represents a ground state expectation value. From Eqs. \eqref{eq:mag} and \eqref{eq:rgf-mag}, we can write the $\alpha$ axis magnetization as the average
\begin{equation}\label{eq:mag-x}
m^\alpha   = \frac{1}{2 L^*} \sum_{j \in |L^*|} \braket{\sigma^\alpha_j} \ ,
\end{equation}
where $|L^*|$ is the reduced set of chain sites $\{9, ... , 26\}$. In Eq.\eqref{eq:mag-x}, we considered an average over $18$ interior sites to avoid edge-effects. Numerical results for DMRG-CV, DMRG+Trotter and QPU methods are illustrated in Figure \ref{fig:magnetization}. For $m^z$ hardware error bars are the shot-noise standard deviation of the site average $\sigma = 0.0024$, whereas for $p^x=|m^x|$ the error bars in terms of the standard deviations $\sigma = (0.0037, \ 0.0017, \ 0.0016)$ at fields $h_x/J=(0,\ 0.764, \ 1.146)$, are not explicitly visible in the $x\in \{0,0.5\}$ scale.

At $h_x=0$, the chain is in a critical Luttinger liquid phase with antiferromagnetic correlations. The DMRG-CV and DMRG+Trotter results agree with theoretical predictions of $m^\alpha=0$ for all components, whereas error mitigated hardware results give $p^x=0.0027 \pm 0.0037$ and $m^z=-0.0013 \pm 0.0024$.
By the $P_x$ parity of the Hamiltonian, the magnetization along $y$ and $z$ axes are must remain zero even for finite fields. As we increase the external field to $h_x=0.764J $, the system is described by a N\'eel-y state canted towards the transverse field in the negative x-axis. The slowly increasing tilt of the spins towards the external field becomes evident by the increase of the $p^x$ to  $0.07\pm 0.0017$ (for error mitigated hardware results), whereas the $z$ component QPU results are $m^z=0.0054 \pm 0.0024$. At $h_x=1.146J$, the system is stronger canted towards the x-axis, and becomes more magnetized with hardware results of  $p^x=0.1359\pm0.0016$.  At the critical field, $h_x=1.604J$, the chain becomes strongly polarized with magnetization in the $x$ direction from DMRG-CV data equal to $p^x=0.4429$ (nearly $90\%$ of the saturation limit) and magnetization in the $z$ direction from hardware data $m^z=0.0014 \pm 0.0024$.

\begin{figure*}[t!]
    \centering
    \includegraphics[width=1\linewidth]{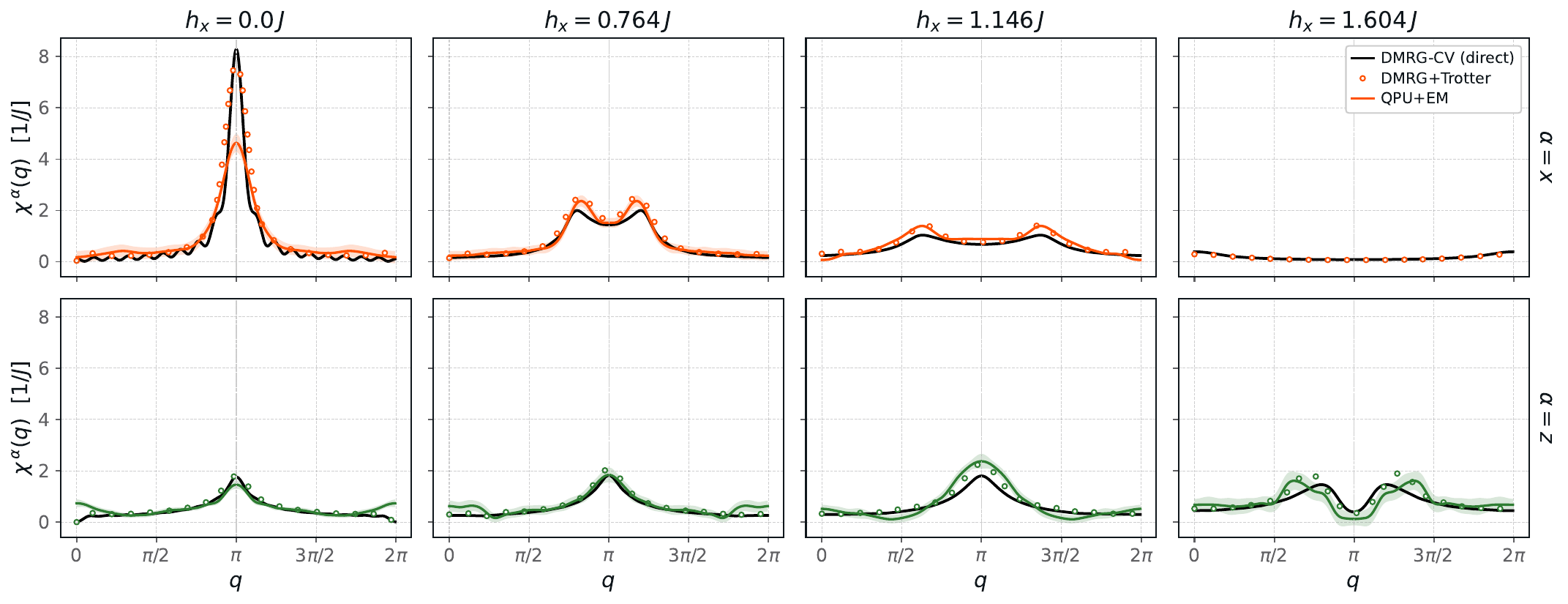}
    \caption{Normalized static susceptibility $\chi^\alpha (q)=2 \int_0^\infty d\omega \frac{S^{\alpha \alpha}(q, \omega)}{\omega}$ of the $L=36$ sites anisotropic $(\Delta=0.25)$ XXZ chain. Rows: different components $\alpha=(x,z)$. Columns: different values of the transverse fields from $h_x/J=(0, \ 0.764J, \ 1.146J, \ 1.604J)$. Each panel is normalized by the sum-rule constant \eqref{eq:sum-rule}. We denote DMRG-CV, DMRG+Trotter, and error mitigated hardware data by solid black, solid gray, and solid colored curves respectively.}
    \label{fig:static-sus}
\end{figure*}

The static susceptibility measures the response of the system to an external magnetic field.At zero temperature $(T=0K)$, it is defined by the Kramers-Kronig relation via 
\begin{equation}\label{eq:sus}
\chi^\alpha(q) = 2 \int_0^\infty d\omega \frac{S^{\alpha \alpha}(q,\omega)}{\omega} \ .
\end{equation}

For the numerical results, we considered the following cutoff frequencies: a minimum frequency $\omega_{min}=0.02 J$ for all transverse fields, and a maximum frequency $\omega_{max}=5.24J$ for $h_x=0$ or $\omega_{max}=2.62J$ for $h_x \neq 0$.
Each channel is then normalized by dividing $\chi^a(q)$ with the ``sum-rule" normalization constant $\mathcal{N}_{sr}$ 
\begin{equation}\label{eq:sum-rule}
\mathcal{N}_{sr}=\frac{\frac{1}{L} \sum_q \int_0^\infty d\omega S^{\alpha\alpha}(q,\omega)}{\frac{1}{4} \ - \ (m^\alpha_{order})^2} \ ,
\end{equation}
with $(m^\alpha_{order})^2= (m^\alpha)^2 +(m^\alpha_{st})^2 $ and the results are summarized in Figure \ref{fig:static-sus}.
 
At zero field $h_x=0$, we observe a peak at $q=\pi$ for both components $(x,z)$ confirming the antiferromagnetic nature of the system in the Luttinger liquid phase, as the system gives a stronger response to a staggered external field. Due to the anisotropy factor $(\Delta=0.25)$, the antiferromagnetic correlations along the $z$ axis are weaker and the response of the $\chi^z$ is smaller compared to the in-plane $(x,y)$ components. For the hard axis ($\chi^z(q=\pi)$-component) we find peak with values $1.578/J$, $1.802/J$, and  $1.461/J\pm 0.146/J$ for the DMRG-CV, DMRG+Trotter, and error mitigated hardware methods.  On the other hand, for the easy plane  $\chi^{x}(q=\pi)$, we find response strength peaks $5.227/J$, $7.749/J$,  and  $4.630/J \pm 0.346/J$ respectively, with larger QPU hardware accumulated error due to late times (low frequency) dependency. Both the compression and the QPU performance are qualitative accurate with a Pearson correlation coefficient (between DMRG-CV and hardware) above $0.998$ for the $x$ axis and $0.828$ for the $z$ axis.

As we increase the field, at value $h_x=0.764 J$ and $h_x=1.146J$, the system becomes magnetized in the $x$ direction with a N\'eel-y canted ground state. The central peak of the $x$ component splits into two peaks with positions and shifts at $q=(1\pm 0.17)\pi $ and $q=(1\pm 0.32) \pi$, respectively, with a $0.01 \pi$ mismatch between DMRG-CV and hardware results. The displacement of the two peaks from $q=\pi$ is set by the magnetization magnitude $p_x$, with peak locations $q=(1\pm2p_x)\pi$. The spike of $z$ component at $q=\pi$, confirms the N\'eel $y$ order of the system and the antiferromagnetic correlations. Across all channels we find qualitative agreement of hardware, DMRG+Trotter, and DMRG-CV results with Pearson coefficient above $0.951$.

At the critical field, $h_x=1.604J$, $\chi^x(q)$ becomes nearly flat with two small peaks at $q=0$ and $q=2\pi$. The spins are strongly polarized along negative $x$, with the magnetization magnitude $p_x$ approaching the saturation value $1/2$. At criticality, the staggered magnetization along $y$ decreases and the susceptibility diverges (at the thermodynamic limit) \cite{dmitriev2002gap}. The $\chi^z(q)$ component has a dip at $q=\pi$ as a staggered field is more difficult to induce spin fluctuations in the $z$ axis compared to the $y$. Hardware runs for the $z$ component capture these features with Pearson coefficient with the DMRG-CV method equal to $0.929$. 

\subsection*{Quantum Entanglement Measures}\label{subsec:QFI}
We examine the field dependence of quantum correlations through the quantum Fisher information (QFI), the one-tangle, and the two-tangle~\cite{laurell2021quantifying}. For an equilibrium state, the QFI density in spin channel $\alpha$ is related to the dissipative susceptibility by
\begin{equation}\label{eq: qfi}
    f_Q^\alpha(q,T)=\frac{4}{\pi}\int_0^\infty d\omega\,
    \tanh\!\left(\frac{\omega}{2k_BT}\right)
    \chi''_{\alpha\alpha}(q,\omega),
\end{equation}
where, in the frequency convention used here,
\begin{equation}\label{eq: dyn-sus}
    \chi''_{\alpha\beta}(q,\omega)
    =\pi\left(1-e^{-\omega/(k_BT)}\right)
    S^{\alpha\beta}(q,\omega).
\end{equation}
At $T=0$, this gives
\begin{equation}\label{eq: qfi-zero}
    f_Q^\alpha(q)=4\int_0^\infty d\omega\,
    S^{\alpha\alpha}(q,\omega).
\end{equation}
For the numerical results, we used the previously defined frequency thresholds; $\omega_{min}=0.02 J$ for all transverse fields, and $\omega_{max}=5.24J$ for $h_x=0$ or $\omega_{max}=2.62J$ for $h_x \neq 0$.
For the spin $1/2$ collective generators at $q=0$ and $\pi$, a correctly normalized QFI density $f_Q^\alpha>m$ witnesses at least $(m+1)$-partite entanglement. This is a lower bound on entanglement depth, and falling below a threshold does not establish its absence. The reconstructed spectra are normalized using the channel sum rule in Eq.~\eqref{eq:sum-rule} before evaluating Eq.~\eqref{eq: qfi-zero}. This removes a common multiplicative amplitude factor, but does not remove uncertainty from the spectral reconstruction or the normalization inputs. We therefore interpret the results below as estimates inferred from the reconstructed ground state response; they do not by themselves certify the entanglement of the noisy states prepared on hardware. 

\begin{figure*}[t!]
    \centering
    \includegraphics[width=\linewidth]{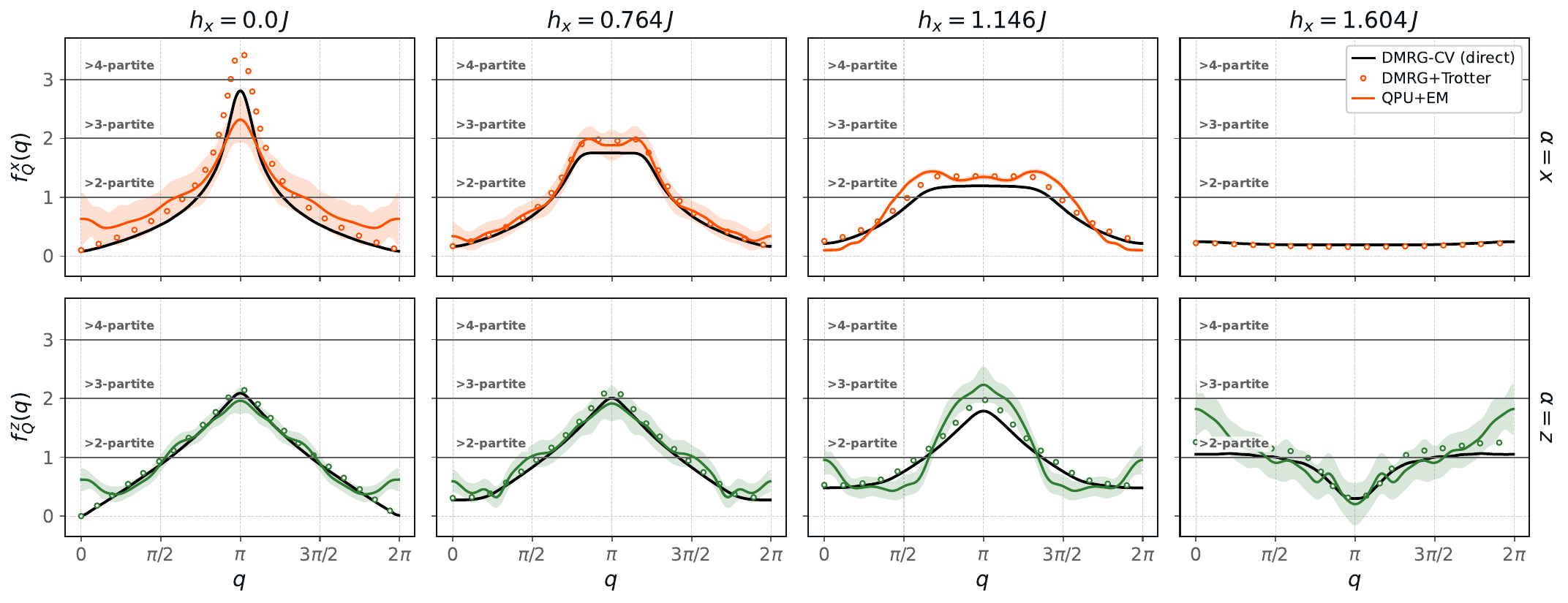}
    \caption{Quantum Fisher information density
$f_Q^{\alpha}(q)=4\int_{0}^{\infty}d\omega\,S^{\alpha\alpha}(q,\omega)$ normalized by the ``sum-rule" constant of Eq.~\eqref{eq:sum-rule} for the
$L=36$ anisotropic ($\Delta=0.25$) XXZ chain. Rows show the $x$ and $z$ components; columns correspond to $h_x/J=(0,\,0.764,\,1.146,\,1.604)$.  We denote DMRG-CV, DMRG+Trotter, and error mitigated hardware data by solid black, solid gray, and solid colored  curves respectively. Hardware results include bands with the quoted one standard deviation. Horizontal gray lines mark $f_Q^{zz}=1$ and $2$, the bipartite and tripartite witness thresholds under the normalization and state assumptions discussed in the text. The field dependence of the curves shows the suppression of the response at $q=\pi$ and its redistribution toward the zone center.}
    \label{fig:qfi}
\end{figure*}

Figure~\ref{fig:qfi} compares the $x$ and $z$ components across the field sweep. At zero field, the response peaks at the antiferromagnetic wavevector. For DMRG-CV, DMRG+Trotter, and error mitigated hardware, respectively, $f_Q^x(\pi)=(3.074,\ 3.521, \ 2.321 \pm 0.363)$ and $f_Q^z(\pi)=(2.159,\ 2.160, \ 1.961 \pm 0.193)$. Under the witness assumptions above, these central values correspond to lower bounds of $(4,4,3)$ and $(3,3,2)$ entangled spins, respectively. For $h_x/J=(0.764,\,1.146, \ 1.604)$ the DMRG-CV $f_Q^x(\pi)=(1.982,\,1.402,\,0.175)$ and DMRG+Trotter results $f_Q^x(\pi)=(1.961,\,1.362,\,0.157)$ decrease with field: the first two exceed the bipartite threshold, whereas the last does not witness entanglement at $q=\pi$. 

The hardware estimates of $f^x_Q$ at $q=\pi$ are $2.321\pm0.363$, $1.884\pm0.270$
and $1.349\pm0.050$ for $h_x/J=(0,\,0.764,\,1.146)$. All three exceed $f^x_Q=1$ (by $3.6\sigma$, $3.3\sigma$, $7.0\sigma$), and the
zero-field value exceeds $f^x_Q=2$ by $0.9\sigma$. The QPU+EM and DMRG-CV discrepancy peak at $h_x=0$ mostly comes from low-frequency (late-times) as $84\%$ of the gap comes from energies $0.025 J < \omega <0.5 J$.
The hardware estimates of $f_Q^z$ at $q=\pi$ are $1.961\pm0.193$, $1.911\pm0.335$, $2.231\pm0.313$, and $0.205\pm0.397$ for $h_x/J=(0,\,0.764,\,1.146,\,1.604)$, respectively. The quoted uncertainties include only shot noise and no other systematic errors.  The first three exceed $f_Q=1$ relative to their uncertainties (by $5\sigma, \ 2.7 \sigma, \ 3.9 \sigma $), while only the $h_x=1.146J$ central value exceeds $f_Q=2$, by about $0.7$ quoted standard deviations. Near the expected critical field, the hardware maximum moves to $q=0$, where it reaches $1.82$. The ratio $f_Q^{zz}(0)/f_Q^{zz}(\pi)$ changes from $0.32$, $0.31$, and $0.43$ at the three lower fields to $8.7$ at the highest field. These results track a redistribution of the inferred QFI from the antiferromagnetic wavevector toward the zone center.  

The concurrence $\mathcal{C}(r)$ is an entanglement monotone measure used to characterize the system by focusing on correlators of two spins. For the Hamiltonian Eq. \eqref{eq:xxzhamiltonian}, it takes the form
\begin{equation} \label{eq:con0}
\mathcal{C}(r) = 2 \max\{{0, \mathcal{C}^{'}_r, \mathcal{C}^{''}_r }\} \ ,
\end{equation}
with
\begin{equation}\label{eq:con1}
\mathcal{C}^{'}_r = \vert{s^y_r+s^z_r }\vert -\sqrt{\left({\frac{1}{4}+s^x_r}\right)^2- (m^x)^2} \ ,
\end{equation}
\begin{equation} \label{eq:con2}
\mathcal{C}^{''}_r =s^x_r + \vert{s^y_r-s^z_r }\vert -\frac{1}{4} \ .
\end{equation}
To calculate Eqs. \eqref{eq:con0}--\eqref{eq:con2}, we introduce the real-space static spin correlators
\begin{equation}
s^\alpha_r = \braket{S_j^\alpha(0) S_{j+r}^\alpha(0)} \ .
\end{equation}
The one-tangle is a property that characterizes the entanglement between one spin site with the rest of the system. Mathematically, its defined as
\begin{figure*}[t!]
    \centering
    \includegraphics[width=0.8\linewidth]{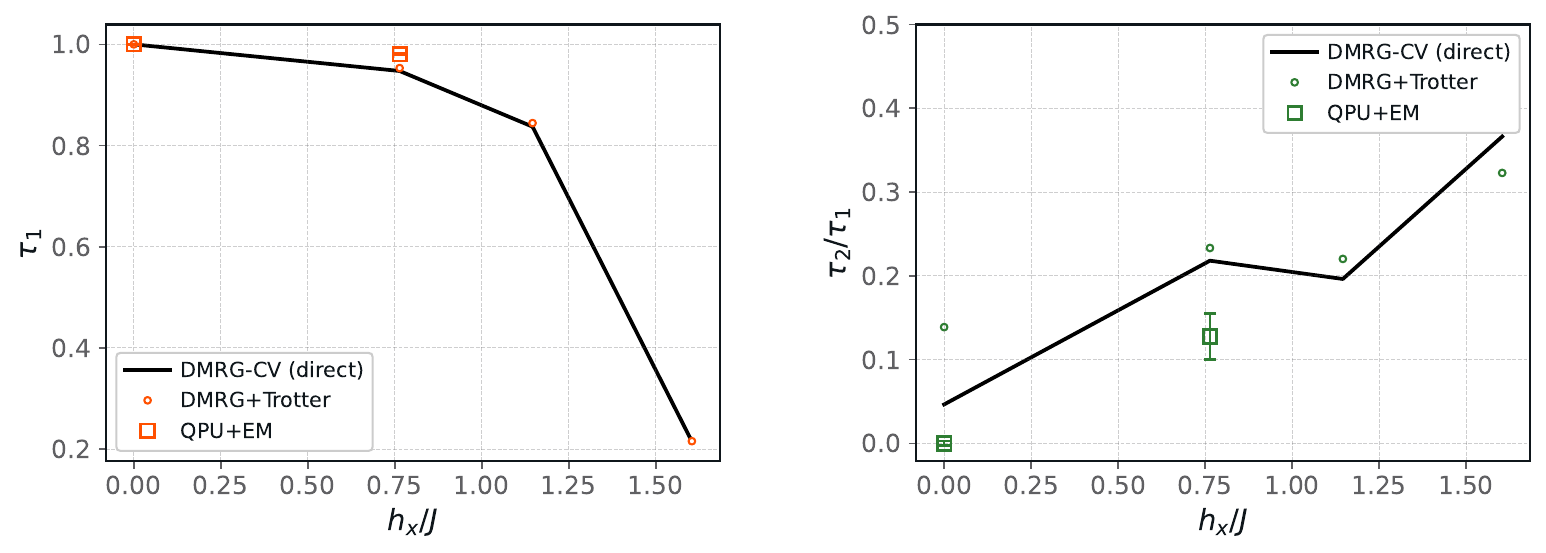}
    \caption{One-tangle (left panel) and ratio of two-tangle over one-tangle (right panel) against the field, evaluated from the equal-time correlator at $t=0$. We denote DMRG-CV, DMRG+Trotter, and error mitigated hardware data by solid black, solid gray, and solid colored curves respectively. As we increase the field , we notice a drop of $\tau_1$ and an increase of $\tau_2/\tau_1$ signaling a transition from multi-partite to bi-partite correlations. The hardware data qualitatively agrees with the DMRG-CV and DMRG+Trotter methods showing the same increasing trend of the ratio $\tau_2/\tau_1$ with the field.}
    \label{fig:tangles}
\end{figure*}
\begin{equation}
\tau_1 = 1-4 \sum_\alpha \braket{S^\alpha_j} ^2 \ .
\end{equation}

The previous definition of the one-tangle is exact in the thermodynamic limit. For a system with open boundary conditions, we compute the one-tangle by choosing $j=j_c$ and by averaging over many translations. Next, we can also compute the two-tangle, which tracks the bi-partite correlations.
\begin{equation}
    \tau_2= 2 \sum_{r=1}  \mathcal{C}^2(r) \ .
\end{equation} 
Numerical results for the one-tangle and two-tangle are shown in Figure \ref{fig:tangles}. The system starts from $\tau_1=1$ at a maximally entangled state and each individual spin is highly entangled with the rest of the system. As we increase the transverse field, the one-tangle degrades and the two-tangle starts increasing, and the entanglement structure transforms from multi-partite to bi-partite. This trend is also supported by the QFI minimum bound from four-particle to two-particle correlations. We observe that hardware data reliable calculate the one-tangle with values ($1 \pm 0.0$, $0.9804\pm 0.0009$) compared to DMRG-CV results ($1.0$, $0.9474$) at fields $h_x/J=(0, \ 0.764)$. On the other hand, the two-tangle is more fragile and cannot be used for reliable characterization of quantum entanglement in near-term quantum devices. Here, we inspect the rise of the two-tangles by plotting the repair ratio $\tau_2/\tau_1$ with hardware values ($0.0001\pm0.0029$, $0.1276\pm0.0275$) as opposed to DMRG-CV values ($0.1413$, $0.2752$). 

\subsection*{Local density of states}

\begin{figure*}[htbp!]
    \centering
    \includegraphics[width=1\linewidth]{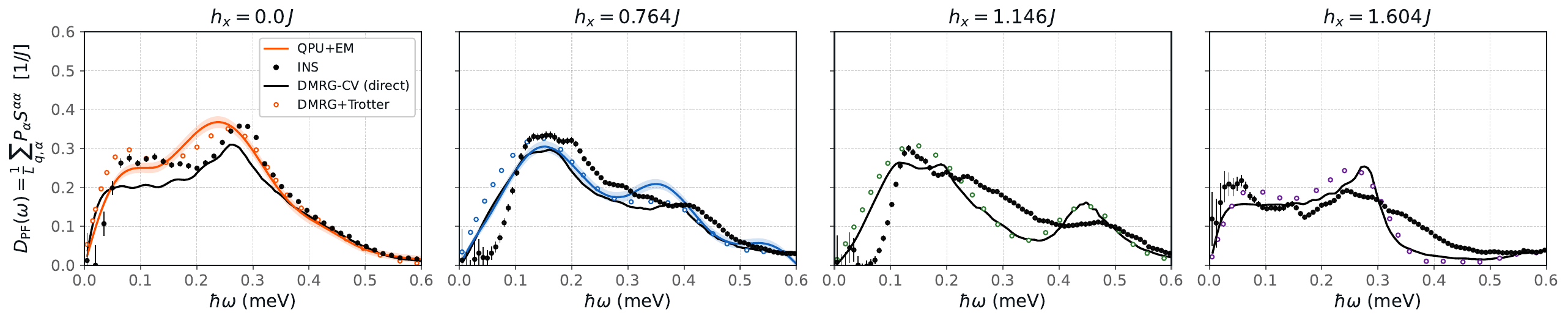}
    \caption{Polarization-weighted local density of states $D_{\rm PF}(\omega)=\frac{1}{L}\sum_{q}\sum_{\alpha}P_\alpha(q,\omega)\,S^{\alpha\alpha}(q,\omega)$ of the $L=36$ site anisotropic ($\Delta=0.25$) XXZ chain at field values $h_x/J=(0, \ 0.764J, \ 1.146J, \ 1.604J)$. The polarization factors
$P_\alpha$ are those of the neutron experiment, so $D_{\rm PF}$ is the
$k$-averaged intensity a neutron probe sees rather than the plain sum of the three channels; each
$S^{\alpha\alpha}$ is normalized to its sum rule before weighting. We denote DMRG-CV, DMRG+Trotter, and error mitigated hardware data by solid black, solid gray, and  solid colored curves respectively. Hardware results include bands with the quoted one standard deviation. Inelastic neutron scattering data from \cite{laurell2021quantifying} are represented by black bullet points.}
    \label{fig:dos}
\end{figure*}

To further study the spectrum of Cs$_2$CoCl$_4$ we study the energy distribution of its magnetic excitations. The kick operator mimics neutron scattering on the quantum magnet, and the polarization-weighted density of states (DOS) is given by
\begin{equation} \label{eq: pol-loc-dos}
D_{\rm PF}^\alpha(\omega) =  \frac{1}{L} \sum_q \sum_\alpha P_\alpha(q,\omega ) S^{\alpha \alpha}(q, \omega)
\ , 
\end{equation}
where for polarization factor $P_\alpha$ we used the data defined in Ref. \cite{laurell2021quantifying}. A detailed comparison between INS, DMRG+Trotter, DMRG-CV  and hardware data can be found in Figure
\ref{fig:dos}. 

At $h_x=0$, the chain is a gapless Luttinger liquid and the kick operator (neutron scattering) creates a two-spinon continuum in the $S_{\text{tot}}(q,\omega)$. By summing over all momenta $q$, we find two a smooth curve with two peaks at $0.12\pm 0.02  \ meV$ and $ 0.281 \pm 0.004 \ m eV$ from the INS data. The envelope of the two peaks $(0,0.26 \ meV)$ and $(0.26 \ meV, 0.53 \ m eV)$ correspond to the lower and upper edge of the spinon continuum. Hardware results cannot resolve the position of the 1st peak and the 2nd peak is found at $0.241\pm 0.006 \ meV$, whereas the Pearson similarity between QPU+EM and DMRG-CV data is greater than $0.99\pm 0.006$. As we increase the field, and below criticality, the two peaks become further separated as the continuum disperses. In the this regime, our hardware data have great qualitative overlap with the INS and DMRG-CV data with a Pearson coefficient of $0.978\pm 0.01$ between QPU+EM and DMRG-CV. At the critical point, all collected data start from a finite valued DOS at zero-frequency reflecting to the gapless critical state.

\end{document}